\documentclass[12pt,fleqn,notitlepage]{report}

\newcommand{\ie}{\emph{i.e.}}
\newcommand{\eg}{\emph{e.g.}}

\usepackage[backend=biber,style=ieee]{biblatex}
\usepackage{fullpage}
\usepackage{amsmath,amssymb}
\usepackage{amsthm}
\usepackage{accents}
\usepackage{latexsym}
\usepackage{mathrsfs}
\usepackage{mathtools}
\usepackage{bm}
\usepackage{siunitx}
\usepackage{mdframed}\mdfsetup{innertopmargin=0pt}
\usepackage{tabularx}
\usepackage{booktabs}
\usepackage{multirow}
\usepackage{subcaption}
\usepackage{stmaryrd}
\usepackage{url}

\DeclareMathOperator*{\minimize}{minimize}
\newcommand*{\diff}{\mathinner{}\!\mathrm{d}}
\newcommand*{\placeholder}{\mathinner{\cdot}}
\newcommand*{\suchthat}{\mathinner{\vert}}

\newcommand*{\im}{i}
\newcommand*{\vol}{V}
\newcommand*{\surf}{S}
\newcommand*{\dotproduct}{\cdot}
\newcommand*{\wavenumber}{k}
\newcommand*{\sphharm}{Y}

\theoremstyle{definition}
\newtheorem{theorem}{Theorem}
\newtheorem*{problem*}{Problem}
\newtheorem*{remark*}{Remark}

\begin{document}

\title{Sound Field Estimation: Theories and Applications}
\author{Natsuki Ueno \and Shoichi Koyama}
\maketitle

\begin{abstract}
  The spatial information of sound plays a crucial role in various situations, ranging from daily activities to advanced engineering technologies. 
  To fully utilize its potential, numerous research studies on spatial audio signal processing have been carried out in the literature. 
  Sound field estimation is one of the key foundational technologies that can be applied to a wide range of acoustic signal processing techniques, including sound field reproduction using loudspeakers and binaural playback through headphones. 
  The purpose of this paper is to present an overview of sound field estimation methods. 
  After providing the necessary mathematical background, two different approaches to sound field estimation will be explained. 
  This paper focuses on clarifying the essential theories of each approach, while also referencing state-of-the-art developments. 
  Finally, several acoustic signal processing technologies will be discussed as examples of the application of sound field estimation. 
\end{abstract}

\chapter{Introduction}
\label{sec:introduction}

\section{Background}
\label{sec:background}

Sound is one of the most commonly used media in all kinds of human activities, including  human (or human--robot) communication, the analysis of materials and environments, and art-related activities. 
In many of these situations, the spatial information of sound plays an essential role as well as temporal information. 
For example, in the localization of sound sources, humans benefit from the interaural difference between sound signals received by both ears without depending much on the temporal waveform of the source signal, and further theoretical and experimental investigations demonstrated the additional effectiveness of head movement during sound source localization~\cite{Wallach:JASA1939, Thurlow:JASA1967, Wightman:PsychologyPress1997, Stern:Wiley2006}. 
Such importance of the spatial information of sound has stimulated the widespread research and development of audio signal processing technologies for the analysis and control of spatial acoustics. 

\begin{figure}
  \centering
  \includegraphics[width=0.7\linewidth]{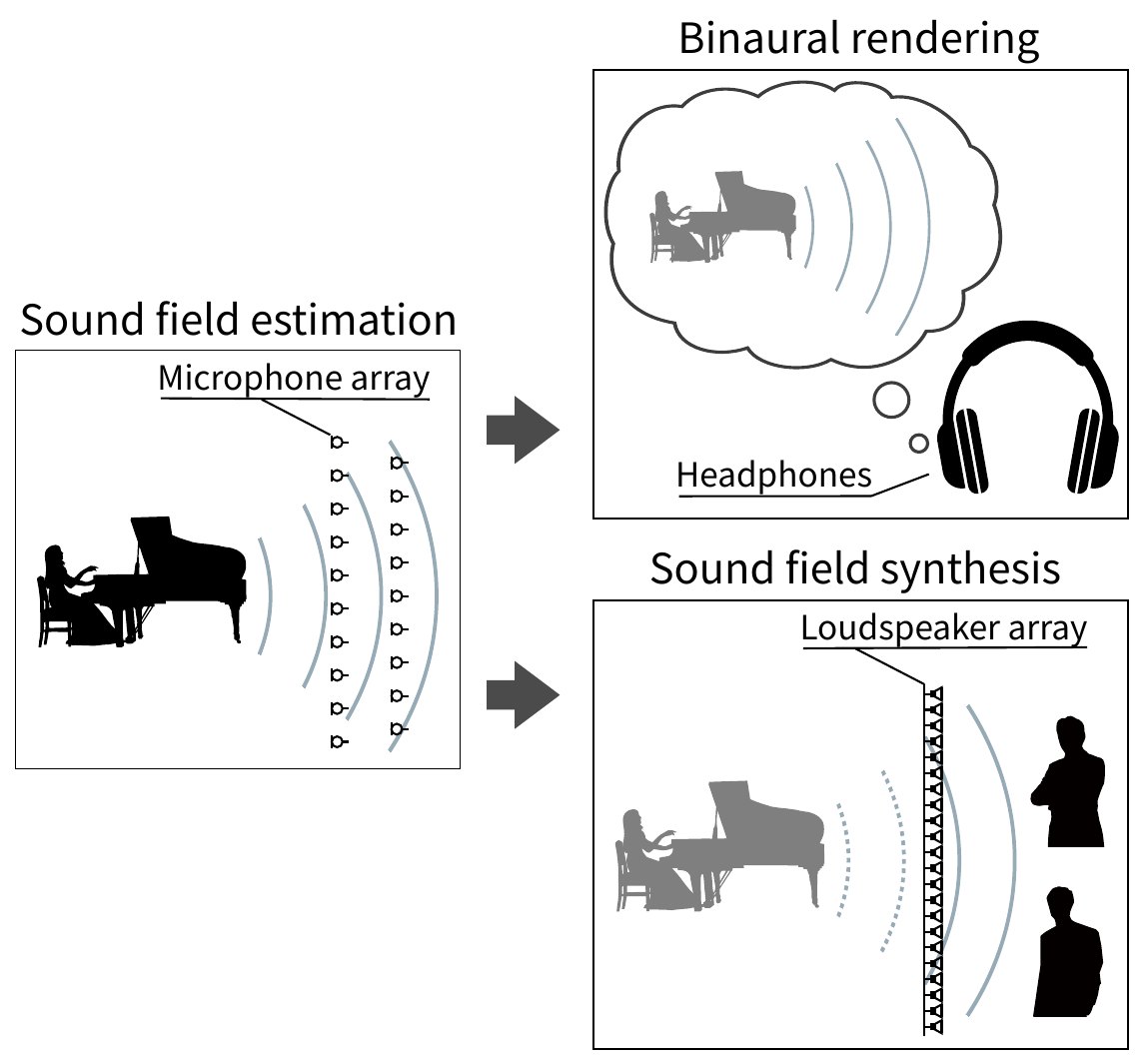}
  \caption{Sound field estimation and its applications.}
  \label{fig:application_example}
\end{figure}

As a direct attempt to obtain the spatial information of sound, sound field estimation, also called sound field reconstruction, measurement, or recording depending on the context, has been a fundamental technique under intensive investigation. 
The purpose of sound field estimation is to estimate the spatio-temporal distribution of sound pressure within a target region from the data obtained by multiple sensors, \ie, microphones. 
In this context, it should be emphasized that the parameters related to the process of generating a sound field, such as the positions of sound sources or the strength of reverberation, are not subject to estimation; a sound field can be directly represented as it is without strong assumptions on or simplifications of these factors. 
By combining other signal processing applications, sound field estimation enables more than just the reconstruction of the sound pressure at an arbitrary position (see Fig~\ref{fig:application_example}). 
One such example is binaural reproduction~\cite{Duraiswami:AES2005, Menzies:JASA2007, Ben-Hur:IEEE2021, Iijima:JASA2021, Ahrens:JASA2019, Rafaely:Acta2022}, which aims to reproduce the sound that someone would hear if they were present in the target sound field, including the complex effects of reflection and diffraction caused by the head. 
Using this technique, one can appreciate, for example, an orchestra music recorded in a concert hall anywhere by a headphone with a high (ideally complete) degree of fidelity. 
Compared with the direct playback of binaural signals recorded with a dummy head, the binaural reproduction from the estimated sound field allows for various post-processing steps \emph{after} the recording, such adapting individual head-related transfer functions (HRTFs) and rendering with head tracking. 
There are also other applications, such as sound field synthesis using multiple loudspeakers~\cite{Berkhout:JASA1993, Daniel:AESConvention2003, Daniel:AES2003, Poletti:JAES2005, Betlehem:JASA2005, Ahrens:Acta2008, Wu:IEEE2009, Ueno:IEEE2019}, spatial active noise control~\cite{JZhang:IEEE2018, Bu:IEEE2018, WZhang:IEEE2018, Maeno:IEEE2020, Koyama:IEEE2021}, the analysis or visualization of room acoustic condition~\cite{Donovan:ICASSP2008, Khaykin:JASA2012, Tervo:IEEE2015, Ribeiro:IEEE2022}. 

A sound field can be essentially regarded as a scalar field, \ie, a function from the space and time/frequency variables to sound pressure.
However, the estimation of a sound field is distinguished from a simple estimation or interpolation of a function in the common context of machine learning in several aspects.
The most distinctive aspect of the sound field estimation problem is the existence of the constraint due to the physical properties of a sound field.
This constraint is described typically by the acoustic wave equation or the Helmholtz equation~\cite{Williams:AcademicPress1999}. 
In addition, there are also distinctive characteristics in the observation of a sound field.
First, unlike a common interpolation problem of a function, the observation of a sound field is not necessarily limited to the sampling of sound pressure. 
This is because a microphone generally has a nonuniform frequency response or directivity.
Actually, a specific (nonuniform) directivity can even have its advantages over the uniform directivity in some cases, which will be described in Section~\ref{sec:boundary} in detail.
On the other hand, the observation of a sound field can be regarded generally as a linear time-invariant system, regardless of the frequency response or directivity of the microphone. 
This linear time-invariant property makes it easy to analyze an observation and estimation of a sound field in the frequency domain. 
Thus, how to deal with these generalities and specificities in a technically tractable way is and will always be a critical problem in sound field estimation. 

From the historical viewpoint, the exact origin of sound field estimation methods is difficult to identify, to the best of our knowledge.
In 1985 at the latest, the pioneering idea of sound field estimation referred to as the near-field acoustic holography was investigated in the field of audio engineering by Maynard et al.~\cite{Maynard:JASA1985}, which was also developed as a sound field synthesis technique called the wave field synthesis~\cite{Berkhout:JASA1993}, although similar theories were developed earlier for the optical field~\cite{Wolf:Elsevier1967, Lalor:AIP1968}.
Their approach was based on the Kirchhoff--Helmholtz integral theorem and Rayleigh's formula, both of which are boundary integral representations that relate a sound field---or, more precisely, a solution of the Helmholtz equation---to its values on a given surface. 
The use of these theorems has so far formed the basis for many sound field estimation methods~\cite{Berkhout:JASA1999, Hulsebos:JAES2002, Daniel:AESConvention2003}.
Ambisonics, developed by Gerzon~\cite{Gerzon:JAES1973} (referred to as ``periphony'' therein), is another type of pioneering approach to sound field estimation.
The original work~\cite{Gerzon:JAES1973} dealt with the directional component of a sound at one position, not a sound field in the direct form.
However, it was later redeveloped as the higher-order ambisonics with theoretical and practical improvements~\cite{Meyer:ICASSP2002, Abhayapala:ICASSP2002, Daniel:AESConvention2003, Rafaely:IEEE2005, Poletti:JAES2005}, where a sound field was explicitly reconstructed from the signals observed by a spherical microphone array. 
In the higher-order ambisonics, a sound field is analyzed via its local expansion using the spherical wave functions, which is seemingly different from the approaches based on the Kirchhoff--Helmholtz integral theorem and Rayleigh's formula. 
However, both approaches basically require the boundary measurement of the sound field by microphones with specific directivities, and indeed they can be interpreted in a unified manner, as pointed out by Daniel et al.~\cite{Daniel:AESConvention2003} and Poletti~\cite{Poletti:JAES2005}. 
These unified theories are also called acoustic holography~\cite{Williams:AcademicPress1999}. 
In 2003, a new approach was proposed by Laborie et al.~\cite{Laborie:AESConvention2003}, which no longer requires boundary measurement but allows arbitrary positions and directivities of the microphones. 
The main idea of this approach lies in the vector and matrix representations of the sound field and observation, respectively, and the same or a similar idea is used in many of the current sound field estimation methods~\cite{Poletti:JAES2005, Samarasinghe:IEEE2014, Ueno:IEEE2018}.
Whereas all methods mentioned above do not rely on any specific assumptions on the target sound field, several methods that utilize prior information on a target sound field, such as the approximate source direction~\cite{Ueno:IEEE2021} or sparsity of the source distribution~\cite{Mignot:IEEE2013, Antonello:IEEE2017, Verburg:JASA2018, Murata:IEEE2018, Bertin:Springer2015}, have also been proposed to improve the estimation accuracy. 
Moreover, several studies on learning-based sound field estimation have been conducted in recent years~\cite{Lluis:JASA2020, Shigemi:IWAENC2022, Cobos:EURASIP2022, Koyama:IEEE2024} to further improve performance, the details of which are beyond the scope of this paper. 

Practical situations in sound field estimation have also changed significantly over the last few decades. 
In many of the early studies, the estimation methods were demonstrated primarily through numerical simulations, with few exceptions~\cite{Maynard:JASA1985, Berkhout:JASA1999} because of the difficulties in the implementation of a large number of microphones and analog-to-digital converters with a large number of synchronized channels. 
In recent years, however, the practical implementation and commercialization of microphone arrays having a large number of channels have been realized~\cite{Duraiswami:AES2005, Jin:IEEE2014} owing to the development of multichannel analog-to-digital converters and the improvement of the miniaturization and integration technologies for microphones. 
One example of a commercially available microphone array is the em64 Eigenmike$^\text{\textregistered}$, which is a spherical microphone array of \SI{84}{mm} diameter equipped with 64 microphones. 
Such an industrial background implies that sound field estimation techniques and their applications mentioned above are now available in almost any situation, in principle, and studies on sound field estimation methods are gaining increasing attention towards further performance improvement.

\section{Purpose of This Paper}

The purpose of this tutorial paper is to provide basic and advanced theories on sound field estimation, focusing on how the physical constraints of sound fields are incorporated into the estimation methods, as well as to introduce several signal processing applications of sound field estimation. 
In this paper, the sound field estimation methods are grouped into two types: one with boundary measurement and the other with discrete measurement, and they are described with the results of numerical experiments. 
It should be noted that the latter is not a simple generalization of the former because they also differ in terms of their background theories. 
Even though the estimation methods with discrete measurement have an advantage in practical feasibility, those with boundary measurement are also beneficial in terms of their rich theoretical implications. 
For the full understanding of these approaches, we provide the required mathematics, especially on the Helmholtz equation, with due care of technical correctness. 
Our main focus lies in the fundamental ideas of the above approaches, but the state-of-the-art methods are also discussed briefly with extensive references. 
Finally, three applications of sound field estimation are presented: binaural sound reproduction, sound field synthesis with loudspeakers, and active noise cancellation. 

\section{Outline of This Paper}

This paper is organized as follows.
We begin in Section~\ref{sec:overview} with the definition of the sound field estimation problem of interest in this paper.
Section~\ref{sec:preliminaries} provides mathematical preliminaries used in later sections. 
Readers who are not interested in theoretical details can skip this section and refer to it later if required. 
In Sections~\ref{sec:boundary} and \ref{sec:discrete}, we describe two different approaches to sound field estimation based on boundary measurement and discrete measurement, respectively.
We present signal processing applications of sound field estimation in Section~\ref{sec:applications}, and finally in Section~\ref{sec:conclusions}, we conclude this paper.

\section{Symbols and Notations}

Basic mathematical symbols and notations are listed in Table~\ref{tab:notations}.
By convention, $\mathbb{R}^{n \times 1}$ and $\mathbb{C}^{n \times 1}$ are regarded as identical to $\mathbb{R}^{n}$ and $\mathbb{C}^n$, respectively ($n \in \mathbb{N}$); for instance, for $\bm{a} \in \mathbb{C}^n$, $\bm{a}^\mathsf{T}$ and $\bm{a}^\mathsf{H}$ are $1 \times n$ matrices.
Similarly, $\mathbb{R}^{1}$ and $\mathbb{C}^{1}$ are regarded as identical to $\mathbb{R}$ and $\mathbb{C}$, respectively.
For an inner product space over $\mathbb{C}$, the inner product is defined such that it is antilinear with respect to the first variable and linear with respect to the second variable. 

\begin{table}
  \caption{Notations}
  \label{tab:notations}
  \begin{tabularx}{\textwidth}{p{0.15\textwidth}X}
    \toprule
    \multicolumn{2}{l}{\underline{Numbers}:}                                                                                                                              \\
    $\mathbb{N}$              & set of natural numbers (including $0$)                                                                                                    \\
    $\mathbb{Z}$              & set of integers                                                                                                                           \\
    $\mathbb{R}$              & set of real numbers                                                                                                                       \\
    $\mathbb{C}$              & set of complex numbers                                                                                                                    \\
    $\im$                     & imaginary unit in $\mathbb{C}$                                                                                                            \\
    $z^\ast$                  & complex conjugate of $z \in \mathbb{C}$                                                                                                            \\
    $\llbracket a, b \rrbracket$                  & set of all integers between $a$ and $b$ included ($a, b \in \mathbb{Z}$)            \\
    \multicolumn{2}{l}{\underline{Linear algebra} ($\mathbb{K} \in \{\mathbb{R},\mathbb{C}\}, \ m,n\in\mathbb{N}$):}                                                  \\
    $\mathbb{K}^n$            & $n$-dimensional coordinate space over $\mathbb{K}$                                                                                        \\
    $\mathbb{K}^{m \times n}$ & set of $m \times n$ matrices over $\mathbb{K}$                                                                                            \\
    $\bm{A}^\mathsf{T}$   & transpose of $\bm{A} \in \mathbb{K}^{m \times n}$                                                                                     \\
    $\bm{A}^\mathsf{H}$   & conjugate transpose of $\bm{A} \in \mathbb{K}^{m \times n}$                                                                           \\
    $\bm{A}^\mathsf{-1}$  & inverse of $\bm{A} \in \mathbb{K}^{n \times n}$ (if exists)                                                                           \\
    $\delta_{a,b}$  & Kronecker's delta ($a, b \in \mathbb{Z}$)                         \\
    \multicolumn{2}{l}{\underline{Vector analysis}:}                                                                                                                      \\
    $\bm{x} \dotproduct \bm{y}$ & dot product between $\bm{x} \in \mathbb{R}^3$ and $\bm{y} \in \mathbb{R}^3$                                                               \\
    $\|\bm{x}\|$              & Euclidean norm of $\bm{x} \in \mathbb{R}^3$                                                                                               \\
    $\mathbb{S}_2$            & unit sphere in $\mathbb{R}^3$                                                                                                         \\
    $\mathrm{SO}(3)$          & rotation group over $\mathbb{R}^3$                                                                                                             \\
    $\mathcal{C}_n(\Omega)$   & set of $n$th continuously differentiable functions from an open set $\Omega \subseteq \mathbb{R}^3$ to $\mathbb{C}$ ($n \in \mathbb{N}$) \\
    $\nabla$                  & vector differential operator (gradient)                                                                                                   \\
    $\Delta$                  & Laplace operator                                                                                                                          \\
    $\partial \Omega$         & topological boundary of $\Omega \subseteq \mathbb{R}^3$                                                                                   \\
    $B(\bm{r},R)$             & open ball centered at $\bm{r} \in \mathbb{R}^3$ with radius $R \in (0, \infty)$ \\
    $\overline{B}(\bm{r},R)$             & closed ball centered at $\bm{r} \in \mathbb{R}^3$ with radius $R \in (0, \infty)$ \\
    $V$                       & volume measure ($3$-dimensional Lebesgue measure)                                                                                         \\
    $S$                       & surface measure ($2$-dimensional Hausdorff measure)                                                                                       \\
    \bottomrule
  \end{tabularx}
\end{table}

\chapter{Overview of Sound Field Estimation Problem}
\label{sec:overview}

To classify and describe the sound field estimation methods, we should first provide an overview of the sound field estimation problem, including fundamental definitions and assumptions. 
In Sections~\ref{sec:sound_field} and \ref{sec:observation}, we define the terms ``sound field'' and its ``observation'' using microphones, respectively, from a mathematical perspective. 
Subsequently, in Section~\ref{sec:problem_statement}, we provide a statement of the sound field estimation problem of interest in this paper.

\section{What Is a Sound Field?}
\label{sec:sound_field}

Vibrating particles in a medium (\eg, air and water) cause the variation in pressure.
The displacement between the instantaneous and equilibrium pressures at a specific position is called the \emph{sound pressure}, and its spatio-temporal distribution is called the \emph{sound field}. 
In mathematical terms, a sound field in the region of interest $\Omega \subset \mathbb{R}^3$ is defined as a function from $\Omega \times \mathbb{R}$ to $\mathbb{R}$, for instance, $U$, where $U(x,y,z,t) \in \mathbb{R}$ denotes the sound pressure at the position $(x,y,z) \in \Omega$ and time $t \in \mathbb{R}$.
We also refer to its frequency-domain representation as the sound field without any special distinction unless clarification is needed. 
In this representation, a sound field is represented as a function from $\Omega \times \mathbb{R} \to \mathbb{C}$, for instance, $u$, where $u(x,y,z,\omega) \in \mathbb{C}$ denotes the sound pressure at the position $(x,y,z) \in \Omega$ and (angular) frequency $\omega \in \mathbb{R}$.
The relationship between the time- and frequency-domain representations, $U$ and $u$, is given by the inverse Fourier transform as 
\begin{align}
  U(x,y,z,t)
  =
  \frac{1}{2\pi}
  \int_{-\infty}^\infty
  u(x,y,z,\omega) \exp(-\im \omega t)
  \diff \omega. 
  \label{eq:inverse_Fourier_transform}
\end{align}
Note that the sign in the term $\exp(-\im \omega t)$ is opposite to that commonly used in signal processing, following conventions in acoustics~\cite{Williams:AcademicPress1999}. 

\subsection{Wave equation and Helmholtz equation}
\label{sec:wave_Helmholtz_equation}
\begin{figure}
  \centering
  \includegraphics[width=0.4\linewidth]{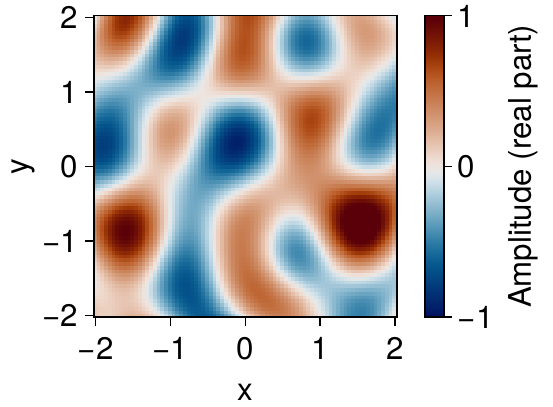}
  \caption{Example of solution of the Helmoltz equation plotted for $xy$-plane.}
  \label{fig:example_sound_field}
\end{figure}

\begin{figure}
  \centering
  \begin{minipage}{0.3\linewidth}
    \includegraphics[width=\linewidth]{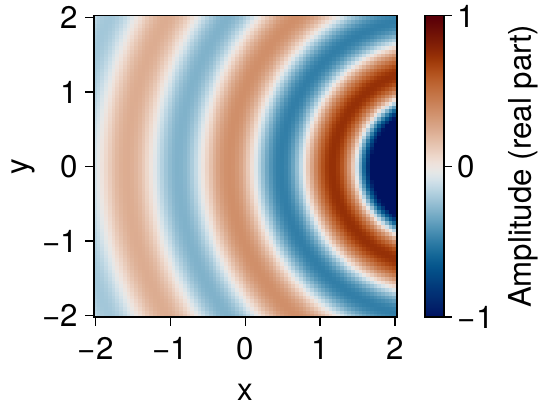}
  \end{minipage}
  $\vcenter{\vspace{-1.3em}\hbox{+}}$
  \begin{minipage}{0.3\linewidth}
    \includegraphics[width=\linewidth]{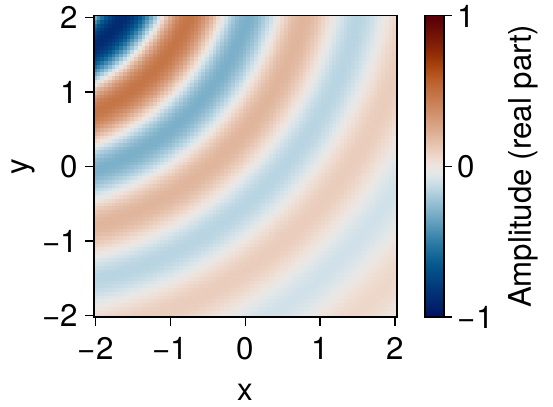}
  \end{minipage}
  $\vcenter{\vspace{-1.3em}\hbox{=}}$
  \begin{minipage}{0.3\linewidth}
    \includegraphics[width=\linewidth]{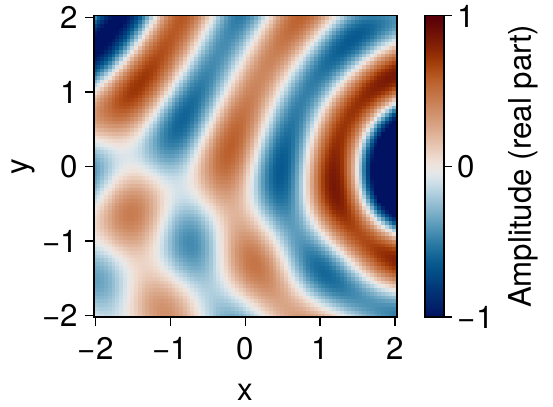}
  \end{minipage}
  \caption{Superposition of sound fields.}
  \label{fig:superposition_sound_field}
\end{figure}

A sound field, as a wave phenomenon, is governed by the physical constraint, which is typically described by partial differential equations as follows. 
Consider a sound field $U$ in the time-domain representation defined in the region $\Omega \subset \mathbb{R}^3$ having no sound sources or scatterers therein. 
For mathematical convenience, we assume that $\Omega$ is an open set.
Then, the sound field $U$ is known to satisfy well\footnote{In a strict sense, this partial differential equation is valid for infinitesimal sound fields since it is derived from several linearized physical equations. In the context of this paper, however, such linearization does not cause significant practical issues. Studies involving sound fields with large sound pressures, where nonlinear terms cannot be ignored, fall under the scope of nonlinear acoustics, which is beyond the scope of this paper.} the partial differential equation
\begin{align}
  \left(
  \frac{\partial^2}{{\partial x}^2}
  +
  \frac{\partial^2}{{\partial y}^2}
  +
  \frac{\partial^2}{{\partial z}^2}
  -
  \frac{1}{c^2}
  \frac{\partial^2}{{\partial t}^2}
  \right)
  U(x,y,z,t)
  =
  0
  \label{eq:wave_equation}
\end{align}
for every $(x,y,z) \in \Omega$ and $t \in \mathbb{R}$.
Here, $c \in (0,\infty)$ denotes the speed of sound, which is assumed to be constant within $\Omega$ and time-invariant.
This partial differential equation is called the \emph{acoustic wave equation}. 
On the other hand, a sound field $u$ in the frequency-domain representation satisfies well the partial differential equation
\begin{align}
  \left(
  \frac{\partial^2}{{\partial x}^2}
  +
  \frac{\partial^2}{{\partial y}^2}
  +
  \frac{\partial^2}{{\partial z}^2}
  +
  \frac{\omega^2}{c^2}
  \right)
  u(x,y,z,\omega)
  =
  0
  \label{eq:Helmholtz_equation_frequency}
\end{align}
for every $(x,y,z) \in \Omega$ and $\omega \in \mathbb{R}$.
This equation is obtained by the Fourier transform of \eqref{eq:wave_equation}\footnote{Solutions of \eqref{eq:wave_equation} and \eqref{eq:Helmholtz_equation_frequency} do not have one to one correspondence in the sense of ``strong solution''; for example, the inverse Fourier transform of a solution of \eqref{eq:Helmholtz_equation_frequency} is not necessarily differentiable. However, we use these formulations for simplicity since they cause no loss of practical significance.}. 
For reasons mentioned later, we often consider a sound field, its observation, and its estimation at a single fixed frequency.
In such contexts, we omit the frequency $\omega$ for notational simplicity, and a sound field is represented as a spatial (instead of spatio-temporal) function from $\Omega$ to $\mathbb{C}$ satisfying
\begin{align}
  \left(
  \frac{\partial^2}{{\partial x}^2}
  +
  \frac{\partial^2}{{\partial y}^2}
  +
  \frac{\partial^2}{{\partial z}^2}
  +
  \wavenumber^2
  \right)
  u(x,y,z)
  =
  0,
  \label{eq:Helmholtz_equation}
\end{align}
or more simply by
\begin{align}
  (\Delta + \wavenumber^2) u(\bm{r})
  =
  0,
\end{align}
for every $\bm{r} = (x,y,z) \in \Omega$.
Here, $\wavenumber = \omega / c$ is a constant called the wavenumber.
This partial differential equation is referred to as the \emph{Helmholtz equation}. 
Throughout this paper, the set of solutions of the Helmholtz equation in $\Omega$ at the wavenumber $\wavenumber$ is denoted by $\mathcal{H}(\Omega; \wavenumber)$, where $k$ is assumed to be nonzero real number. 
An example of a solution of the Helmholtz equation is plotted in Fig.~\ref{fig:example_sound_field}; one can see waveforms therein. 
Note that $\mathcal{H}(\Omega; \wavenumber)$ is a vector space over $\mathbb{C}$, which means that the superposition of sound fields is also interpreted as a sound field (see Fig.~\ref{fig:superposition_sound_field}). 

Several important functions following the Helmholtz equation are listed below\footnote{Note that $\bm{r} = \bm{0}$ is a removable singularity for the regular spherical wave function; in particular, $\varphi_{\nu,\mu}(\bm{0})$ is defined as 
\begin{align}
  \varphi_{\nu,\mu}(\bm{0}) = 
  \begin{cases}
    1 & (\nu = 0)
    \\
    0 & (\text{otherwise})
  \end{cases}
\end{align}
}. 
\begin{description}
  \item[Free-field Green's function:]
  \begin{align}
    \hspace{-2em}
    G(\bm{r}; \bm{r}_\mathrm{s}) 
    \coloneqq 
    \frac{\exp(\im \wavenumber \|\bm{r} - \bm{r}_\mathrm{s}\|)}{4\pi \|\bm{r} - \bm{r}_\mathrm{s}\|}
    \quad (\bm{r} \in \mathbb{R}^3 \setminus \{\bm{r}_\mathrm{s}\}).
    \label{eq:def_Green_function}
  \end{align}
  \item[Plane wave function:]
  \begin{align}
    \hspace{-2em}
    u_{\mathrm{pw}}(\bm{r}; \bm{x}) 
    \coloneqq 
    \exp(-\im \wavenumber \bm{x} \dotproduct \bm{r})
    \quad (\bm{r} \in \mathbb{R}^3).
    \label{eq:def_plane_wave_function}
  \end{align}
  \item[Regular spherical wave function:]
  \begin{align}
    \hspace{-2em}
    \varphi_{\nu,\mu}(\bm{r})
    \coloneqq 
    \frac{1}{\im^\nu} j_\nu(\wavenumber \|\bm{r}\|) \hat{\sphharm}_{\nu,\mu}\left(\frac{\bm{r}}{\|\bm{r}\|}\right)
    \quad (\bm{r} \in \mathbb{R}^3).
    \label{eq:def_regular_spherical_wave_function}
  \end{align}
  \item [Singular spherical wave function:]
  \begin{align}
    \hspace{-2em}
    \psi_{\nu,\mu}(\bm{r})
    \coloneqq 
    \frac{i\wavenumber}{4\pi}
    \im^{\nu} h_\nu(\wavenumber \|\bm{r}\|) \hat{\sphharm}_{\nu,\mu} \left(\frac{\bm{r}}{\|\bm{r}\|}\right)^\ast
    \quad (\bm{r} \in \mathbb{R}^3\setminus\{\bm{0}\}). 
    \label{eq:def_singular_spherical_wave_function}
  \end{align}
\end{description}
Here, $\bm{r}_\mathrm{s} \in \mathbb{R}^3$, $\bm{x} \in \mathbb{S}_2$, $\nu \in \mathbb{N}$, and $\mu \in \llbracket -\nu, \nu \rrbracket$ are parameters in these functions, $j_\nu(\placeholder): \mathbb{R} \to \mathbb{R}$ and $h_\nu(\placeholder): \mathbb{R} \setminus \{0\} \to \mathbb{C}$ are respectively the $\nu$th-order spherical Bessel and spherical Hankel functions of the first kind, and $\hat{\sphharm}_{\nu,\mu}(\placeholder): \mathbb{S}_2 \to \mathbb{C}$ is the unnormalized spherical harmonic function of order $\nu$ and degree $\mu$ defined as 
\begin{align}
  \hat{\sphharm}_{\nu,\mu}(\bm{x}) \coloneqq \sqrt{4\pi} \sphharm_{\nu,\mu}(\bm{x}) \quad (\bm{x} \in \mathbb{S}_2)
  \label{eq:def_unnormalized_spherical_harmonics}
\end{align} 
using the normalized spherical harmonic function $Y_{\nu,\mu}(\placeholder): \mathbb{S}_2 \to \mathbb{C}$. 
See \cite{Martin:Cambridge2006} for the definitions of these special functions. 
The scaling factors in \eqref{eq:def_regular_spherical_wave_function}, \eqref{eq:def_singular_spherical_wave_function}, and \eqref{eq:def_unnormalized_spherical_harmonics} are introduced for theoretical convenience, especially in Sections~\ref{sec:preliminaries}, \ref{sec:discrete}, and \ref{sec:applications}. 
From a physical perspective, the free-field Green's function models the sound field originating from a point source located at $\bm{r}_\mathrm{s}$ in an anechoic environment, and the plane wave function models a sound field with the incident direction $\bm{x}$ caused by the vibration of an infinite plate or approximates a sound field at an infinite distance from the source. 
The regular/singular spherical wave functions are usually used as basis patterns for incoming/outgoing sound fields and also simply called the spherical wave functions unless they need to be distinguished. 

When we consider a sound field in a region including sound sources, the right-hand sides of \eqref{eq:wave_equation}, \eqref{eq:Helmholtz_equation_frequency}, and \eqref{eq:Helmholtz_equation} are replaced by nonzero functions called source distributions~\cite{Williams:AcademicPress1999}. 
In this case, the mathematical treatment becomes much more difficult owing to the inhomogeneity of the partial differential equations. 
In this paper, we focus on sound fields within a source-free region; see \cite{Badia:IOP2011,Dogan:SPIE2011,Koyama:IEEE2019} for estimation problems of a sound field in a region including sound sources.

\subsection{Interior and exterior sound fields}

In many practical situations, a target sound field falls into one of the following two categories or their combination. 
First, when the region of interest $\Omega$ is bounded and simply connected, a sound field in $\Omega$ is referred to as an interior sound field. 
On the other hand, when $\Omega$ is the exterior of a bounded and simply connected set and a sound field $u:\Omega \to \mathbb{C}$ satisfies the Sommerfeld radiation condition~\cite{Williams:AcademicPress1999} 
\begin{align}
  \lim_{\|\bm{r}\| \to \infty}
  \|\bm{r}\| \left(\frac{\partial}{\partial \|\bm{r}\|} - \im \wavenumber\right) 
  u(\bm{r}) = 0
\end{align}
uniformly in all directions, it is referred to as an exterior sound field. 
Here, $\frac{\partial}{\partial \|\bm{r}\|}$ denotes the partial derivative with respect to the radial coordinate in the spherical coordinate system. 
For example, the free-field Green's function and singular spherical wave functions satisfy the Sommerfeld radiation condition. 
This condition physically means an outgoing sound field in an anechoic environment. 

From an engineering perspective, the purpose of interior sound field estimation is to analyze how a listener perceives the surrounding sound, while that of exterior sound field estimation is to analyze how a sound source produces the sound. 
Although this paper primarily focuses on the estimation of interior sound fields, several works on exterior sound fields share some underlying theories with those on interior ones. 
Relevant methods of exterior sound field estimation will also be discussed at the end of Sections~\ref{sec:boundary} and \ref{sec:discrete}. 

\subsection{Representations of sound field}
We have stated in Section~\ref{sec:wave_Helmholtz_equation} that the superposition of sound fields is also interpreted as a sound field. 
Conversely, under certain appropriate conditions, a sound field can be expanded or approximated by the superposition of certain collections of simple sound fields, such as plane wave functions or spherical wave functions, which provides us a powerful tool for the analysis of sound fields. 
For example, the following representations are widely used for an interior sound field: 
\begin{description}
  \item[Expansion using plane wave functions:]
  \begin{align}
    \hspace{-2em}
    u(\bm{r}) 
    = 
    \int_{\bm{x} \in \mathbb{S}_2} \tilde{u}(\bm{x};\bm{r}_0) \exp(-\im \wavenumber \bm{x} \dotproduct (\bm{r} - \bm{r}_0)) \diff \surf.
    \label{eq:expansion_plane_wave_function}
  \end{align}
  \item[Expansion using regular spherical wave functions:]
  \begin{align}
    \hspace{-2em}
    u(\bm{r}) = \sum_{\nu,\mu} \ring{u}_{\nu,\mu}(\bm{r}_0) \varphi_{\nu,\mu}(\bm{r}-\bm{r}_0). 
    \label{eq:expansion_regular_spherical_wave_function}
  \end{align}
\end{description}
Here and hereafter, $\tilde{u}(\placeholder;\bm{r}_0):\mathbb{S}_2 \to \mathbb{C}$ and $\ring{u}_{\nu,\mu}(\bm{r}_0) \in \mathbb{C}$ ($\nu \in \mathbb{N}, \ \mu \in \llbracket -\nu, \nu \rrbracket$) are the expansion coefficients around $\bm{r}_0 \in \mathbb{R}^3$ in \eqref{eq:expansion_plane_wave_function} and \eqref{eq:expansion_regular_spherical_wave_function}, respectively, and $\sum_{\nu,\mu}$ is the abbreviation of $\sum_{\nu=0}^\infty \sum_{\mu=-\nu}^\nu$. 
These expansion coefficients are related as follows: 
\begin{align}
  \ring{u}_{\nu,\mu}(\bm{r}_0) 
  = 
  \int_{\bm{x} \in \mathbb{S}_2} 
  \tilde{u}(\bm{x};\bm{r}_0) 
  \hat{\sphharm}_{\nu,\mu}(\bm{x})^\ast 
  \diff \surf, 
  \label{eq:coef_pw2sw}
\end{align}
\begin{align}
  \tilde{u}(\bm{x};\bm{r}_0) 
  = 
  \frac{1}{4\pi}
  \sum_{\nu, \mu} 
  \ring{u}_{\nu,\mu}(\bm{r}_0) 
  \hat{\sphharm}_{\nu,\mu}(\bm{x}), 
  \label{eq:coef_sw2pw}
\end{align}
when $(\ring{u}_{\nu,\mu}(\bm{r}_0))_{\nu,\mu}$ is square summable, or equivalently, $\tilde{u}(\placeholder;\bm{r}_0)$ is square integrable. 
These relations are derived by the following formulae~\cite{Martin:Cambridge2006}: 
\begin{align}
  \varphi_{\nu,\mu}(\bm{r})
  = 
  \frac{1}{4\pi}
  \int_{\bm{x} \in \mathbb{S}_2} 
  \hat{\sphharm}_{\nu,\mu}(\bm{x})
  \exp(-\im \wavenumber \bm{x} \dotproduct \bm{r})  
  \diff \surf, 
  \label{eq:sw2pw} 
\end{align}
\begin{align}
  \exp(-\im \wavenumber \bm{x} \dotproduct \bm{r})
  = 
  \sum_{\nu,\mu} 
  \hat{\sphharm}_{\nu,\mu}(\bm{x})^\ast \varphi_{\nu,\mu}(\bm{r}). 
  \label{eq:Jacobi_Anger_expansion}
\end{align}

\begin{figure}
  \centering
  \begin{minipage}{0.3\linewidth}
    \includegraphics[width=\linewidth]{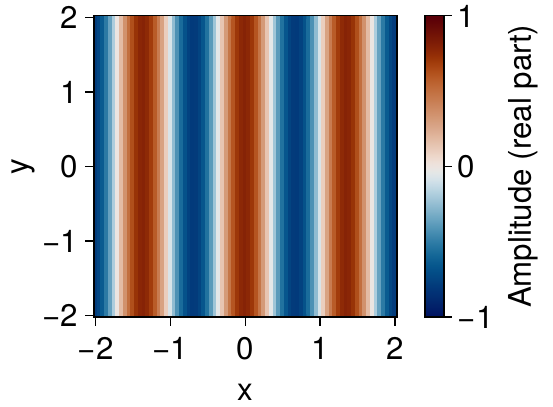}
    \subcaption{Original sound field}
  \end{minipage}
  \begin{minipage}{0.3\linewidth}
    \includegraphics[width=\linewidth]{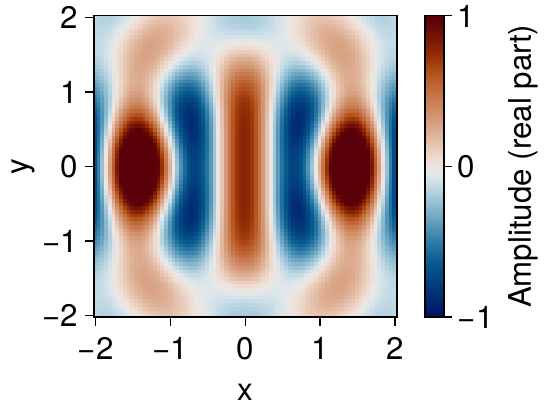}
    \subcaption{Fourth-order expansion}
  \end{minipage}
  \begin{minipage}{0.3\linewidth}
    \includegraphics[width=\linewidth]{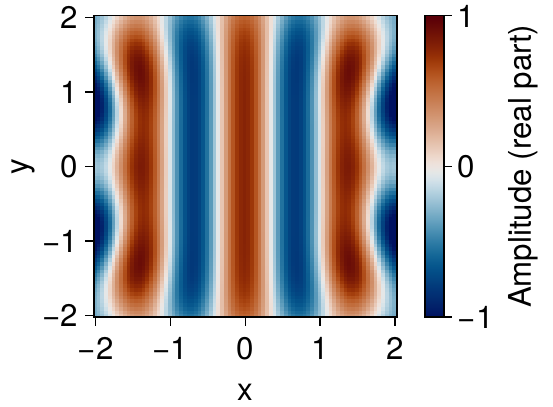}
    \subcaption{Eighth-order expansion}
  \end{minipage}
  \caption{Expansion of sound field by sperical wave functions.}
  \label{fig:expansion_example}
\end{figure}

One may naturally ask whether it is possible to decompose any sound field into these basis functions. 
Indeed, every $u \in \mathcal{H}(\Omega; k)$ allows the expansion of \eqref{eq:expansion_regular_spherical_wave_function} around arbitrary $\bm{r}_0 \in \Omega$. 
Moreover, it is known that the series on the right-hand side, similar to the power series expansion of a holomorphic function, provides increasingly accurate approximations over a wider range as the order $\nu$ increases, starting from the vicinity of the expansion center. 
For example, Fig.~\ref{fig:expansion_example} illustrates the expansion of a plane wave function by the spherical wave functions, truncated at different orders.
On the other hand, not all $u \in \mathcal{H}(\Omega; k)$ can necessarily be expanded as in \eqref{eq:expansion_plane_wave_function}. 
However, any interior sound field can be approximated arbitrarily by the form of both \eqref{eq:expansion_plane_wave_function} and \eqref{eq:expansion_regular_spherical_wave_function}, in the sense of uniform convergence on compact sets. 
Theories on these representations and approximations will be described in detail in Section~\ref{sec:preliminaries}. 

\section{What Is the Observation of a Sound Field?}
\label{sec:observation}

\begin{figure}
  \centering
  $\vcenter{\hbox{\includegraphics[width=0.32\linewidth]{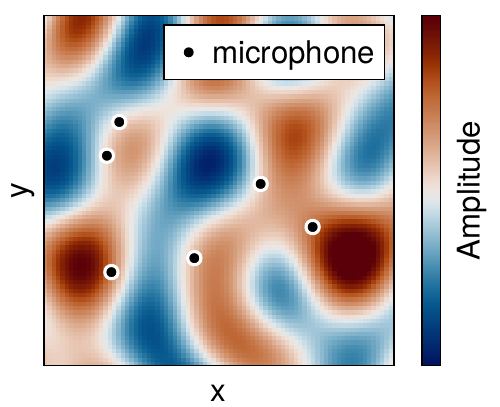}}}$
  $\vcenter{\hbox{\includegraphics[width=0.05\linewidth]{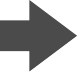}}}$
  $\vcenter{\hbox{\includegraphics[width=0.4\linewidth]{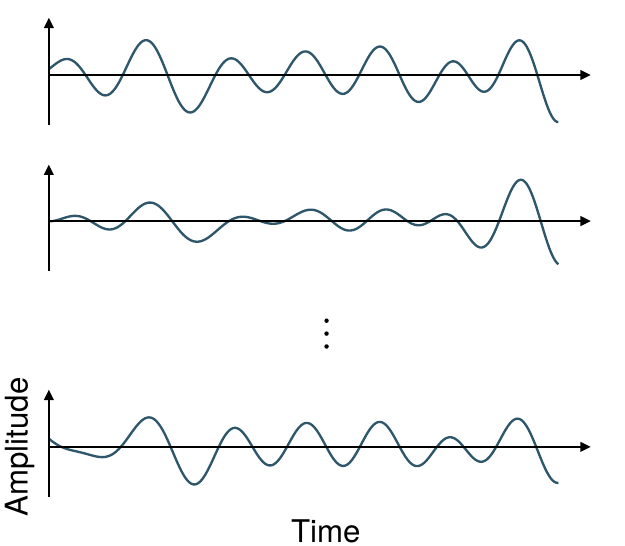}}}$
  \caption{Observation of sound field by microphones.}
  \label{fig:observation}
\end{figure}

The observation of a sound field means the acquisition of some physical quantity in the sound field by sensors, \ie, microphones. 
To estimate a sound field, it is necessary to observe the spatial information of sound, which typically requires the use of multiple microphones, as shown in Fig.~\ref{fig:observation}.
A typical example is the sampling of sound pressure at certain positions with omnidirectional microphones (also called pressure microphones). 
In general cases including a nonuniform frequency response or directivity, a mathematical representation of an observation becomes more complex than the sampling of sound pressure.
For example, a bidirectional microphone (also called a velocity microphone) observes the directional derivative of the sound field at its position.
However, most types of microphone have a linear time-invariant property regardless of the frequency response or directivity.
Therefore, we refer to the term ``observation'' in this paper as any type of linear time-invariant transform from the sound field to single-channel or multichannel time-series signals. 

\subsection{Representation of single-microphone characteristics}
\label{sec:single_microphone}

First, we consider the mathematical modeling of a single-channel observation, \ie, characteristics of a single microphone. 
From the linear time-invariant property of the observation, a single-channel observation can be represented in the frequency domain with the observation noise taken into account as 
\begin{align}
  s(\omega) = \mathcal{F}(\omega) u(\placeholder,\omega) + \epsilon(\omega)
  \label{eq:observation_model_frequency}
\end{align}
for $\omega \in \mathbb{R}$, where $s(\omega) \in \mathbb{C}$ denotes the observed signal, $\mathcal{F}(\omega) : \mathcal{H}(\Omega; \omega/c) \to \mathbb{C}$ is the linear functional referred to as an \emph{observation operator} in this paper, and $\epsilon(\omega) \in \mathbb{C}$ denotes the observation noise.
When considering a single fixed frequency, we simplify \eqref{eq:observation_model_frequency} as
\begin{align}
  s = \mathcal{F} u + \epsilon
  \label{eq:observation_model}
\end{align}
by omitting the frequency variable $\omega$. 
For example, the observation operator for an omnidirectional, bidirectional, and first-order microphone located at $\bm{r}_0$ in a sound field $u \in \mathcal{H}(\Omega;\wavenumber)$ is modeled with the parameters $\bm{y} \in \mathbb{S}_2$ and $a \in [0,1]$ as follows. 
\begin{description}
  \item[Omnidirectional:]
    \begin{align}
      \hspace{-2em}
      \mathcal{F} u = u(\bm{r}_0).
      \label{eq:derivative_omnidirectional}
    \end{align}
  \item[Bidirectional:]
    \begin{align}
      \hspace{-2em}
      \mathcal{F} u = \frac{\im}{\wavenumber} \bm{y} \dotproduct \nabla u(\bm{r}_0).
      \label{eq:derivative_bidirectional}
    \end{align}
  \item[First-order:]
    \begin{align}
      \hspace{-2em}
      \mathcal{F} u = a u(\bm{r}_0) + (1 - a) \frac{\im}{\wavenumber} \bm{y} \dotproduct \nabla u(\bm{r}_0).
      \label{eq:derivative_first_order}
    \end{align}
\end{description}

As an alternative way, the characteristics of a microphone are often represented by its \emph{directivity}. 
In the mathematical expression, the directivity is a directional function, for instance, $\gamma: \mathbb{S}_2 \to \mathbb{C}$, where $\gamma(\bm{x})^\ast$ denotes the response of the microphone to the plane wave field traveling from the direction $\bm{x} \in \mathbb{S}_2$, \ie, $u_\mathrm{pw}(\bm{r};\bm{x}) = \exp(-\im \wavenumber \bm{x} \dotproduct (\bm{r}-\bm{r}_0))$. 
Here, we consider $\gamma(\bm{x})^\ast$ instead of $\gamma(\bm{x})$ for theoretical convenience. 
This expression is often useful because the directivity of the microphone is directly measurable using loudspeakers at a sufficient distance in an anechoic room. 
Furthermore, the spherical harmonic expansion of $\gamma$ defined by 
\begin{align}
  \gamma(\bm{x})
  = 
  \sum_{\nu,\mu} d_{\nu,\mu} \hat{\sphharm}_{\nu,\mu}(\bm{x})
  \label{eq:directivity_isht}
\end{align}
is also used so that the directivity of the microphone can be represented by the scalar sequence $(d_{\nu,\mu})_{\nu,\mu}$. 
Such coefficients are obtained explicitly as 
\begin{align}
  d_{\nu,\mu} 
  = 
  \frac{1}{4 \pi}
  \int_{\bm{x} \in \mathbb{S}_2} 
  \gamma(\bm{x}) 
  \hat{\sphharm}_{\nu,\mu}(\bm{x})^\ast 
  \diff \surf 
  \label{eq:directivity_sht}
\end{align}
from the orthogonality of the spherical harmonic functions~\cite{Williams:AcademicPress1999} given by 
\begin{align}
  \frac{1}{4\pi}
  \int_{\bm{x}\in\mathbb{S}_2} 
  \hat{\sphharm}_{\nu,\mu}(\bm{x})^\ast 
  \hat{\sphharm}_{\nu',\mu'}(\bm{x})
  \diff \surf
  = 
  \delta_{\nu, \nu'} \delta_{\mu, \mu'}. 
  \label{eq:orthogonality_spherical_harmonic_function}
\end{align}
When the sound field $u$ is given by \eqref{eq:expansion_plane_wave_function} around the microphone position $\bm{r}_0$, the observation operator $\mathcal{F}$ can be represented using its directivity $\gamma$ as 
\begin{align}
  \mathcal{F} u 
  = 
  \int_{\bm{x} \in \mathbb{S}_2} 
  \gamma(\bm{x})^\ast 
  \tilde{u}(\bm{x};\bm{r}_0) 
  \diff \surf. 
  \label{eq:observation_directivity}
\end{align}
On the other hand, when $u$ is given by \eqref{eq:expansion_regular_spherical_wave_function} around $\bm{r}_0$, $\mathcal{F}$ can be represented using the spherical harmonic coefficients $(d_{\nu,\mu})_{\nu,\mu}$ as 
\begin{align}
  \mathcal{F} u 
  = 
  \sum_{\nu,\mu} 
  d_{\nu,\mu}^\ast 
  \ring{u}_{\nu,\mu}(\bm{r}_0), 
  \label{eq:observation_sphharm}
\end{align}
which is derived by \eqref{eq:coef_pw2sw}, \eqref{eq:directivity_sht}, and \eqref{eq:observation_directivity}. 

\begin{figure}
  \centering
  \begin{minipage}{0.32\linewidth}
    \includegraphics[width=\linewidth]{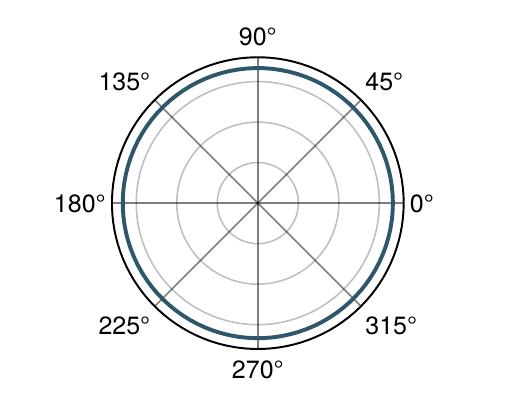}
    \subcaption{Omnidirectional}
  \end{minipage}
  \begin{minipage}{0.32\linewidth}
    \includegraphics[width=\linewidth]{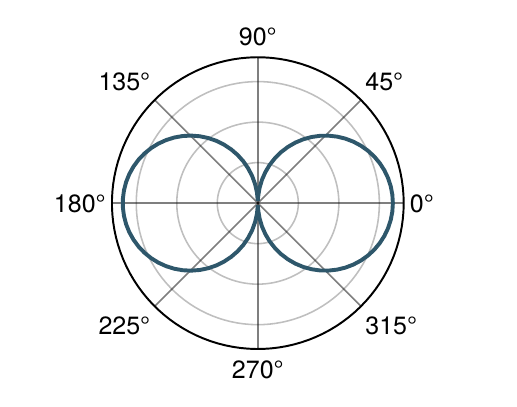}
    \subcaption{Bidirectional}
  \end{minipage}
  \begin{minipage}{0.32\linewidth}
    \includegraphics[width=\linewidth]{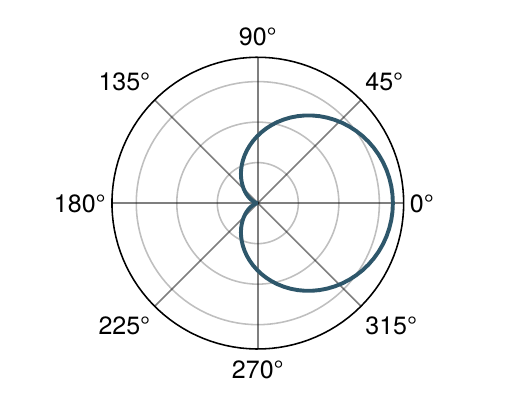}
    \subcaption{First-order ($a = 0.5$)}
  \end{minipage}
  \caption{Examples of directivity patterns. Absolute values of the directivity function $\gamma(\bm{x})$ are plotted against the angle between $\bm{x}$ and $\bm{y}$.}
  \label{fig:directivity}
\end{figure}

As seen in \eqref{eq:observation_directivity} and \eqref{eq:observation_sphharm}, the characteristics of the microphone can be represented equivalently by different domains. 
Several examples of their correspondences are listed in Table~\ref{tab:microphone_characteristics}, and their directivity patterns are shown in Fig.~\ref{fig:directivity}, where the absolute values of the directivity function $\gamma(\bm{x})$ are plotted in the polar coordinate system with $\bm{y}$ defined as the polar axis. 

\begin{table}
  \caption{Representation of characteristics of microphone.}
  \label{tab:microphone_characteristics}
  \footnotesize
  \centering
  \begin{tabular}{lll}
    \toprule 
    & Directivity & Spherical harmonic coefficient \\
    \midrule
    Omnidirectional
    & $\gamma(\bm{x}) = 1$ & $d_{\nu,\mu} \! = \! \begin{cases} 1 & (\nu = 0) \\ 0 & (\text{otherwise}) \end{cases}$ \\ 
    \addlinespace
    Bidirectional
    & $\gamma(\bm{x}) = \bm{y} \dotproduct \bm{x}$ & $d_{\nu,\mu} \! = \! \begin{cases} \frac{1}{3} \hat{\sphharm}_{1,\mu}(\bm{y})^\ast & (\nu = 1) \\ 0 & (\text{otherwise}) \end{cases}$ \\
    \addlinespace
    First-order
    & $\gamma(\bm{x}) = a + (1 \! - \! a) \bm{y} \dotproduct \bm{x}$ & $d_{\nu,\mu} \! = \! \begin{dcases} a & (\nu = 0) \\ \! \frac{1 \! - \! a}{3} \hat{\sphharm}_{1,\mu}(\bm{y})^\ast & (\nu = 1) \\ 0 & (\text{otherwise}) \end{dcases}$ \\
    \bottomrule
  \end{tabular}
\end{table}

\subsection{Modeling of multichannel observation}
When $M \in \mathbb{N}$ microphones are located in a sound field $u$, the observed signal can be represented similarly to the single-channel case as 
\begin{align}
  s_m = \mathcal{F}_m u + \epsilon_m \quad (m \in \mathbb \llbracket 1, M \rrbracket),
  \label{eq:observation_multichannel}
\end{align}
where $s_m \in \mathbb{C}$, $\mathcal{F}_m : \mathcal{H}(\Omega; \wavenumber) \to \mathbb{C}$, and $\epsilon_m \in \mathbb{C}$ respectively denotes the observed signal, observation operator, and observation noise of the $m$th microphone. 
The observation operator of each microphone is not affected by the existence of other microphones when their acoustic interaction, such as the effects of scattering, diffraction, and absorption, can be regarded small enough. 
This condition is well satisfied when the microphones are acoustically transparent or located at a sufficient distance from each other; this condition is imposed throughout this paper unless otherwise stated. 
When this assumption is not satisfied, the accurate modeling of the observation operators $\mathcal{F}_1,\ldots,\mathcal{F}_M$ becomes much more complex because the multibody problem called \emph{multiple scattering}~\cite{Martin:Cambridge2006} has to be considered; however, the linear time-invariant property of the observation, and therefore the formulation of \eqref{eq:observation_multichannel}, are still valid in this case.

\section{Statement of Sound Field Estimation Problem}
\label{sec:problem_statement}

The goal of sound field estimation is to estimate an unknown sound field from the observed signals on the basis of a given observation model in \eqref{eq:observation_multichannel}. 
In this paper, we focus on the estimation methods applied to each frequency independently. 
Such an approach means that no interfrequency assumption on the target sound field is used, and most of the current sound field estimation methods, especially those not based on the machine learning approach, are classified into this type. 
Then, the sound field estimation problem is summarized as follows. 

\begin{figure}
  \centering
  \includegraphics[width=0.8\linewidth]{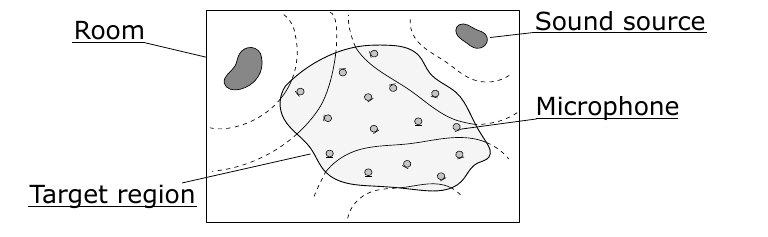} 
  \caption{Sound field estimation problem.}
  \label{fig:sound_field_estimation}
\end{figure}

\newpage

\begin{mdframed}
  \begin{problem*}
    Let $\Omega \subseteq \mathbb{R}^3$ be an open and simply connected set, $u\in \mathcal{H}(\Omega; \wavenumber)$ be a sound field at a fixed wavenumber $\wavenumber \in \mathbb{R} \setminus \{0\}$, $M\in\mathbb{N}$ be the number of microphones, and $s_1, \ldots, s_M \in \mathbb{C}$ denote the observed signals modeled by
    \begin{align}
      s_m = \mathcal{F}_m u + \epsilon_m \quad (m \in \mathbb \llbracket 1, M \rrbracket)
    \end{align}
    with the observation operators $\mathcal{F}_1, \ldots, \mathcal{F}_M: \mathcal{H}(\Omega; \wavenumber) \to \mathbb{C}$ and the observation noises $\epsilon_1,\ldots,\epsilon_M \in \mathbb{C}$ (see Fig.~\ref{fig:sound_field_estimation}). 
    Here, the observation operators are assumed to be given. 
    Then, the objective of the sound field estimation problem is to estimate the unknown $u$ from the given $s_1,\ldots,s_M$.
  \end{problem*}
\end{mdframed}

When the sound field has been estimated for each frequency, one can obtain the estimated sound field in the time domain representation via the inverse Fourier transform \eqref{eq:inverse_Fourier_transform}. 
In practice, the (inverse) discrete Fourier transform is often used with analysis at discrete frequencies. 
Here, if the estimator is linear with respect to the observed signal at each frequency, it can be regarded as a linear time-invariant system from the observed signal to the estimated sound field. 
This property is important especially when the sound field estimation is combined with another linear time-invariant post-stage signal processing because the entire system becomes a single linear time-invariant system, which can be implemented efficiently using finite impulse response (FIR) filters. 
There are many such cases including the applications described in Section~\ref{sec:applications}. 
Otherwise, the estimation is performed typically in the time--frequency domain, for example, by the short-time Fourier transform. 

\section{Classification of Sound Field Estimation Methods}

There are many sound field estimation methods for the problem defined in Section~\ref{sec:problem_statement}, and they can be further classified on the basis of their underlying theories. 
One is based on boundary integral representations of the Helmholtz equation and also called near-field acoustic holography (NAH)~\cite{Williams:AcademicPress1999}. 
We refer to this approach as the \emph{boundary measurement approach} and describe its theory in Section~\ref{sec:boundary}. 
Although this approach requires continuous measurement on the boundary of a target region, the complete reconstruction is guaranteed, except for the discretization effect of the microphones' placement. 
Another is based on the inverse problem for a finite number of discrete measurements and also closely related to a function interpolation problem such as the kernel method~\cite{Murphy:MIT2014}. 
We refer to this approach as the \emph{discrete measurement approach} and describe its theory in Section~\ref{sec:discrete}. 
An almost arbitrary placement of microphones is allowed in this approach, although comprehensive theories on accuracy guarantee and numerical stability are still to be fully established, compared with the boundary measurement approach. 
While beyond the scope of this paper, we should also mention the learning-based approach~\cite{Lluis:JASA2020, Shigemi:IWAENC2022, Cobos:EURASIP2022, Koyama:IEEE2024} especially developed in recent years. 
Although this approach requires a large-scale dataset of sound fields or room impulse responses to train the estimator, active efforts have also been made in the construction and use of datasets of room impulse responses~\cite{Koyama:WASPAA2021, Karakonstantis:DTU2021}. 
Learning-based methods are expected to not only improve the estimation performance but also achieve a broader application range than before, such as the estimation of a sound field including sources within the region of interest.

\chapter{Preliminaries: Theories on Helmholtz Equation}
\label{sec:preliminaries}

This section provides mathematical preliminaries on the Helmholtz equation. 
The specific goal of this section is to explain how an expansion or approximation of a solution of the Helmholtz equation using plane wave or spherical wave functions is verified. 
Therefore, readers not interested in theoretical details can skip this section or focus on only the statements of Theorems~\ref{thm:expansion_spherical_wave_function}, \ref{thm:approximation_spherical_wave_function}, and \ref{thm:approximation_plane_wave_function}. 
All the mathematical facts presented in this section can be found in the literature~\cite{Williams:AcademicPress1999, Martin:Cambridge2006, Kirsch:Springer2015}; however, we summarize them in a manner as unified and complete as possible. 
The organization of this section is based on Brown and Churchill's textbook on complex analysis~\cite{Brown:McGrawn2014} because several similarities can be found between theories on the Helmholtz equation and the Cauchy--Riemann equations. 

To avoid repetition of terms, the following conditions are applied throughout this section and all theorems therein: 
\begin{itemize}
  \item $\Omega \subset \mathbb{R}^3$ is a nonempty open set. 
  \item $\wavenumber \in \mathbb{R} \setminus \{0\}$. 
  \item $\mathcal{H}(\Omega; \wavenumber)$ denotes the set of all solutions of the Helmholtz equation, \ie, $\mathcal{H}(\Omega; \wavenumber) = \left\{u \in \mathcal{C}_2(\Omega) \suchthat (\Delta + \wavenumber^2)u = 0 \right\}$. 
  \item For a bounded open set $D \subset \mathbb{R}^3$ with $\mathcal{C}_1$-smooth boundary, $\bm{n}_D: \partial D \to \mathbb{S}_2$ denotes the outward normal vector. 
\end{itemize}

\section{Integral Identities}

There are several integral representations related to the Helmholtz equation. 
The two integral identities described below are derived using the Stoke's theorem: 
\begin{align}
  \int_{\bm{r} \in D} \Delta f(\bm{r}) \diff \vol 
  = 
  \int_{\bm{r}_\mathrm{s} \in \partial D} \nabla_\mathrm{s} f (\bm{r}_\mathrm{s}) \dotproduct \bm{n}_D(\bm{r}_\mathrm{s}) \diff \surf, 
\end{align}
which holds for a bounded open set $D \subset \Omega$ with a $\mathcal{C}_1$-smooth boundary $\partial D \subset \Omega$ and $f \in \mathcal{C}_2(\Omega)$\footnote{In fact, the geometrical conditions for $\partial D$ can be relaxed to a more generalized class, such as a piecewise smooth boundary, by using several advanced concepts, e.g., the \emph{standard domain} introduced by Whitney~\cite{Whitney:Princeton1957}.}. 
Here and hereafter, the notation $\nabla_\mathrm{s}$ is used as the vector differential operator with respect to $\bm{r}_\mathrm{s}$ to avoid confusion. 

\begin{mdframed}
  \begin{theorem}
    \label{thm:Helmholtz_integral}
    Let $D \subset \Omega$ be a bounded open set with a $\mathcal{C}_1$-smooth boundary $\partial D \subset \Omega$. 
    Then, the following equality holds for $u, v \in \mathcal{H}(\Omega; \wavenumber)$: 
    \begin{align}
      \int_{\bm{r}_\mathrm{s} \in \partial D}
      \left(
      v(\bm{r}_\mathrm{s}) \nabla_\mathrm{s} u(\bm{r}_\mathrm{s}) - u(\bm{r}_\mathrm{s}) \nabla_\mathrm{s} v(\bm{r}_\mathrm{s})
      \right)
      \dotproduct
      \bm{n}_D(\bm{r}_\mathrm{s})
      \diff \surf
      =
      0.
    \end{align}
  \end{theorem}
\end{mdframed}
\begin{proof}
  From Stokes' theorem, we have
  \begin{align}
    &
    \int_{\bm{r}_\mathrm{s} \in \partial D}
    \left(
    v(\bm{r}_\mathrm{s}) \nabla_\mathrm{s} u(\bm{r}_\mathrm{s}) - u(\bm{r}_\mathrm{s}) \nabla_\mathrm{s} v(\bm{r}_\mathrm{s})
    \right)
    \dotproduct
    \bm{n}_D(\bm{r}_\mathrm{s})
    \diff \surf
    \nonumber
    \\
    & =
    \int_{\bm{r} \in D}
    \left(
    v(\bm{r}) \Delta u(\bm{r}) - u(\bm{r}) \Delta v(\bm{r})
    \right)
    \diff \vol
    \nonumber
    \\
    & =
    \int_{\bm{r} \in D}
    \left(
      v(\bm{r}) (\Delta + \wavenumber^2) u(\bm{r}) - u(\bm{r}) (\Delta + \wavenumber^2) v(\bm{r})
    \right)
    \diff \vol
    \nonumber
    \\
    & =
    0.
  \end{align}
\end{proof}

\begin{mdframed}
  \begin{theorem}
    \label{thm:Kirchhoff_Helmholtz}
    Let $D \subset \Omega$ be a bounded open set with a $\mathcal{C}_1$-smooth boundary $\partial D \subset \Omega$.
    Then, the following equality holds for $u \in \mathcal{H}(\Omega; \wavenumber)$ with $G(\placeholder; \placeholder)$ defined by \eqref{eq:def_Green_function}: 
    \begin{align}
      u(\bm{r}) 
      & = 
      \int_{\bm{r}_\mathrm{s} \in \partial D}
      \bigl(
      G(\bm{r}; \bm{r}_\mathrm{s}) \nabla_\mathrm{s} u(\bm{r}_\mathrm{s})
      \nonumber 
      \\ 
      & \hspace{3.5em} - 
      u(\bm{r}_\mathrm{s}) \nabla_\mathrm{s} G(\bm{r}; \bm{r}_\mathrm{s}) 
      \bigr)
      \dotproduct
      \bm{n}_D(\bm{r}_\mathrm{s})
      \diff \surf 
      \quad \forall \bm{r} \in D.
      \label{eq:Kirchhoff_Helmholtz}
    \end{align}
  \end{theorem}
\end{mdframed}

\begin{proof}
  Let $\bm{r} \in D$ be fixed and $v \in \mathcal{H}(\Omega \setminus \{\bm{r}\}; \wavenumber)$ be defined as 
  \begin{align}
    v(\bm{r}') 
    \coloneqq 
    G(\bm{r}; \bm{r}') \quad (\bm{r}' \in \Omega \setminus \{\bm{r}\}). 
  \end{align}
  Since $D$ is an open set, there exists some $\delta_{\bm{r}} \in (0, \infty)$ satisfying 
  \begin{align}
    \overline{B}(\bm{r}, \epsilon) \subset D \quad \forall \epsilon \in (0, \delta_{\bm{r}}). 
  \end{align}
  For $\epsilon \in (0, \delta_{\bm{r}})$, by defining  
  \begin{align}
    D_{\bm{r}, \epsilon} 
    \coloneqq 
    D \setminus \overline{B}(\bm{r}, \epsilon),
  \end{align}
  we see that $D_{\bm{r}, \epsilon} \subset \Omega \setminus \{\bm{r}\}$ is a bounded open set with a $\mathcal{C}_1$-smooth boundary $\partial D_{\bm{r}, \epsilon} \subset \Omega \setminus \{\bm{r}\}$. 
  Therefore, applying Theorem~\ref{thm:Helmholtz_integral} to $u, v \in \mathcal{H}(\Omega \setminus \{\bm{r}\}; \wavenumber)$ yields 
  \begin{align}
    \int_{\bm{r}_\mathrm{s} \in \partial D_{\bm{r},\epsilon}} 
    \left( 
    v(\bm{r}_\mathrm{s}) \nabla_\mathrm{s}  u(\bm{r}_\mathrm{s}) 
    - 
    u(\bm{r}_\mathrm{s}) \nabla_\mathrm{s} u(\bm{r}_\mathrm{s})
    \right) 
    \dotproduct 
    \bm{n}_{D_{\bm{r},\epsilon}} (\bm{r}_\mathrm{s})
    \diff \surf
    = 
    0. 
  \end{align}
  Since this equality holds for every $\epsilon \in (0,\delta_{\bm{r}})$, we have 
  \begin{align}
    \lim_{\epsilon \to 0}
    \int_{\bm{r}_\mathrm{s} \in \partial D_{\bm{r},\epsilon}} \!
    \left(
      v(\bm{r}_\mathrm{s}) \nabla_\mathrm{s} u(\bm{r}_\mathrm{s}) - u(\bm{r}_\mathrm{s}) \nabla_\mathrm{s} v(\bm{r}_\mathrm{s})
    \right)
    \dotproduct
    \bm{n}_{D_{\bm{r},\epsilon}}(\bm{r}_\mathrm{s})
    \diff \surf
    = 
    0. 
    \label{eq:KH_proof_1}
  \end{align}
  Here, the left-hand side can be rewritten as 
  \begin{align}
    & 
    \lim_{\epsilon \to 0}
    \int_{\bm{r}_\mathrm{s} \in \partial D_{\bm{r},\epsilon}} \! 
    \left(
      v(\bm{r}_\mathrm{s}) \nabla_\mathrm{s} u(\bm{r}_\mathrm{s}) - u(\bm{r}_\mathrm{s}) \nabla_\mathrm{s} v(\bm{r}_\mathrm{s})
    \right)
    \dotproduct
    \bm{n}_{D_{\bm{r},\epsilon}}(\bm{r}_\mathrm{s})
    \diff \surf
    \nonumber 
    \\ 
    & = 
    \int_{\bm{r}_\mathrm{s} \in \partial D}
    \left(
      G(\bm{r}; \bm{r}_\mathrm{s}) \nabla_\mathrm{s} u(\bm{r}_\mathrm{s}) - u(\bm{r}_\mathrm{s}) \nabla_\mathrm{s} G(\bm{r}; \bm{r}_\mathrm{s})
    \right)
    \dotproduct
    \bm{n}_D(\bm{r}_\mathrm{s})
    \diff \surf
    \nonumber 
    \\ 
    & \quad - 
    \lim_{\epsilon \to 0}
    \int_{\bm{r}_\mathrm{s} \in \partial B(\bm{r}, \epsilon)}
    G(\bm{r}; \bm{r}_\mathrm{s}) \nabla_\mathrm{s} u(\bm{r}_\mathrm{s})
    \dotproduct
    \frac{\bm{r}_\mathrm{s} - \bm{r}}{\|\bm{r}_\mathrm{s} - \bm{r}\|}
    \diff \surf
    \nonumber 
    \\ 
    & \quad + 
    \lim_{\epsilon \to 0}
    \int_{\bm{r}_\mathrm{s} \in \partial B(\bm{r}, \epsilon)}
    u(\bm{r}_\mathrm{s}) \nabla_\mathrm{s} G(\bm{r}; \bm{r}_\mathrm{s}) 
    \dotproduct
    \frac{\bm{r}_\mathrm{s} - \bm{r}}{\|\bm{r}_\mathrm{s} - \bm{r}\|}
    \diff \surf, 
  \end{align}
  and we have 
  \begin{align}
    \lim_{\epsilon \to 0} 
    \int_{\bm{r}_\mathrm{s} \in \partial B(\bm{r}, \epsilon)}
    G(\bm{r}; \bm{r}_\mathrm{s}) \nabla_\mathrm{s} u(\bm{r}_\mathrm{s}) 
    \dotproduct
    \frac{\bm{r}_\mathrm{s} - \bm{r}}{\|\bm{r}_\mathrm{s} - \bm{r}\|}
    \diff \surf 
    = 
    0, 
  \end{align}
  \begin{align}
    \lim_{\epsilon \to 0} 
    \int_{\bm{r}_\mathrm{s} \in \partial B(\bm{r}, \epsilon)}
    u(\bm{r}_\mathrm{s}) \nabla_\mathrm{s} G(\bm{r}; \bm{r}_\mathrm{s}) 
    \dotproduct
    \frac{\bm{r}_\mathrm{s} - \bm{r}}{\|\bm{r}_\mathrm{s} - \bm{r}\|}
    \diff \surf 
    = 
    - u(\bm{r}), 
    \label{eq:KH_proof_2}
  \end{align}
  from the definition of $G(\placeholder)$ and the continuity of $u$ and $\nabla u$ around $\bm{r}$. 
  By summarizing \eqref{eq:KH_proof_1} to \eqref{eq:KH_proof_2}, we obtain \eqref{eq:Kirchhoff_Helmholtz}. 
\end{proof}

Theorem~\ref{thm:Kirchhoff_Helmholtz} is called the \emph{Kirchhoff--Helmholtz integral theorem}. 
The role of this theorem in the Helmholtz equation is similar to that of the Cauchy integral formula~\cite{Brown:McGrawn2014} in the Cauchy--Riemann equation, in the sense that they relate the partial differential and integral equations. 

\section{Expansion Using Spherical Wave Functions}

From the well-known method of \emph{separation of variable}~\cite{Williams:AcademicPress1999}, we can see the spherical wavefunctions and their superpositions satisfy the Helmholtz equation. 
Conversely, it will be shown that any solution of the Helmholtz equation can be expanded by the spherical wave function. 

\subsection{Identities for Green's function}
\label{sec:Green_function}

First, the expansion formula (also called the \emph{addition theorem}) and the differentiation formula for Green's function are provided below. 

\begin{mdframed}
  \begin{theorem}
    \label{thm:addition_theorem_Green_function}
    The following equality holds for every $\bm{r}_1, \bm{r}_2 \in \mathbb{R}^3$ satisfying $\|\bm{r}_1\| < \|\bm{r}_2\|$ with $\varphi_{\nu,\mu}(\placeholder)$ and $\psi_{\nu,\mu}(\placeholder)$ defined by \eqref{eq:def_regular_spherical_wave_function} and \eqref{eq:def_singular_spherical_wave_function}, respectively: 
    \begin{align}
      G(\bm{r}_1; \bm{r}_2) =
      \sum_{\nu,\mu}
      \psi_{\nu,\mu}(\bm{r}_2) \varphi_{\nu,\mu}(\bm{r}_1), 
      \label{eq:addition_theorem_Green_function}
    \end{align}
    \begin{align}
      \frac{\partial}{\partial\|\bm{r}_2\|}
      G(\bm{r}_1; \bm{r}_2) = 
      \sum_{\nu,\mu}
      \frac{\partial}{\partial\|\bm{r}_2\|}\psi_{\nu,\mu}(\bm{r}_2) \varphi_{\nu,\mu}(\bm{r}_1). 
      \label{eq:addition_theorem_Green_function_derivative}
    \end{align}
    These equalities hold in the sense of uniform convergence with respect to $\bm{r}_1, \bm{r}_2 \in \mathbb{R}^3$ such that $\|\bm{r}_1\| < R_1 < R_2 < \|\bm{r}_2\| < R_3$ for any $R_1, R_2, R_3 \in (0, \infty)$ satisfying $R_1 < R_2 < R_3$. 
  \end{theorem}
\end{mdframed}
\begin{proof}
  See Theorem~$2$.$40$ in \cite{Kirsch:Springer2015} for the proof of \eqref{eq:addition_theorem_Green_function}, and \eqref{eq:addition_theorem_Green_function_derivative} can also be similarly proved. 
\end{proof}

\begin{mdframed}
  \begin{theorem}
    \label{thm:differentiation_theorem_Green_function}
    Let $\mathcal{Y}_{\nu,\mu}^\ast$ be the $\nu$th-differential 
    operator corresponding to the spherical harmonic function $\hat{Y}_{\nu,\mu}(\placeholder)^\ast$ (see ``Remark'' below for the definition). 
    Then, the following equality holds: 
    \begin{align}
      \frac{1}{(-ik)^\nu}
      \mathcal{Y}_{\nu,\mu}^\ast 
      G(\bm{r}) =
      \psi_{\nu,\mu}(\bm{r}) 
      \quad \forall \bm{r} \in \mathbb{R}^3 \setminus \{\bm{0}\}.
      \label{eq:differentiation_theorem_Green_function}
    \end{align}
  \end{theorem}
\end{mdframed}
\begin{remark*}
  As described in \cite{Martin:Cambridge2006}, for each $\nu \in \mathbb{N}$ and $\mu \in \llbracket-\nu, \nu\rrbracket$, there exists a $\nu$th-order homogeneous polynomial $\mathcal{P}_{\nu,\mu}^\ast(\placeholder):\mathbb{R}^3 \to \mathbb{C}$ satisfying 
  \begin{align}
    \mathcal{P}_{\nu,\mu}^\ast(x, y, z) 
    = 
    \|\bm{r}\|^\nu
    \hat{Y}_{\nu,\mu}\left(\frac{\bm{r}}{\|\bm{r}\|}\right)^\ast 
    \quad \forall \bm{r} = (x, y, z) \in \mathbb{R}^3. 
  \end{align}
  The differential operator $\mathcal{Y}_{\nu,\mu}^\ast$ is defined by replacing $x$, $y$, and $z$ in the polynomial $\mathcal{P}^\ast_{\nu,\mu}(x, y, z)$ with $\frac{\partial}{\partial x}$, $\frac{\partial}{\partial y}$, and $\frac{\partial}{\partial z}$, respectively. 
\end{remark*}
\begin{proof}
  See Theorem~$3$.$10$ in \cite{Martin:Cambridge2006}. 
\end{proof}

\subsection{Expansion theorem}

From the theorems in Section~\ref{sec:Green_function}, we can verify the expansion of any solution of the Helmholtz equation using the spherical wave functions. 

\begin{mdframed}
  \begin{theorem}
    \label{thm:expansion_spherical_wave_function}
    Let $\bm{r}_0 \in \Omega$ and $R \in (0, \infty)$ satisfy $\overline{B}(\bm{r}_0, R) \subset \Omega$. 
    Then, the following equality holds for $u \in \mathcal{H}(\Omega; \wavenumber)$ in the sense of uniform convergence, 
    \begin{align}
      u(\bm{r}) 
      = 
      \sum_{\nu,\mu}
      \ring{u}_{\nu,\mu}(\bm{r}_0) 
      \varphi_{\nu,\mu}(\bm{r}-\bm{r}_0) 
      \quad 
      \forall \bm{r} \in \overline{B}(\bm{r}_0, R). 
    \end{align}
    Here, $\ring{u}_{\nu,\mu}(\bm{r}_0) \in \mathbb{C}$ is obtained by 
    \begin{align}
      \ring{u}_{\nu,\mu}(\bm{r}_0) 
      = 
      \frac{1}{(-ik)^\nu}
      \mathcal{Y}_{\nu,\mu}^\ast 
      u (\bm{r}_0). 
      \label{eq:coef_derivative}
    \end{align}
    Furthermore, let $D \subset \Omega$ be a bounded open set including $\bm{r}_0$ with a $\mathcal{C}_1$-smooth boundary $\partial D \subset \Omega$. 
    Then, the following equality holds: 
    \begin{align}
      \ring{u}_{\nu,\mu}(\bm{r}_0) 
      & = 
      \int_{\bm{r}_\mathrm{s} \in \partial D} 
      \bigl( \psi_{\nu,\mu}(\bm{r}_\mathrm{s} - \bm{r}_0) \nabla_\mathrm{s} u(\bm{r}_\mathrm{s}) 
      \nonumber 
      \\ 
      & \hspace{3.5em}
      - u(\bm{r}_\mathrm{s}) \nabla_\mathrm{s} \psi_{\nu,\mu}(\bm{r}_\mathrm{s} - \bm{r}_0) \bigr) 
      \dotproduct 
      \bm{n}_D(\bm{r}_\mathrm{s})
      \diff \surf. 
      \label{eq:coef_integral}
    \end{align}
  \end{theorem}
\end{mdframed}

\begin{proof}
  Since $\Omega$ is an open set and $\overline{B}(\bm{r}_0,R) \subset \Omega$, there exists some $R^+ \in (R, \infty)$ satisfying $\overline{B}(\bm{r}_0,R^+) \subset \Omega$.
  From Theorem~\ref{thm:Kirchhoff_Helmholtz}, we have 
  \begin{align}
    u(\bm{r}) 
    & = 
    \int_{\bm{r}_\mathrm{s} \in \partial B(\bm{r}_0,R^+)}
    \bigl(
    G(\bm{r}; \bm{r}_\mathrm{s}) \nabla_\mathrm{s} u(\bm{r}_\mathrm{s})
    \nonumber 
    \\ 
    & \hspace{1.5em} - 
    u(\bm{r}_\mathrm{s}) \nabla_\mathrm{s} G(\bm{r}; \bm{r}_\mathrm{s}) 
    \bigr)
    \dotproduct
    \bm{n}_{B(\bm{r}_0,R^+)}(\bm{r}_\mathrm{s})
    \diff \surf
    \quad 
    \forall \bm{r} \in \overline{B}(\bm{r}_0, R). 
  \end{align}
  Then, applying Theorem~\ref{thm:addition_theorem_Green_function} to the integrand yields 
  \begin{align}
    u(\bm{r}) 
    & = 
    \int_{\bm{r}_\mathrm{s} \in \partial B(\bm{r}_0, R^+)} 
    \sum_{\nu,\mu} \Bigl( \psi_{\nu,\mu}(\bm{r}_\mathrm{s} - \bm{r}_0) \varphi_{\nu,\mu}(\bm{r} - \bm{r}_0) \nabla_\mathrm{s} u (\bm{r}_\mathrm{s}) 
    \nonumber\\
    & \hspace{2em} - 
    u (\bm{r}_\mathrm{s}) \varphi_{\nu,\mu}(\bm{r} - \bm{r}_0) \nabla_\mathrm{s} \psi_{\nu,\mu}(\bm{r}_\mathrm{s} - \bm{r}_0) \Bigr) 
    \dotproduct 
    \bm{n}_{B(\bm{r}_0, R^+)}(\bm{r}_\mathrm{s}) 
    \diff \surf 
    \nonumber\\
    & \hspace{16em}
    \quad 
    \forall \bm{r} \in \overline{B}(\bm{r}_0, R). 
  \end{align}
  Here, from the continuity of $u$, $\nabla u$, and $\bm{n}_{B(\bm{r},R^+)}$ on the bounded closed set $\partial B(\bm{r}_0,R^+)$ and the uniform convergence in Theorem~\ref{thm:addition_theorem_Green_function}, the integrand of the right-hand side converges with respect to $\bm{r}_\mathrm{s}$ uniformly on $\partial B(\bm{r}_0, R^+)$.
  Therefore, the summation and integration can be interchanged, and we have 
  \begin{align}
    u(\bm{r}) = \sum_{\nu,\mu} \alpha_{\nu,\mu} \varphi_{\nu,\mu}(\bm{r} - \bm{r}_0)
    \quad 
    \forall \bm{r} \in \overline{B}(\bm{r}_0, R) 
    \label{eq:proof_u_expansion}
  \end{align}
  with 
  \begin{align}
    \alpha_{\nu,\mu} 
    & = 
    \int_{\bm{r}_\mathrm{s} \in \partial B(\bm{r}_0, R^+)} 
    \Bigl( \psi_{\nu,\mu}(\bm{r}_\mathrm{s} - \bm{r}_0) \nabla_\mathrm{s} u (\bm{r}_\mathrm{s}) 
    \nonumber\\
    & \hspace{5em} - 
    u (\bm{r}_\mathrm{s}) \nabla_\mathrm{s} \psi_{\nu,\mu}(\bm{r}_\mathrm{s} - \bm{r}_0) \Bigr) 
    \dotproduct 
    \bm{n}_{B(\bm{r}_0, R^+)}(\bm{r}_\mathrm{s}) 
    \diff \surf. 
  \end{align}
  Then, by repeatedly applying Theorem~\ref{thm:Helmholtz_integral} in a manner similar to the proof of Theorem~\ref{thm:Kirchhoff_Helmholtz}, we have 
  \begin{align}
    \alpha_{\nu,\mu} 
    & = 
    \int_{\bm{r}_\mathrm{s} \in \partial D} 
    \Bigl( \psi_{\nu,\mu}(\bm{r}_\mathrm{s} - \bm{r}_0) \nabla_\mathrm{s} u (\bm{r}_\mathrm{s}) 
    \nonumber\\
    & \hspace{5em} - 
    u (\bm{r}_\mathrm{s}) \nabla_\mathrm{s} \psi_{\nu,\mu}(\bm{r}_\mathrm{s} - \bm{r}_0) \Bigr) 
    \dotproduct 
    \bm{n}_D(\bm{r}_\mathrm{s}) 
    \diff \surf. 
    \label{eq:proof_expansion_integral}
  \end{align}
  Here, we have
  \begin{align}
    & 
    \max_{\bm{r} \in \overline{B}(\bm{r}_0, R)}
    \left|
      \sum_{\nu = N}^\infty \sum_{\mu = -\nu}^\nu 
      \alpha_{\nu,\mu} \varphi_{\nu,\mu}(\bm{r}-\bm{r}_0) 
    \right|
    \nonumber\\ 
    & \leq 
    \surf(\partial B(\bm{r}_0, R^+)) \cdot 
    \max_{\bm{r}_\mathrm{s} \in \partial B(\bm{r}_0, R^+)} 
    \left| 
      \frac{\partial}{\partial \|\bm{r}_\mathrm{s} - \bm{r}_0\|} u(\bm{r}_\mathrm{s})
    \right|
    \nonumber\\
    & \quad \cdot \max_{\substack{\bm{r}_\mathrm{s} \in \partial B(\bm{r}_0, R^+),\\ \bm{r} \in \overline{B}(\bm{r}_0, R)}}
    \left| 
      \sum_{\nu = N}^\infty \sum_{\mu = -\nu}^\nu \psi_{\nu,\mu}(\bm{r}_\mathrm{s}-\bm{r}_0) \varphi_{\nu,\mu}(\bm{r}-\bm{r}_0)
    \right|
    \nonumber\\ 
    & \quad + 
    \surf(\partial B(\bm{r}_0, R^+)) \cdot 
    \max_{\bm{r}_\mathrm{s} \in \partial B(\bm{r}_0, R^+)} 
    \left| 
      u(\bm{r}_\mathrm{s})
    \right|
    \nonumber\\
    & \quad \cdot 
    \max_{\substack{\bm{r}_\mathrm{s} \in \partial B(\bm{r}_0, R^+),\\ \bm{r} \in \overline{B}(\bm{r}_0, R)}} 
    \left| 
      \sum_{\nu = N}^\infty \sum_{\mu = -\nu}^\nu \frac{\partial}{\partial \|\bm{r}_\mathrm{s} - \bm{r}_0\|} \psi_{\nu,\mu}(\bm{r}_\mathrm{s}-\bm{r}_0) \varphi_{\nu,\mu}(\bm{r}-\bm{r}_0)
    \right|,
  \end{align}
  and from the uniform convergence for \eqref{eq:addition_theorem_Green_function} and \eqref{eq:addition_theorem_Green_function_derivative} in Theorem~\ref{thm:addition_theorem_Green_function}, we have 
  \begin{align}
    \lim_{N \to \infty} 
    \max_{\bm{r} \in \overline{B}(\bm{r}_0, R+)}
    \left|
      \sum_{\nu = N}^\infty \sum_{\mu = -\nu}^\nu 
      \alpha_{\nu,\mu} \varphi_{\nu,\mu}(\bm{r}-\bm{r}_0) 
    \right|
    = 
    0.
  \end{align} 
  This means that the right-hand side of \eqref{eq:proof_u_expansion} converges uniformly on $\overline{B}(\bm{r}_0,R)$. 
  On the other hand, from Theorem~\ref{thm:Kirchhoff_Helmholtz}, we have 
  \begin{align}
    u(\bm{r}) 
    & 
    = 
    \int_{\bm{r}_\mathrm{s} \in \partial D}
    (
    G(\bm{r}; \bm{r}_\mathrm{s}) \nabla_\mathrm{s} u(\bm{r}_\mathrm{s}) 
    \nonumber\\
    & \hspace{4em} 
    - 
    u(\bm{r}_\mathrm{s}) \nabla_\mathrm{s} G(\bm{r}; \bm{r}_\mathrm{s}) 
    )
    \dotproduct 
    \bm{n}_D(\bm{r}_\mathrm{s})
    \diff \surf 
    \quad \forall \bm{r} \in D. 
  \end{align}
  By applying $\frac{1}{(-ik)^\nu} \mathcal{Y}_{\nu,\mu}^\ast$ to both sides of \eqref{eq:Kirchhoff_Helmholtz}, we have 
  \begin{align}
    & \frac{1}{(-ik)^\nu} \mathcal{Y}_{\nu,\mu}^\ast u(\bm{r}) 
    \nonumber 
    \\
    & 
    = 
    \frac{1}{(-ik)^\nu} 
    \int_{\bm{r}_\mathrm{s} \in \partial D}
    \Bigl(
    \mathcal{Y}_{\nu,\mu}^\ast G(\bm{r}; \bm{r}_\mathrm{s}) \nabla_\mathrm{s} u (\bm{r}_\mathrm{s})
    \nonumber 
    \\
    & \hspace{8em} 
    - 
    u (\bm{r}_\mathrm{s}) \nabla_\mathrm{s} \mathcal{Y}_{\nu,\mu}^\ast G(\bm{r}; \bm{r}_\mathrm{s})  
    \Bigr)
    \dotproduct 
    \bm{n}_D(\bm{r}_\mathrm{s})
    \diff \surf 
    \quad \forall \bm{r} \in D. 
  \end{align}
  Here, the integration and partial differentiation can be interchanged because of the compactness of $\partial D$ and the infinite differentiability of $G(\placeholder)$. 
  Finally, applying Theorem~\ref{eq:differentiation_theorem_Green_function} to the integrand yields 
  \begin{align}
    & \frac{1}{(-ik)^\nu} \mathcal{Y}_{\nu,\mu}^\ast u(\bm{r}) 
    \nonumber 
    \\
    & = 
    \int_{\bm{r}_\mathrm{s} \in \partial D} 
    \Bigl( \psi_{\nu,\mu}(\bm{r}_\mathrm{s} - \bm{r}_0) \nabla_\mathrm{s} u (\bm{r}_\mathrm{s}) 
    \nonumber\\
    & \hspace{4em} - 
    u (\bm{r}_\mathrm{s}) \nabla_\mathrm{s} \psi_{\nu,\mu}(\bm{r}_\mathrm{s} - \bm{r}_0) \Bigr) 
    \dotproduct 
    \bm{n}_D(\bm{r}_\mathrm{s}) 
    \diff \surf
    \quad \forall \bm{r} \in D. 
    \label{eq:proof_expansion_derivative}
  \end{align}
  Equations \eqref{eq:proof_u_expansion}, \eqref{eq:proof_expansion_integral}, and \eqref{eq:proof_expansion_derivative} conclude the proof. 
\end{proof}

\subsection{Translation and rotation of spherical wave function}
\label{eq:preliminary_translation_rotation}

The translation and rotation of the spherical wave functions are given by the following formulae, which relate the expansions in Theorem~\ref{thm:expansion_spherical_wave_function} for different coordinate systems: 
\begin{description}
  \item[Translation of spherical wave function:]
  \begin{align}
    \hspace{-2em}
    \varphi_{\nu',\mu'}(\bm{r} + \bm{r}')
    & = 
    \sum_{\nu,\mu} 
    T_{\nu',\mu'}^{\nu,\mu}(\bm{r}') \varphi_{\nu,\mu}(\bm{r}) 
    \nonumber\\
    & \qquad 
    \forall \bm{r}, \bm{r}' \in \mathbb{R}^3, \ \nu' \in \mathbb{N}, \ \mu' \in \llbracket -\nu', \nu' \rrbracket.
    \label{thm:tranlsation_spherical_wave_function}
   \end{align}
  \item[Rotation of spherical wave function:] 
  \begin{align}
    \hspace{-2em}
    \varphi_{\nu',\mu'}(\mathcal{R} \bm{r})
    & = 
    \sum_{\nu,\mu} 
    \delta_{\nu,\nu'}
    D_{\mu,\mu'}^{(\nu)}(\mathcal{R})^\ast \varphi_{\nu,\mu}(\bm{r}) 
    \nonumber\\
    & \qquad 
    \forall \bm{r} \in \mathbb{R}^3, \ \mathcal{R} \in \mathrm{SO}(3), \ \nu' \in \mathbb{N}, \ \mu' \in \llbracket -\nu', \nu' \rrbracket.
    \label{thm:rotation_spherical_wave_function}
  \end{align}
\end{description}
Here, $T_{\nu',\mu'}^{\nu,\mu}(\placeholder): \mathbb{R}^3 \to \mathbb{C}$ and $D^{(\nu)}_{\mu,\mu'}(\placeholder): \mathrm{SO}(3) \to \mathbb{C}$ are called the \emph{translation operator}~\cite{Martin:Cambridge2006} and \emph{Wigner $D$-matrix}~\cite{Edmonds:Princeton1996}, respectively.
The translation operator is given by 
\begin{align}
  T_{\nu',\mu'}^{\nu,\mu}(\bm{r}) 
  & \coloneqq 
  \sum_{\nu'',\mu''}^{\nu+\nu'} 
  \mathcal{G}(\nu, \mu; \nu', \mu'; \nu'', \mu'') 
  \varphi_{\nu'',\mu''}(\bm{r})
  \nonumber\\
  & \qquad 
  (\bm{r} \in \mathbb{R}^3, \ \nu, \nu' \in \mathbb{N}, \ \mu \in \llbracket -\nu, \nu \rrbracket, \ \mu' \in \llbracket -\nu', \nu' \rrbracket)
  \label{eq:def_translation_operator}
\end{align}
with the modified Gaunt coefficient $\mathcal{G}(\placeholder,\placeholder;\placeholder,\placeholder;\placeholder,\placeholder)$, defined as 
\begin{align}
  & \mathcal{G}(\nu, \mu; \nu', \mu'; \nu'', \mu'') 
  \nonumber\\
  & \coloneqq 
  \frac{1}{4\pi}
  \int_{\bm{x} \in \mathbb{S}_2}
  \hat{\sphharm}_{\nu,\mu}(\bm{x})^\ast 
  \hat{\sphharm}_{\nu',\mu''}(\bm{x})
  \hat{\sphharm}_{\nu'',\mu''}(\bm{x})^\ast 
  \diff \surf
  \nonumber\\
  & \qquad 
  (\nu, \nu', \nu'' \in \mathbb{N}, \ \mu \in \llbracket -\nu, \nu \rrbracket, \ \mu' \in \llbracket -\nu', \nu' \rrbracket, \ \mu'' \in \llbracket -\nu'', \nu'' \rrbracket). 
  \label{eq:def_Gaunt_coefficient}
\end{align}
The Wigner $D$-matrix is given by 
\begin{align}
  D_{\mu,\mu'}^{(\nu)}(\mathcal{R}) 
  \coloneqq
  \frac{1}{4\pi}
  \int_{\bm{x} \in \mathbb{S}_2}
  \hat{\sphharm}_{\nu,\mu}(\mathcal{R}\bm{x})^\ast 
  \hat{\sphharm}_{\nu,\mu'}(\bm{x})
  \diff \surf.
  \label{eq:def_Wigner_D_matrix}
\end{align}
The closed-form expressions of \eqref{eq:def_Gaunt_coefficient} and \eqref{eq:def_Wigner_D_matrix} can be obtained using the formulae in \cite{Martin:Cambridge2006} and \cite{Edmonds:Princeton1996}, respectively. 

\section{Approximation Using Spherical/Plane Wave Functions}

Finally, we introduce two approximation theorems for the solutions of the Helmholtz equation. 
\begin{mdframed}
  \begin{theorem}
    \label{thm:approximation_spherical_wave_function}
    Let $u \in \mathcal{H}(\Omega; \wavenumber)$, $\bm{r}_0 \in \mathbb{R}^3$, and $C \subseteq \Omega$ be a simply connected bounded closed set. 
    Then, for any $\epsilon \in (0,\infty)$, there exists some $u_{\mathrm{approx}} \in \mathcal{H}(\Omega; \wavenumber)$ satisfying 
    \begin{align}
      | u(\bm{r}) - u_{\mathrm{approx}}(\bm{r}) | < \epsilon \quad \forall \bm{r} \in C
    \end{align}
    in the form of 
    \begin{align}
      u_{\mathrm{approx}}(\bm{r})
      = 
      \sum_{\nu,\mu}^N \alpha_{\nu,\mu} \varphi_{\nu,\mu}(\bm{r}-\bm{r}_0) 
      \quad (\bm{r} \in \Omega)
    \end{align}
    with some $N \in \mathbb{N}$ and $\alpha_{\nu,\mu} \in \mathbb{C}$ ($\nu \in \mathbb{N}, \mu \in \llbracket -\nu, \nu \rrbracket$). 
  \end{theorem}
  \begin{theorem}
    \label{thm:approximation_plane_wave_function}
    Let $u \in \mathcal{H}(\Omega; \wavenumber)$, $\bm{r}_0 \in \Omega$, and $C \subseteq \Omega$ be a simply connected bounded closed set. 
    Then, for any $\epsilon \in (0,\infty)$, there exists some $u_{\mathrm{approx}} \in \mathcal{H}(\Omega; \wavenumber)$ satisfying 
    \begin{align}
      | u(\bm{r}) - u_{\mathrm{approx}}(\bm{r}) | < \epsilon \quad \forall \bm{r} \in C
    \end{align}
    in the form of 
    \begin{align}
      u_{\mathrm{approx}}(\bm{r}) 
      = 
      \int_{\bm{x} \in \mathbb{S}_2} 
      w(\bm{x}) 
      \exp(-\im \wavenumber \bm{x} \dotproduct (\bm{r} - \bm{r}_0)) 
      \diff \surf 
    \end{align}
    with some square-integrable function $w: \mathbb{S}_2 \to \mathbb{C}$. 
  \end{theorem}
\end{mdframed}
\begin{proof}
  See \cite{Ueno:IEEE2019}. 
\end{proof}

The relationship between Theorems~\ref{thm:expansion_spherical_wave_function} and \ref{thm:approximation_spherical_wave_function} is similar to that between Taylor's theorem~\cite{Brown:McGrawn2014} and Runge's theorem~\cite{Greene:AMS2006} for holomorphic functions in complex analysis. 
In contrast to the expansion theorem, which is valid only within a specific radius of convergence, the above approximation theorems hold for a broader class of regions. 

\chapter{Sound Field Estimation with Boundary Measurement Approach}
\label{sec:boundary}

Many of the early studies on sound field estimation, such as the near-field acoustic holography and higher-order ambisonics, are based on the theory of the boundary integral, which aims to reconstruct the sound field from the boundary measurement. 
This boundary measurement approach is still one of the commonly used approaches today, particularly in sound field estimation using a spherical microphone array. 

The goal in this section is to review the boundary measurement approach. 
Throughout this section, we focus specifically on sound field estimation for spherically bounded regions. However, theories concerning other types of boundary, such as planes, cylinders, and other general geometries, are also explored in the literature~\cite{Williams:AcademicPress1999}. 
From Sections~\ref{sec:boundary_motivation} to \ref{sec:boundary_expansion}, we derive essentially the same formulae for determining a sound field from its boundary pressure values using two different approaches. 
Since these formulae are invalid at specific frequencies, we introduce two improved array designs to address this problem. 
Discretization methods for microphone positions are also discussed in Section~\ref{sec:boundary_discretization}. 
The formulations for each array design are summarized and compared in Section~\ref{sec:boundary_summary}. 
The results of simulations to evaluate estimation performance are presented in Section~\ref{sec:boundary_simulation}. 
Finally, related studies are reviewed in Section~\ref{sec:boundary_related}. 

\section{Theoretical Motivation}
\label{sec:boundary_motivation}

In Section~\ref{sec:preliminaries}, we introduced the Kirchhoff--Helmholtz integral theorem (Theorem~\ref{thm:Kirchhoff_Helmholtz}): 
\begin{align}
  u(\bm{r}) 
  & = 
  \int_{\bm{r}_\mathrm{s} \in \partial D}
  \bigl(
  G(\bm{r}; \bm{r}_\mathrm{s}) \nabla_\mathrm{s} u(\bm{r}_\mathrm{s})
  \nonumber 
  \\ 
  & \hspace{4em} - 
  u(\bm{r}_\mathrm{s}) \nabla_\mathrm{s}  G(\bm{r}; \bm{r}_\mathrm{s})
  \bigr)
  \dotproduct
  \bm{n}_D(\bm{r}_\mathrm{s})
  \diff \surf 
  \quad \forall \bm{r} \in D, 
  \label{eq:Kirchhoff_Helmholtz_reiteration}
\end{align}
where $u \in \mathcal{H}(\Omega; \wavenumber)$ and $D \subset \Omega$ is a bounded open set with a $\mathcal{C}_1$-smooth boundary $\partial D \subset \Omega$. 
This identity implies that the sound field $u$ can be determined within the region $D$ by observing its pressure values $u(\bm{r}_\mathrm{s})$ and normal derivatives $\nabla u(\bm{r}_\mathrm{s}) \dotproduct \bm{n}_D(\bm{r}_\mathrm{s})$ on the boundary $\partial D$, which can be achieved theoretically by using continuously aligned omnidirectional and bidirectional microphones. 
However, it is burdensome to realize such an observation because these two types of microphone, omnidirectional and bidirectional ones, are required, or alternatively, two layers of microphone arrays have to be aligned closely to approximate these two types of directivity. 
A certain modification for Green's function is applied to omit either of the two terms in \eqref{eq:Kirchhoff_Helmholtz_reiteration}, \ie, to estimate the sound field only from a single layer array of microphones. 
As will be shown, in the case of a spherical boundary, this problem can be solved in a closed form except for specific frequencies. 

On the other hand, the spherical wave function expansion (Theorem~\ref{thm:expansion_spherical_wave_function}), given by 
\begin{align}
  u(\bm{r}) 
  & = 
  \sum_{\nu,\mu} 
  \ring{u}_{\nu,\mu}(\bm{r}_0) 
  \varphi_{\nu,\mu}(\bm{r}-\bm{r}_0) 
  \quad 
  \forall \bm{r} \in \overline{B}(\bm{r}_0, R)
\end{align}
with $\bm{r}_0 \in \Omega$ and $R \in (0, \infty)$ satisfying $\overline{B}(\bm{r}_0, R) \subset \Omega$, provides another clue to sound field estimation. 
From the orthogonality of the spherical harmonic functions in $\varphi_{\nu,\mu}(\placeholder)$, one can determine the expansion coefficients on the right-hand side from the pressure values of $u$ on the spherical boundary. 

Interestingly, essentially the same formulae are derived from the above two different approaches, as will be shown in Sections~\ref{sec:boundary_Green} and \ref{sec:boundary_expansion}. 
These underlying theories form the fundamentals of the boundary measurement approach. 

\section{Green's Function Approach}
\label{sec:boundary_Green}

First, we consider a method for estimating the sound field from only the sound pressure values on the boundary by employing the Dirichlet Green's functions. 

\subsection{Derivation of estimation formula}

For a bounded open set $D \subset \Omega$ with a $\mathcal{C}_1$-smooth boundary $\partial D \subset \Omega$, the Dirichlet Green's function $G_\mathrm{D}(\placeholder;\placeholder)$ is given (if exists) by 
\begin{align}
  G_\mathrm{D}(\bm{r};\bm{r}') 
  \coloneqq 
  G(\bm{r};\bm{r}') 
  + 
  v_\mathrm{D}(\bm{r};\bm{r}')
  \quad (\bm{r} \in D, \ \bm{r}' \in D^{+}_{\bm{r}} \setminus \{\bm{r}\}),
\end{align}
where for each fixed $\bm{r} \in D$, $D^{+}_{\bm{r}} \subset \Omega$ is an open set satisfying $D \cap \partial D \subset D^{+}_{\bm{r}}$ and $v_\mathrm{D}(\bm{r}; \placeholder) \in \mathcal{H}(D^{+}_{\bm{r}}; \wavenumber)$ satisfies the following boundary condition: 
\begin{align}
  G_\mathrm{D}(\bm{r};\bm{r}_\mathrm{s})
  = 
  0 
  \quad \forall \bm{r}_\mathrm{s} \in \partial D. 
  \label{eq:Dirichlet_Green_boundary}
\end{align} 
For such $G_\mathrm{D}(\placeholder;\placeholder)$, from the above boundary condition and Theorems~\ref{thm:Helmholtz_integral} and \ref{thm:Kirchhoff_Helmholtz}, we have
\begin{align}
  u(\bm{r}) 
  = 
  -
  \int_{\bm{r}_\mathrm{s} \in \partial D} 
  u(\bm{r}_\mathrm{s}) \nabla_\mathrm{s} G_\mathrm{D}(\bm{r};\bm{r}_\mathrm{s}) 
  \dotproduct
  \bm{n}_D(\bm{r}_\mathrm{s})
  \diff \surf
  \quad 
  \forall \bm{r} \in D. 
  \label{eq:Dirichlet_Kirchhoff_Helmholtz}
\end{align}
This relation means that the sound field $u$ can be determined only from its boundary value on $\partial D$ if the Dirichlet Green's function exists. 

In the case of the spherical region (assumed to be centered at the origin without loss of generality), \ie, $D = B(\bm{0}, R)$ with $R \in (0,\infty)$, $v_\mathrm{D}(\placeholder; \placeholder)$ is given by 
\begin{align}
  v_\mathrm{D}(\bm{r}; \bm{r}') 
  = 
  -
  \frac{\im \wavenumber}{4\pi}
  \sum_{\nu,\mu}
  \frac{h_\nu(\wavenumber R)}{j_\nu(\wavenumber R)}
  \varphi_{\nu,\mu}(\bm{r}')^\ast \varphi_{\nu,\mu}(\bm{r})
  \quad (\bm{r} \in D, \ \bm{r}' \in D^{+}_{\bm{r}})
  \label{eq:v_Dirichlet_sphere}
\end{align}
with $D^{+}_{\bm{r}} \coloneqq B(\bm{0}, R^2 / \|\bm{r}\|)$, if $j_\nu(\wavenumber R) \neq 0$ for all $\nu \in \mathbb{N}$. 
Here, the convergence of the right-hand side of \eqref{eq:v_Dirichlet_sphere} can be verified by using Theorem~2.31 in \cite{Kirsch:Springer2015}. 
\begin{figure}
  \centering
  \includegraphics[width=0.4\linewidth]{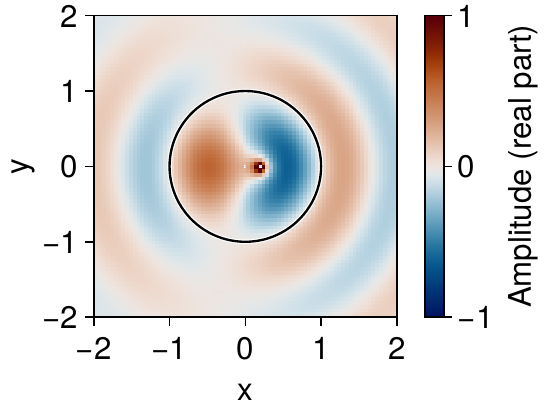}
  \caption{Dirichlet Green's function $G_D(\bm{r};\bm{r}_\mathrm{s})$ plotted with respect to $\bm{r}_\mathrm{s}$ for $xy$-plane ($R = 1$, $\bm{r} = (2, 0, 0)$, $k = 1.5 \pi$).}
  \label{fig:Dirichlet_Green_function}
\end{figure}
Figure~\ref{fig:Dirichlet_Green_function} shows an example of the Dirichlet Green's function. 
One can see the boundary condition in \eqref{eq:Dirichlet_Green_boundary} from this figure. 
Indeed, this boundary condition can be easily confirmed from 
\begin{align}
  G_\mathrm{D}(\bm{r}; \bm{r}_\mathrm{s})
  & = 
  \frac{\im \wavenumber}{4\pi}
  \sum_{\nu,\mu}
  \im^\nu h_\nu(\wavenumber R) \hat{\sphharm}_{\nu,\mu}\left(\frac{\bm{r}_\mathrm{s}}{R}\right)^\ast 
  \varphi_{\nu,\mu}(\bm{r})
  \nonumber\\ 
  & \quad 
  - 
  \frac{\im \wavenumber}{4\pi}
  \sum_{\nu,\mu}
  \frac{h_\nu(\wavenumber R)}{j_\nu(\wavenumber R)}
  \im^\nu j_\nu(\wavenumber R) \hat{\sphharm}_{\nu,\mu}\left(\frac{\bm{r}_\mathrm{s}}{R}\right)^\ast 
  \varphi_{\nu,\mu}(\bm{r}) 
  \nonumber\\ 
  & 
  = 
  0 \quad \forall \bm{r} \in D, \ \bm{r}_\mathrm{s} \in \partial D. 
\end{align}
Furthermore, from the Wronskian relationship~\cite{Williams:AcademicPress1999} 
\begin{align}
  j_\nu(x)h'_\nu(x) - j'_\nu(x)h_\nu(x) 
  = 
  \frac{i}{x^2} \quad \forall x \in \mathbb{R} \setminus \{0\},
  \label{eq:Wronskian_relationship}
\end{align}
we have 
\begin{align}
  & 
  \nabla_\mathrm{s} G(\bm{r}; \bm{r}_\mathrm{s}) 
  \dotproduct
  \bm{n}_D(\bm{r}_\mathrm{s})
  \nonumber\\
  & = 
  \frac{\im \wavenumber}{4\pi}
  \sum_{\nu,\mu}
  \im^\nu \wavenumber
  \left( 
    h'_\nu(kR) 
    -
    \frac{h_\nu(\wavenumber R)}{j_\nu(kR)} j'_\nu(kR) 
  \right)
  \hat{\sphharm}_{\nu,\mu}\left(\frac{\bm{r}_\mathrm{s}}{R}\right)^\ast 
  \varphi_{\nu,\mu}(\bm{r})
  \nonumber\\ 
  & = 
  -\frac{1}{4\pi R^2}
  \sum_{\nu,\mu}
  \frac{\im^\nu}{j_\nu(\wavenumber R)}
  \hat{\sphharm}_{\nu,\mu}\left(\frac{\bm{r}_\mathrm{s}}{R}\right)^\ast 
  \varphi_{\nu,\mu}(\bm{r})
  \quad 
  \forall \bm{r} \in D, \ \bm{r}_\mathrm{s} \in \partial D. 
\end{align}
Thus, by substituting this relation into \eqref{eq:Dirichlet_Kirchhoff_Helmholtz}, we can obtain the estimation as 
\begin{align}
  u(\bm{r})
  & = 
  \frac{1}{4\pi R^2}
  \int_{\bm{r}_\mathrm{s} \in \partial D}
  u(\bm{r}_\mathrm{s})
  \sum_{\nu,\mu}
  \frac{i^\nu}{j_\nu(\wavenumber R)}
  \hat{\sphharm}_{\nu,\mu}\left(\frac{\bm{r}_\mathrm{s}}{R}\right)^\ast 
  \varphi_{\nu,\mu}(\bm{r})
  \diff \surf
  \nonumber\\ 
  & \quad \hspace{17em}
  \quad \forall \bm{r} \in D.
  \label{eq:estimation_formula_Green}
\end{align}

A similar modification of Green's function can also be applied to eliminate the second term in \eqref{eq:Kirchhoff_Helmholtz_reiteration} instead of the first term. 
Such Green's function is called the Neumann Green's function and obtained by imposing the boundary condition 
\begin{align}
  \nabla_\mathrm{s} G_\mathrm{N}(\bm{r};\bm{r}_\mathrm{s}) \cdot \bm{n}_D(\bm{r}_\mathrm{s})
  = 
  0 
  \quad \forall \bm{r}_\mathrm{s} \in \partial D 
  \label{eq:Neumann_Green_boundary}
\end{align}
instead of \eqref{eq:Dirichlet_Green_boundary}, where $G_\mathrm{N}(\placeholder; \placeholder)$ denotes the Neumann Green's function. 
See \cite{Williams:AcademicPress1999} for the explicit representation of the Neumann Green's function in cases of spherically bounded regions.
In this formulation, the sound field can be estimated using bidirectional microphones oriented along the normal derivative of the boundary surface. 

For several other classes of shapes of $D$, such as a rectangle and cylinder, the Dirichlet and Neumann Green's functions can be derived in a closed form~\cite{Williams:JASA1997,Williams:AcademicPress1999, Okoyenta:WSP2020}. 
These Green's functions in a half space, whose boundary is an infinite plane, can also be derived in a closed form as the limit as $D$ becomes infinitely large, which are also known as Rayleigh's formulae~\cite{Williams:AcademicPress1999}.
On the other hand, it is difficult to derive the Dirichlet Green's function for arbitrarily shaped regions; in such cases, numerical methods, such as the boundary element methods and equivalent source methods, are required to solve the relationship between $u$ and its boundary value~\cite{Valdivia:IntechOpen2011}. 

\subsection{Forbidden frequency problem}
\label{sec:boundary_forbidden_frequency}

One can immediately see that \eqref{eq:estimation_formula_Green} is invalid when there exists some $\nu \in \mathbb{N}$ satisfying $j_\nu(\wavenumber R) = 0$ owing to zeros of the denominator. 
From another viewpoint, this condition corresponds to the case where there exists some nonzero $u_\text{mode} \in \mathcal{H}(\Omega; \wavenumber)$ such that 
\begin{align}
  u_\text{mode}(\bm{r}_\mathrm{s}) = 0 \quad \forall \bm{r}_\mathrm{s} \in \partial D, 
\end{align}
which means for any $u \in \mathcal{H}(\Omega, \wavenumber)$, the sound fields $u$ and $u + u_\text{mode}$ cannot be distinguished only from the boundary pressure measurement. 
In particular, $u_\text{mode}(\bm{r}) = \varphi_{\nu,\mu}(\bm{r}) \ (\bm{r} \in \Omega)$ for $\mu \in \llbracket -\nu, \nu \rrbracket$ satisfies this condition. 
This problem is called the \emph{forbidden frequency problem}~\cite{Williams:AcademicPress1999}, and it is known that the forbidden frequency problem occurs for a general class of boundary shapes when $D$ is bounded~\cite{Stakgold:SIAM2000}. 
For a spherical microphone array, several methods are proposed to overcome the forbidden frequency problem, which will be reviewed in Section~\ref{sec:boundary_array_design}.

\section{Spherical Wave Function Expansion Approach}
\label{sec:boundary_expansion}

Essentially the same formula as \eqref{eq:estimation_formula_Green} can also be derived by the spherical wave function expansion. 

\subsection{Derivation of estimation formula}

Again, we use the notations in Section~\ref{sec:boundary_Green}, and we further define $D^+ \coloneqq B(\bm{0}, R^+)$ with $R^+ \in (R, \infty)$ satisfying $\overline{B}(\bm{0}, R^+) \subseteq \Omega$. 
Then, from Theorem~\ref{thm:expansion_spherical_wave_function}, $u$ can be represented around the origin as 
\begin{align}
  u(\bm{r}) 
  & = 
  \sum_{\nu,\mu}
  \alpha_{\nu,\mu} \varphi_{\nu,\mu}(\bm{r})
  \quad 
  \forall \bm{r} \in D^+, 
  \label{eq:expansion_regular_spherical_wave_function_origin}
\end{align}
where $\alpha_{\nu,\mu} \coloneqq \ring{u}_{\nu,\mu}(\bm{0})$ is the expansion coefficient. 
In this approach, this coefficient is estimated from the observed signal. 
Suppose for every $\bm{x} \in \mathbb{S}_2$, the omnidirectional microphone given in \eqref{eq:derivative_omnidirectional} is located at $R\bm{x} \in \partial D$. 
Then, the observation operator, denoted by $\mathcal{F}^{(\mathrm{omni})}_{\bm{x}}: \mathcal{H}(\Omega; \wavenumber) \to \mathbb{C}$, is given by 
\begin{align}
  \mathcal{F}^{(\mathrm{omni})}_{\bm{x}} u
  & = 
  u(R\bm{x}) = 
  \sum_{\nu,\mu}
  \alpha_{\nu,\mu} 
  \frac{1}{\im^\nu} j_\nu(kR) 
  \hat{\sphharm}_{\nu,\mu}(\bm{x})
  \quad 
  \forall \bm{x} \in \mathbb{S}_2. 
\end{align} 
Therefore, from the orthogonality of the spherical harmonic function given in \eqref{eq:orthogonality_spherical_harmonic_function}, the expansion coefficients are derived as 
\begin{align}
  \alpha_{\nu,\mu}  
  & = 
  \frac{\im^\nu}{4\pi j_\nu(\wavenumber R)}
  \int_{\bm{x}\in\mathbb{S}_2} 
  (\mathcal{F}^{(\mathrm{omni})}_{\bm{x}} u)
  \hat{\sphharm}_{\nu,\mu}(\bm{x})^\ast 
  \diff \surf 
  \nonumber\\ 
  & \quad \hspace{8em}
  \forall \nu \in \mathbb{N}, \ \mu \in \llbracket -\nu, \nu \rrbracket, 
  \label{eq:boundary_estimation_coefficient_omnidirectional}
\end{align}
if $j_\nu(\wavenumber R) \neq 0$ for all $\nu \in \mathbb{N}$. 
By substituting this relation into \eqref{eq:expansion_regular_spherical_wave_function_origin}, we can obtain the estimation formula as 
\begin{align}
  u(\bm{r}) 
  & = 
  \sum_{\nu,\mu}
  \frac{\im^\nu}{4\pi j_\nu(kR)} \! 
  \left(
  \int_{\bm{x}\in\mathbb{S}_2} 
  (\mathcal{F}^{(\mathrm{omni})}_{\bm{x}} u)
  \hat{\sphharm}_{\nu,\mu}(\bm{x})^\ast 
  \diff \surf
  \right) \! 
  \varphi_{\nu,\mu}(\bm{r}) 
  \nonumber\\ 
  & = 
  \frac{1}{4\pi R^2}
  \sum_{\nu,\mu}
  \frac{\im^\nu}{j_\nu(kR)}
  \left(
  \int_{\bm{r}_\mathrm{s} \in \partial D}
  u(\bm{r}_\mathrm{s})
  \hat{\sphharm}_{\nu,\mu}\left(\frac{\bm{r}_\mathrm{s}}{R}\right)^\ast 
  \diff \surf 
  \right)
  \varphi_{\nu,\mu}(\bm{r}) 
  \nonumber\\ 
  & \quad \hspace{16em} 
  \quad \forall \bm{r} \in D^+. 
  \label{eq:estimation_formula_expansion}
\end{align}
By comparing \eqref{eq:estimation_formula_Green} and \eqref{eq:estimation_formula_expansion}, one can see that they are identical inside $D$ by interchanging the summation and integration of their right-hand sides. 
However, the region where \eqref{eq:estimation_formula_expansion} is valid is larger than that for \eqref{eq:estimation_formula_Green}, \ie, $D^+ \supset D$. 
It should be noted that the sound field outside as well as inside the microphone array can be estimated by the boundary measurement using the formula \eqref{eq:estimation_formula_expansion}. 

\subsection{Improved array designs}
\label{sec:boundary_array_design}

Since the forbidden frequency problem arises essentially in the observation, not in the estimation methods, the design of an observation system, \ie, a microphone array, should be modified to avoid this problem. 
For spherical microphone arrays, several approaches have been proposed to avoid the forbidden frequency problem~\cite{Meyer:ICASSP2002, Abhayapala:ICASSP2002, Poletti:JAES2005, Rafaely:Springer2019}. 
We will review two examples of such array designs. 

\subsubsection{First-order microphones}

One simple approach is the use of first-order microphones~\cite{Daniel:AESConvention2003}. 
Suppose for every $\bm{x} \in \mathbb{S}_2$, the first-order microphone given in \eqref{eq:derivative_first_order} is located at $R\bm{x} \in \partial D$ so that it faces in the outward normal direction. 
Then, the observation operator, denoted by $\mathcal{F}^{(\mathrm{first})}_{\bm{x}}: \mathcal{H}(\Omega; \wavenumber) \to \mathbb{C}$, is given by 
\begin{align}
  \mathcal{F}^{(\mathrm{first})}_{\bm{x}} u 
  & = 
  \left.\left( a + (1-a) \frac{\im}{\wavenumber} \frac{\partial}{\partial r}\right)
  \sum_{\nu,\mu}
  \alpha_{\nu,\mu}  
  \frac{1}{\im^\nu} j_\nu(\wavenumber r)
  \hat{\sphharm}_{\nu,\mu}(\bm{x})
  \right|_{r = R}
  \nonumber\\ 
  & = 
  \sum_{\nu,\mu}
  \alpha_{\nu,\mu} 
  \frac{1}{\im^\nu} (a j_\nu(\wavenumber R) + \im (1 - a) j'_\nu(\wavenumber R)) 
  \hat{\sphharm}_{\nu,\mu}(\bm{x}). 
\end{align}
Therefore, we have 
\begin{align}
  \alpha_{\nu,\mu} 
  & = 
  \frac{\im^\nu}{4\pi \left(a j_\nu(\wavenumber R) + \im (1 - a) j_\nu'(\wavenumber R) \right)}
  \nonumber\\
  & \quad \cdot
  \int_{\bm{x} \in \mathbb{S}_2} 
  (\mathcal{F}^{(\mathrm{first})}_{\bm{x}} u) 
  \hat{\sphharm}_{\nu,\mu} (\bm{x})^\ast 
  \diff \surf
  \quad \forall \nu \in \mathbb{N}, \ \mu \in \llbracket -\nu, \nu \rrbracket.
  \label{eq:boundary_estimation_coefficient_first_order}
\end{align}
Unlike in \eqref{eq:boundary_estimation_coefficient_omnidirectional}, the denominator in \eqref{eq:boundary_estimation_coefficient_first_order} has no zeros for $a \in (0, 1)$ since its real part $j_\nu(\cdot)$ and imaginary part $j_\nu'(\cdot)$ do not share zeros for any $\nu \in \mathbb{N}$. 

\subsubsection{Rigid sphere}
Another approach is the use of a microphone array mounted on a scattering object~\cite{Meyer:ICASSP2002,Abhayapala:ICASSP2002}. 
When an acoustically rigid sphere bounded by $\partial D$ is placed in the sound field $u$ and the acoustic interaction between the sphere and other objects or walls, such as interreflection, can be ignored, the resulting sound field $u^{(\text{obs})} \in \mathcal{H}(\Omega \setminus (D \cup \partial D); k)$ is represented as 
\begin{align}
  u^{(\text{obs})}(\bm{r}) 
  = 
  u(\bm{r}) 
  + 
  u^{(\text{sct})}(\bm{r}) 
  \quad 
  (\bm{r} \in \Omega \setminus (D \cup \partial D)), 
\end{align}
where $u^{(\text{sct})} \in \mathcal{H}(\Omega \setminus (D \cup \partial D); k)$ denotes the sound field scattered by the rigid sphere. 
From the Sommerfeld radiation condition and the rigid boundary condition~\cite{Martin:Cambridge2006} on $\partial D$, \ie,
\begin{align}
  \lim_{r \to R+0}
  \frac{\partial}{\partial r} 
  u^{(\text{obs})}(r \bm{x}) 
  = 
  0 \quad \forall \bm{x} \in \mathbb{S}_2, 
\end{align}
the scattered sound field $u^{(\text{sct})}$ is determined as~\cite{Poletti:JAES2005, Martin:Cambridge2006}
\begin{align}
  u^{(\text{sct})}(\bm{r}) 
  & = 
  - \sum_{\nu,\mu} 
  \alpha_{\nu,\mu} 
  \frac{1}{\im^\nu}
  \frac{j_\nu'(\wavenumber R)}{h_\nu'(\wavenumber R)}
  h_\nu(k \|\bm{r}\|) \hat{Y}_{\nu,\mu}\left( \frac{\bm{r}}{\|\bm{r}\|}\right)
  \nonumber\\ 
  & \quad \hspace{11em}
  (\bm{r} \in \Omega \setminus (D \cup \partial D)).  
\end{align}

Suppose for every $\bm{x} \in \mathbb{S}_2$, the omnidirectional microphone given in \eqref{eq:derivative_omnidirectional} is mounted on the rigid sphere at $R \bm{x} \in \partial D$. 
Then, the observation operator, denoted by $\mathcal{F}^{(\mathrm{rigid})}_{\bm{x}}: \mathcal{H}(\Omega; \wavenumber) \to \mathbb{C}$, is given by 
\begin{align}
  \mathcal{F}^{(\text{rigid})}_{\bm{x}} u 
  & 
  = 
  \lim_{r \to R+0} u^{(\text{obs})}(r\bm{x})
  \nonumber\\ 
  & 
  = 
  \sum_{\nu,\mu} 
  \alpha_{\nu,\mu} 
  \frac{1}{\im^\nu} 
  \left( j_\nu(kR) - \frac{j'_\nu(\wavenumber R)}{h'_\nu(\wavenumber R)}h_{\nu}(kR) \right) 
  \hat{\sphharm}_{\nu,\mu}(\bm{x})
  \quad 
  \nonumber\\ 
  & 
  = 
  \sum_{\nu,\mu} 
  \alpha_{\nu,\mu} 
  \frac{1}{\im^\nu} 
  \frac{\im}{(\wavenumber R)^2 h'_\nu(\wavenumber R)}
  \hat{\sphharm}_{\nu,\mu}(\bm{x})
  \quad 
  (\bm{x} \in \mathbb{S}_2), 
\end{align}
where the third line is derived using the Wronskian relation \eqref{eq:Wronskian_relationship}. 
Therefore, we have 
\begin{align}
  \alpha_{\nu,\mu} 
  & = 
  \frac{\im^{\nu-1} (\wavenumber R)^2 h'_\nu(\wavenumber R)}{4\pi}
  \int_{\bm{x} \in \mathbb{S}_2} 
  (\mathcal{F}^{(\text{rigid})}_{\bm{x}} u) 
  \hat{\sphharm}_{\nu,\mu} (\bm{x})^\ast 
  \diff \surf 
  \nonumber\\ 
  & \quad \hspace{8em}
  \forall \nu \in \mathbb{N}, \ \mu \in \llbracket -\nu, \nu \rrbracket. 
  \label{eq:boundary_estimation_coefficient_rigid}
\end{align}
Also in this case, the above formula is well defined for all frequencies because it has no zero division. 

The assumption of no further acoustic interaction with other objects is well satisfied when a small array is employed. 
When using a large rigid sphere in an environment with other reflective objects, however, this assumption may no longer hold; its effect on estimation accuracy is investigated in, for example, \cite{Nakanishi:WASPAA2019,Kaneko:JASA2021}. 

\section{Discretization of Spherical Integral}
\label{sec:boundary_discretization}

Because the estimation formulae \eqref{eq:boundary_estimation_coefficient_omnidirectional}, \eqref{eq:boundary_estimation_coefficient_first_order}, and \eqref{eq:boundary_estimation_coefficient_rigid} involve a spherical integral, it should be discretized in practical situations. 
The discretization of the spherical integral of a general function $f:\mathbb{S}_2 \to \mathbb{C}$ is formulated in general as 
\begin{align}
  \int_{\bm{x} \in \mathbb{S}_2} f(\bm{x}) \diff \surf 
  \approx 
  4\pi \sum_{m=1}^M w_m f(\bm{x}_m), 
  \label{eq:spherical_integral}
\end{align}
where $\bm{x}_1, \ldots, \bm{x}_M \in \mathbb{S}_2$ and $w_1, \ldots, w_M \in \mathbb{R}$ are chosen to approximate well the integral. 
One natural choice of $\bm{x}_1, \ldots, \bm{x}_M$ is vertices of a regular polyhedron; in this case, $w_1, \ldots, w_M$ are all set as $1/M$. 
In the three-dimensional Euclidean space, however, there exist only five different regular polyhedra, \ie, $M \in \{4, 6, 8, 12, 20\}$, unlike in the case of regular polygons in the two-dimensional space. 
In other cases, several sampling schemes are applied to realize a nearly uniform discretization of the sphere. 
One representative method is the use of the \emph{spherical design}~\cite{Delsarte:AcademicPress1991}, which is a set of $\bm{x}_1, \ldots, \bm{x}_M$ such that \eqref{eq:spherical_integral} holds without any error with uniform weights $w_1, \ldots, w_M (= 1/M)$ for any polynomial $f$ of a certain degree $t \in \mathbb{N}$ or less (such a set is particularly called a spherical $t$-design). 
The existence and explicit construction of the spherical design for different values of $t$ and $M$ are still to be fully elucidated; however, there are many spherical designs available in the literature~\cite{Chen:PolyU2004}. 
For other sampling schemes such as the Gaussian sampling, see \cite{Rafaely:IEEE2005, Rafaely:Springer2019} and references therein. 

By discretizing the spherical integral, errors occur in the calculations of \eqref{eq:boundary_estimation_coefficient_omnidirectional}, \eqref{eq:boundary_estimation_coefficient_first_order}, and \eqref{eq:boundary_estimation_coefficient_rigid}. 
The distortion in the estimated sound field caused by these errors is referred to as \emph{spatial aliasing}. 
In general, higher-order integrals require finer discretization, which means a larger number of microphones, for accurate calculation. 
However, theoretical analysis of spatial aliasing is challenging because a completely uniform discretization of the sphere is generally unattainable, as previously mentioned. 
Based on analyses under specific constraints or simply emprical rules, several criteria have been provided regarding how many microphones are required for a given wavenumber and the radius of the microphone array~\cite{Jones:ICASSP2002, Poletti:JAES2005}. 
Following these criteria, the expansion in \eqref{eq:expansion_regular_spherical_wave_function_origin} is typically truncated at $\nu = N$ satisfying $N \geq \lceil k R \rceil$ or $N \geq \lceil \frac{\exp(1)}{2} k R \rceil$, and $M \geq (N + 1)^2$ microphones are required to obtain the expansion orders. 
Conversely, when the number of microphones is insufficient, the expansion is often truncated at $\nu = N$ satisfying $M \geq (N + 1)^2$, and the higher-order components are discarded because the spherical integrals are inaccurate. 

\section{Summary and Comparison}
\label{sec:boundary_summary}

\begin{figure}
  \centering
  \begin{minipage}{0.48\linewidth}
    \centering
    \includegraphics{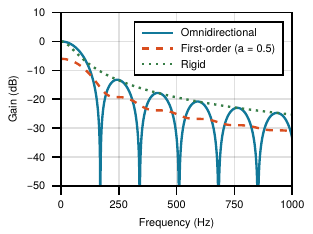}
    \subcaption{$\nu = 0$}
  \end{minipage}
  \begin{minipage}{0.48\linewidth}
    \centering
    \includegraphics{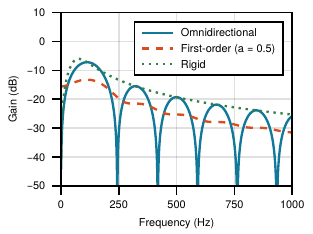}
    \subcaption{$\nu = 1$}
  \end{minipage}
  \begin{minipage}{0.48\linewidth}
    \centering
    \includegraphics{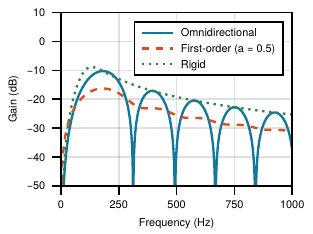}
    \subcaption{$\nu = 2$}
  \end{minipage}
  \begin{minipage}{0.48\linewidth}
    \centering
    \includegraphics{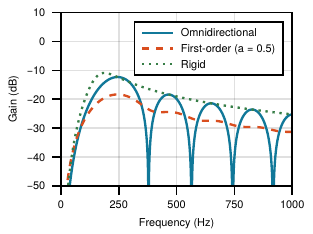}
    \subcaption{$\nu = 3$}
  \end{minipage}
  \caption{Radial response $10 \log_{10} |A_{\nu}^{(\ast)}|^2$ plotted against frequency ($c = \SI{340.65}{m/s}$, $R = \SI{1}{m}$).}
  \label{fig:boundary_measurement_response}
\end{figure}

By summarizing the theories discussed in this section, we can derive the estimated sound field, denoted by $\hat{u} \in \mathcal{H}(D^+, k)$, as 
\begin{align}
  \hat{u}(\bm{r}) 
  = 
  \sum_{\nu,\mu}^N
  \hat{\alpha}_{\nu,\mu} 
  \varphi_{\nu,\mu}(\bm{r}) 
  \quad (\bm{r} \in D^+),
  \label{eq:boundary_estimated_sound_field}
\end{align}
where $N \in \mathbb{N}$ is the truncation order and the estimated coefficients $\hat{\alpha}_{0,0}, \ldots, \hat{\alpha}_{N,N} \in \mathbb{C}$ are obtained as 
\begin{align}
  \hat{\alpha}_{\nu,\mu} 
  = 
  \frac{1}{A^{(\ast)}_{\nu}} 
  \sum_{m = 1}^M w_m \hat{\sphharm}_{\nu,\mu}(\bm{x}_m)^\ast s_m 
  \quad \forall \nu \in \llbracket 0, N \rrbracket, \ \mu \in \llbracket -\nu, \nu \rrbracket
  \label{eq:boundary_estimated_coefficient}
\end{align}
from the observed signal $s_m \in \mathbb{C}$ modeled by 
\begin{align}
  s_m = \mathcal{F}^{(\ast)}_{\bm{x}_m} u + \epsilon_m \quad (m \in \llbracket 0, M \rrbracket)
  \label{eq:boundary_observed_signal}
\end{align} 
with the observation noise $\epsilon_m \in \mathbb{C}$. 
Here, $(\ast) \in \{(\text{omni}), (\text{first}), (\text{rigid})\}$ denotes the array type, and the radial response $A^{(\ast)}_{\nu} \in \mathbb{C}$ is given by 
\begin{align}
  A^{(\ast)}_{\nu} 
  = 
  \begin{dcases}
    \frac{1}{\im^\nu} j_\nu(\wavenumber R) & ((\ast) = (\text{omni}))
    \\ 
    \frac{1}{\im^\nu} \left(a j_\nu(\wavenumber R) + \im (1 - a) j_\nu'(\wavenumber R) \right) & ((\ast) = (\text{first}))
    \\
    \frac{1}{\im^{\nu-1} (\wavenumber R)^2 h'_\nu(\wavenumber R)} & ((\ast) = (\text{rigid}))
  \end{dcases}
\end{align}
for $\nu \in \llbracket 0, N \rrbracket$. 
In a matrix-vector form, the above formula between $\hat{\bm{\alpha}} \coloneqq [\hat{\alpha}_{0,0}, \ldots, \hat{\alpha}_{N,N}] \in \mathbb{C}^{(N+1)^2}$ and $\bm{s} \coloneqq [s_1, \ldots, s_M] \in \mathbb{C}^M$ can be rewritten as 
\begin{align}
  \hat{\bm{\alpha}} 
  = 
  \bm{A}^{-1} \bm{Y}^\mathsf{H} \bm{W} \bm{s}
\end{align}
with $\bm{A} \in \mathbb{C}^{(N+1)^2 \times (N+1)^2}$, $\bm{Y} \in \mathbb{C}^{M \times (N+1)^2}$, and $\bm{W} \in \mathbb{R}^{M \times M}$ defined respectively as 
\begin{align}
  \bm{A} 
  = 
  \begin{bmatrix}
    A_0^{(\ast)} &  & 0
    \\ 
    & \ddots & 
    \\
    0 & & A_N^{(\ast)} 
  \end{bmatrix},
\end{align}
\begin{align}
  \bm{Y} 
  = 
  \begin{bmatrix}
    \hat{\sphharm}_{0,0}(\bm{x}_1) & \hdots & \hat{\sphharm}_{N,N}(\bm{x}_1)
    \\ 
    \vdots & \ddots & \vdots 
    \\
    \hat{\sphharm}_{0,0}(\bm{x}_M) & \hdots & \hat{\sphharm}_{N,N}(\bm{x}_M)
  \end{bmatrix},
\end{align}
and 
\begin{align}
  \bm{W} 
  = 
  \begin{bmatrix}
    w_1 &  & 0
    \\ 
    & \ddots & 
    \\
    0 & & w_M
  \end{bmatrix}.
\end{align}
As seen from the above representation, differences regarding array types appear only in the radial response term $\bm{A}$. 
Figure~\ref{fig:boundary_measurement_response} shows an example of radial responses for the three array types. 
One can see from this figure that the radial response for \eqref{eq:boundary_estimation_coefficient_omnidirectional} has zeros, but those for \eqref{eq:boundary_estimation_coefficient_first_order} and \eqref{eq:boundary_estimation_coefficient_rigid} do not. 

\section{Simulation Results}
\label{sec:boundary_simulation}

Here, we present the simulation results of sound field estimation using the three aforementioned types of spherical microphone array, denoted here by \textbf{Omnidirectional}, \textbf{First-order}, and \textbf{Rigid}. 
For \textbf{First-order}, the directivity parameter in \eqref{eq:derivative_first_order} was set as $a = 0.5$. 
Throughout the simulations, the speed of sound was set as $c = \SI{340.65}{m/s}$. 
In a three-dimensional free field, $M = 64$ microphones were placed on a sphere with a radius of $\SI{1}{m}$ centered at the origin. 
Their positions were determined according to the spherical $t$-design with $t = 7$~\cite{Chen:PolyU2004}. 
For \textbf{Omnidirectional} and \textbf{First-order}, the microphones are placed on an open sphere, while for \textbf{Rigid}, they are placed on a rigid sphere. 
The true sound field to be estimated was $u(\bm{r}) = \exp(- \im \wavenumber \bm{x}_\mathrm{inc} \dotproduct \bm{r})$ with the incident direction $\bm{x}_\mathrm{inc} \coloneqq (1, 0, 0)$. 
The observed signals were calculated using \eqref{eq:boundary_observed_signal}, and the observation noises were sampled independently from the circularly symmetric Gaussian distribution with zero mean and a variance of $10^{-3} \times \frac{1}{M} \sum_{m =1}^M |\mathcal{F}^{(\ast)}_m u|^2$ so that the expected signal-to-noise ratio was $\SI{30}{dB}$.  
According to the concept of the spherical design, the weight $w_m$ in \eqref{eq:boundary_estimated_coefficient} was set as $1 / M$ for all $m \in \llbracket 0, M \rrbracket$, and the truncation order $N$ in \eqref{eq:boundary_estimated_sound_field} was set as $7$. 

As an evaluation criterion, the normalized mean square error (NMSE), defined as 
\begin{align}
  \mathrm{NMSE}(\hat{u}, u)
  = 
  10 \log_{10}
  \left(
    \frac{\sum_{\mathrm{i} \in I_\mathrm{eval}} |\hat{u}(\bm{r}^{(\mathrm{i})}_\mathrm{eval}) - u(\bm{r}^{(\mathrm{i})}_\mathrm{eval})|^2}{\sum_{\mathrm{i} \in I_\mathrm{eval}} |u(\bm{r}^{(\mathrm{i})}_\mathrm{eval})|^2}
  \right),
\end{align}
was used, where $\hat{u}$ denotes the estimated sound field, and the evaluation points $\{\bm{r}^{(\mathrm{i})}_\mathrm{eval}\}_{\mathrm{i} \in I_\mathrm{eval}}$ were set as all grid points at $\SI{0.1}{m}$ intervals within the spherical region with a radius of $\SI{1}{m}$ centered at the origin, \ie, inside the microphone array. 
Since the observation noises were randomly determined, the NMSEs were averaged over 10 trials. 

\begin{figure}
  \centering
  \includegraphics{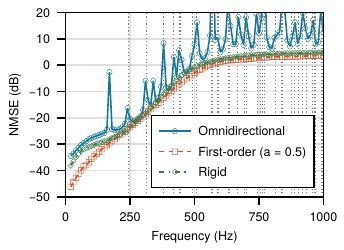}
  \caption{NMSE plotted against frequency. The gray dotted lines denote the forbidden frequencies for an array of omnidirectional microphones.}
  \label{fig:boundary_measurement_NMSE}
\end{figure}

\begin{figure}
  \centering
  \begin{minipage}{\linewidth}
    \centering
    \begin{minipage}{0.4\linewidth}
      \centering
      \includegraphics{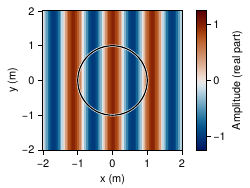}
    \end{minipage}
    \subcaption{True sound field}
    \label{fig:boundary_measurement_true_300Hz}
    \vspace{1em}
  \end{minipage}
  \begin{minipage}{\linewidth}
    \centering
    \begin{minipage}{0.4\linewidth}
      \centering
      \includegraphics{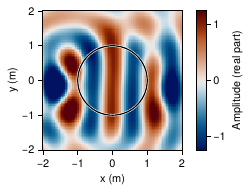}
    \end{minipage}
    \begin{minipage}{0.4\linewidth}
      \centering
      \includegraphics{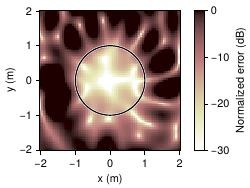}
    \end{minipage}
    \subcaption{Estimated sound field (\textbf{Omnidirectional})}
    \label{fig:boundary_measurement_omnidirectional_300Hz}
    \vspace{1em}
  \end{minipage}
  \begin{minipage}{\linewidth}
    \centering
    \begin{minipage}{0.4\linewidth}
      \centering
      \includegraphics{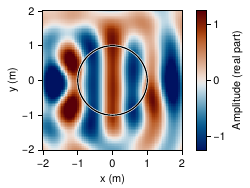}
    \end{minipage}
    \begin{minipage}{0.4\linewidth}
      \centering
      \includegraphics{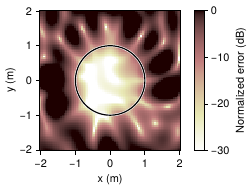}
    \end{minipage}
    \subcaption{Estimated sound field (\textbf{First-order})}
    \label{fig:boundary_measurement_first-order_300Hz}
    \vspace{1em}
  \end{minipage}
  \begin{minipage}{\linewidth}
    \centering
    \begin{minipage}{0.4\linewidth}
      \centering
      \includegraphics{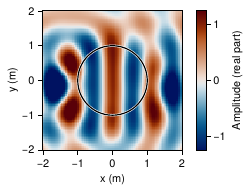}
    \end{minipage}
    \begin{minipage}{0.4\linewidth}
      \centering
      \includegraphics{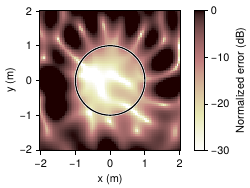}
    \end{minipage}
    \subcaption{Estimated sound field (\textbf{Rigid})}
    \label{fig:boundary_measurement_rigid_300Hz}
  \end{minipage}
  \caption{Results of estimation at \SI{300}{Hz}: Distributions of sound pressure (in (\textbf{a}) and the left panels of (\textbf{b}), (\textbf{c}), and (\textbf{d})) and normalized error (in the right panels of (\textbf{b}), (\textbf{c}), and (\textbf{d})) plotted in the $xy$-plane ($\bm{r} = \SI{(x, y, 0)}{m}$). The black solid line denotes the boundary of the microphone array.}
  \label{fig:boundary_measurement_300Hz}
\end{figure}

\begin{figure}
  \centering
  \begin{minipage}{\linewidth}
    \centering
    \begin{minipage}{0.4\linewidth}
      \centering
      \includegraphics{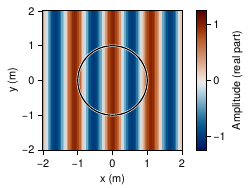}
    \end{minipage}
    \subcaption{True sound field}
    \label{fig:boundary_measurement_true_310Hz}
    \vspace{1em}
  \end{minipage}
  \begin{minipage}{\linewidth}
    \centering
    \begin{minipage}{0.4\linewidth}
      \centering
      \includegraphics{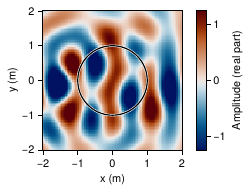}
    \end{minipage}
    \begin{minipage}{0.4\linewidth}
      \centering
      \includegraphics{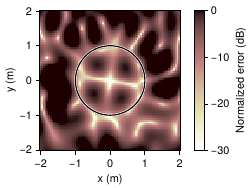}
    \end{minipage}
    \subcaption{Estimated sound field (\textbf{Omnidirectional})}
    \label{fig:boundary_measurement_omnidirectional_310Hz}
    \vspace{1em}
  \end{minipage}
  \begin{minipage}{\linewidth}
    \centering
    \begin{minipage}{0.4\linewidth}
      \centering
      \includegraphics{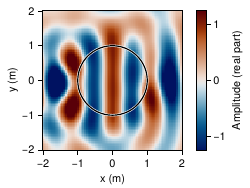}
    \end{minipage}
    \begin{minipage}{0.4\linewidth}
      \centering
      \includegraphics{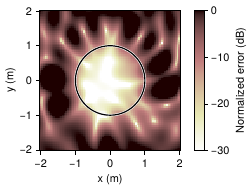}
    \end{minipage}
    \subcaption{Estimated sound field (\textbf{First-order})}
    \label{fig:boundary_measurement_first-order_310Hz}
    \vspace{1em}
  \end{minipage}
  \begin{minipage}{\linewidth}
    \centering
    \begin{minipage}{0.4\linewidth}
      \centering
      \includegraphics{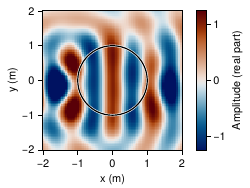}
    \end{minipage}
    \begin{minipage}{0.4\linewidth}
      \centering
      \includegraphics{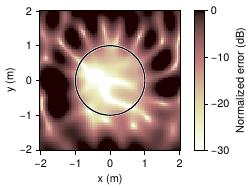}
    \end{minipage}
    \subcaption{Estimated sound field (\textbf{Rigid})}
    \label{fig:boundary_measurement_rigid_310Hz}
  \end{minipage}
  \caption{Results of estimation at \SI{310}{Hz}: Distributions of sound pressure (in (\textbf{a}) and the left panels of (\textbf{b}), (\textbf{c}), and (\textbf{d})) and normalized error (in the right panels of (\textbf{b}), (\textbf{c}), and (\textbf{d})) plotted in the $xy$-plane ($\bm{r} = \SI{(x, y, 0)}{m}$). The black solid line denotes the boundary of the microphone array.}
  \label{fig:boundary_measurement_310Hz}
\end{figure}

The relationship between frequencies and NMSEs is plotted in Fig.~\ref{fig:boundary_measurement_NMSE}. 
In many cases, especially at frequencies below $\SI{500}{Hz}$, the NMSEs for the three methods were close; however, the NMSEs for \textbf{Omnidirectional} increased rapidly at several specific frequencies, \eg, $170$, $240$, and $\SI{310}{Hz}$. 
This is considered to be due to the forbidden frequency problem; in particular, around $170.32$, $243.62$, and $\SI{312.47}{Hz}$ correspond to the forbidden frequencies. 
In contrast, the other two array types showed consistently low NMSEs at low frequencies. 
Examples of the true and estimated sound fields and the pointwise normalized error distributions at two different frequencies are also plotted in Figs.~\ref{fig:boundary_measurement_300Hz} and \ref{fig:boundary_measurement_310Hz}. 
There does not appear to be a significant difference between the three array types at $\SI{300}{Hz}$. 
However, the effect of the forbidden frequency problem can also be seen in Fig.~\ref{fig:boundary_measurement_omnidirectional_310Hz}; a decrease in estimation accuracy for \textbf{Omnidirectional} is observed at \SI{310}{Hz}, whereas the other two array types maintain their high estimation accuracies. 

\section{Related Studies}
\label{sec:boundary_related}

Beyond the methods discussed in this section, other approaches to avoiding the forbidden frequency problem have been proposed, such as the use of dual-radius spherical arrays~\cite{Balmages:IEEE2007, Parthy:ICASSP2009}. 
The design of spherical microphone arrays is an actively researched field (see \cite{Rafaely:Springer2019} and references therein for details) with the aim of improving both practical feasibility and numerical stability. 

Theories on the boundary measurement approach can also be applied to the estimation of exterior sound fields as well as interior sound fields. 
In this scenario, microphones are placed around the sound-emitting object to estimate the exterior sound field. 
This problem is known as the exterior Dirichlet/Neumann problem, and it is also known that the exterior Dirichlet/Neumann problem does not have forbidden frequencies unlike the interior problems~\cite{Stakgold:SIAM2000}. 
Compared with the estimation of interior sound fields, there appear to be few recent works on applying this approach to the estimation of exterior sound fields in acoustic signal processing (see \cite{Zotter:Dissertation2009}, for example). 

\chapter{Sound Field Estimation with Discrete Measurement Approach}
\label{sec:discrete}

One major challenge in sound field estimation using the boundary measurement approach described in Section~\ref{sec:boundary} is the geometric limitations of the microphone array. 
For instance, when estimating a sound field in a large target region, constructing a large-scale spherical array that encloses the entire region is often impractical.
Additionally, it is challenging to utilize additional microphones located off the boundary even if they are available. 
To address these limitations, sound field estimation methods based on arbitrary discrete measurements have been extensively studied, particularly over the past two decades~\cite{Laborie:AESConvention2003, Poletti:JAES2005}. 
This discrete measurement approach does not impose any special assumptions regarding the positions and directivities of the microphones, as long as they are known, allowing for greater flexibility in designing the microphone array. 
Furthermore, the formulation of this approach easily incorporates various types of prior information about the target sound field, such as the approximate source direction or source distribution sparsity, when available, to the improve estimation accuracy. 

The goal in this section is to review the discrete measurement approach. 
From Sections~\ref{sec:discrete_motivation} to \ref{sec:discrete_infinite}, we further divide this approach into two types: the finite-dimensional modeling approach and infinite-dimensional modeling approach, and we outline the basic ideas and formulations of each. 
The relationship between these two types of approach is discussed in Section~\ref{sec:discrete_summary}. 
Performance evaluations using simulations and real-data experiments are presented in Sections~\ref{sec:discrete_simulation} and \ref{sec:discrete_experiment}, respectively.  
Finally, related studies are reviewed in Section~\ref{sec:discrete_related}. 

\section{Theoretical Motivation}
\label{sec:discrete_motivation}

As mentioned in Section~\ref{sec:overview}, a sound field $u \in \mathcal{H}(\Omega; \wavenumber)$ and its observations $s_1, \ldots, s_M \in \mathbb{C}$ with $M \in \mathbb{N}$ microphones are related by the equation 
\begin{align}
  s_m = \mathcal{F}_m u + \epsilon_m \quad (m \in \llbracket1, M\rrbracket), 
  \label{eq:discrete_forward}
\end{align}
with the observation operators $\mathcal{F}_1, \ldots, \mathcal{F}_M: \mathcal{H}(\Omega; \wavenumber) \to \mathbb{C}$ and observation noises $\epsilon_1, \ldots, \epsilon_M \in \mathbb{C}$. 
The objective of sound field estimation can be viewed as the inverse problem of \eqref{eq:discrete_forward}. 
Here, if $u$ were a finite-dimensional vector, meaning it can be decomposed into a finite number of basis functions, this problem would be reduced to a classical linear inverse problem in finite dimensions, commonly encountered in various fields of signal processing. 
In this case, the unknown vector $u$ can be estimated using an appropriate regularization method, such as Tikhonov regularization or truncated singular value decomposition~\cite{Hansen:BIT1987}. 
Laborie et al. proposed this approach~\cite{Laborie:AESConvention2003}, using the spherical wave functions up to a certain finite order as the basis functions.
This method was further refined and developed by Poletti~\cite{Poletti:JAES2005}, Samarasinghe et al.~\cite{Samarasinghe:IEEE2014}, and other researchers. 
Ueno et al. extended this approach to the infinite-dimensional modeling of sound fields~\cite{Ueno:IEEE2018}, where a closed-form solution can be obtained with the aid of Hilbert space theories. 
All of these methods are applicable to (almost) any observation operators, \ie, the positions and directivities of microphones, as long as they are known. 
This flexibility is a key advantage over the boundary measurement approach. 

\section{Finite-dimensional Modeling Approach}
\label{sec:discrete_finite}

First, we describe the sound field estimation methods that use the finite-dimensional modeling of sound fields~\cite{Laborie:AESConvention2003, Poletti:JAES2005, Samarasinghe:IEEE2014}. 

\subsection{Formulation and solution}
\label{sec:discrete_finite_formulation}

In this approach, a sound field $u\in\mathcal{H}(\Omega; \wavenumber)$ is approximated by a finite number of basis functions as follows: 
\begin{align}
  u(\bm{r}) = \sum_{n = 1}^N c_n p_n(\bm{r}), 
  \label{eq:model_finite_dimension}
\end{align}
where $N \in \mathbb{N}$ is the number of basis functions, $c_1, \ldots, c_N \in \mathbb{C}$ are their coefficients, and $p_1, \ldots, p_N \in \mathcal{H}(\Omega; \wavenumber)$ are the basis functions. 
Typical choices of the basis functions include the spherical wave or plane wave functions, which will be discussed in Section~\ref{sec:discrete_finite_basis}. 
Suppose $M \in \mathbb{N}$ microphones are placed arbitrarily in the sound field $u$. 
The observation operator $\mathcal{F}_m: \mathcal{H}(\Omega; \wavenumber)$ of the $m$th microphone is modeled on the basis of its position $\bm{r}_m \in \Omega$ and directivity $\gamma_m: \mathbb{S}_2 \to \mathbb{C}$ as in Section~\ref{sec:single_microphone}. 

The observed signal $\bm{s} \coloneqq [s_1, \ldots, s_M]^\mathsf{T} \in \mathbb{C}^M$ is then represented with the observation noise $\bm{\epsilon} \coloneqq [\epsilon_1, \ldots, \epsilon_M]^\mathsf{T} \in \mathbb{C}^{M}$ as 
\begin{align}
  \bm{s} 
  = 
  \bm{B}
  \bm{c} 
  + 
  \bm{\epsilon}, 
  \label{eq:forward_finite_dimension}
\end{align}
where $\bm{B} \in \mathbb{C}^{M \times N}$ and $\bm{c} \in \mathbb{C}^{N}$ are defined respectively as 
\begin{align}
  \bm{B} 
  \coloneqq 
  \begin{bmatrix}
    \mathcal{F}_1 p_1 & \hdots & \mathcal{F}_1 p_N 
    \\ 
    \vdots & \ddots & \vdots 
    \\ 
    \mathcal{F}_M p_1 & \hdots & \mathcal{F}_M p_N 
  \end{bmatrix}
\end{align}
and 
\begin{align}
  \bm{c} 
  \coloneqq 
  \begin{bmatrix}
    c_1 
    \\
    \vdots 
    \\ 
    c_N 
  \end{bmatrix}. 
\end{align}
The inverse problem of \eqref{eq:forward_finite_dimension} is typically formulated using Tikhonov regularization as 
\begin{align}
  \minimize_{\bm{c} \in \mathbb{C}^N} 
  \ 
  \|\bm{B}\bm{c} - \bm{s}\|_{\bm{\Sigma}^{-1}}^2 + \lambda \|\bm{c}\|_2^2, 
  \label{eq:inverse_finite_dimension}
\end{align}
where $\lambda \in (0, \infty)$ is the regularization parameter, $\|\placeholder\|_{\bm{\Sigma}^{-1}}$ is defined for a positive-definite matrix $\bm{\Sigma} \in \mathbb{C}^{M \times M}$ as 
\begin{align}
  \|\bm{a}\|_{\bm{\Sigma}^{-1}}^2 
  \coloneqq 
  \bm{a}^\mathsf{H} \bm{\Sigma}^{-1} \bm{a} 
  \quad (\bm{a} \in \mathbb{C}^M), 
\end{align}
and $\|\cdot\|_2$ denotes the $\ell_2$-norm. 
The matrix $\bm{\Sigma}$ is designed on the basis of observational uncertainty and is typically defined as a scalar matrix, \ie, $\bm{\Sigma} = \sigma^{2} \bm{I}_M$, with a single parameter $\sigma \in (0, \infty)$. 

This minimization problem can be solved in a closed form, and the estimated coefficients $\hat{\bm{c}} = [\hat{c}_1,\ldots,\hat{c}_M]^\mathsf{T} \in \mathbb{C}^N$, \ie, the optimal solution of \eqref{eq:inverse_finite_dimension}, are given by 
\begin{align}
  \hat{\bm{c}} 
  & = 
  \left(
    \bm{B}^\mathsf{H}\bm{\Sigma}^{-1}\bm{B} + \lambda \bm{I}_N
  \right)^{-1}
  \bm{B}^\mathsf{H} \bm{\Sigma}^{-1}
  \bm{s}
  \nonumber\\ 
  & = 
  \bm{B}^\mathsf{H} 
  \left(
    \bm{B}\bm{B}^\mathsf{H} + \lambda \bm{\Sigma}
  \right)^{-1}
  \bm{s}.
  \label{eq:solution_coefficient_finite_dimension}
\end{align}
The equivalence between the first and second lines in \eqref{eq:solution_coefficient_finite_dimension} follows from the matrix inversion lemma~\cite{Bernstein:Princeton2018}. 
Finally, the estimated sound field is given by 
\begin{align}
  \hat{u}(\bm{r}) 
  = 
  \sum_{m = 1}^M 
  \hat{c}_m p_m(\bm{r}) \quad (\bm{r} \in \Omega). 
  \label{eq:solution_finite_dimension}
\end{align}
Note that the estimated sound field $\hat{u}$ is linear with respect to the observed signal $\bm{s}$, as indicated by \eqref{eq:solution_coefficient_finite_dimension} and \eqref{eq:solution_finite_dimension}, which means the linear time-invariant property of this estimation method. 

\subsection{Choice of basis functions}
\label{sec:discrete_finite_basis}

Although the theoretical framework presented in Section~\ref{sec:discrete_finite_formulation} is applicable to any set of basis functions, the estimation performance obviously depends on their choice. 
Ideally, the sound field in the region of interest should be well represented by the chosen basis functions. 
For interior sound fields in a simply connected region, the spherical wave and plane wave functions are typically employed; the resulting formulations are described in this section. 

Determining the preferable basis functions may seem nontrivial. 
However, these two choices become essentially equivalent as their dimensions approach infinity. This equivalence arises because the spherical wave function can be expressed as a finite number of plane wave functions and vice versa, as implied by \eqref{eq:sw2pw} and \eqref{eq:Jacobi_Anger_expansion}, respectively. 

\subsubsection{Spherical wave functions}
\label{sec:example_sphwavefun}
In this case, a sound field is represented as a superposition of the spherical wave functions, \ie, 
\begin{align}
  u(\bm{r}) = \sum_{\nu,\mu}^{N_0} c_{\nu,\mu} p_{\nu,\mu}(\bm{r}) \quad (\bm{r} \in \Omega)
  \label{eq:model_finite_dimension_spherical}
\end{align}
with 
\begin{align}
  p_{\nu,\mu}(\bm{r}) = \varphi_{\nu,\mu}(\bm{r}-\bm{r}_0) \quad (\bm{r} \in \Omega). 
  \label{eq:def_basis_sphwavefun}
\end{align}
Here, $\bm{r}_0$ refers to the expansion center, also called the \emph{global origin}, and $N_0$ denotes the truncation order. 
For notational simplicity, we use double indices $\nu, \mu$ instead of $n$ in \eqref{eq:model_finite_dimension}.
Thus, the total number of basis functions is $N = (N_0 + 1)^2$. 
Moreover, we assume that the directivity $\gamma_m : \mathbb{S}_2 \to \mathbb{C}$ of each microphone is represented by the finite-order spherical harmonic functions, \ie, 
\begin{align}
  \gamma_m(\bm{x})
  = 
  \sum_{\nu,\mu}^{N_m}
  d_{m,\nu,\mu} \hat{Y}_{\nu,\mu}(\bm{x}), 
  \label{eq:directivity_finite}
\end{align}
where $N_m \in \mathbb{N}$ is the maximum order of the $m$th microphone. 
This assumption is valid for many microphones with simple characteristics, including those discussed in Section~\ref{sec:single_microphone}. 
Since $p_{\nu,\mu}$ is expanded around the microphone position $\bm{r}_m$ using \eqref{eq:sw2pw} as 
\begin{align}
  p_{\nu,\mu}(\bm{r}) 
  & = 
  \frac{1}{4\pi}
  \int_{\bm{x} \in \mathbb{S}_2}
  \hat{\sphharm}_{\nu,\mu}(\bm{x})
  \exp(-\im \wavenumber \bm{x} \dotproduct (\bm{r}_m-\bm{r}_0)) 
  \nonumber\\ 
  & \quad \hspace{7em} \cdot
  \exp(-\im \wavenumber \bm{x} \dotproduct (\bm{r}-\bm{r}_m)) 
  \diff \surf, 
\end{align}
each elements of $\bm{B}$ can then be represented on the basis of \eqref{eq:observation_directivity} as 
\begin{align}
  \mathcal{F}_m p_{\nu,\mu} 
  & = 
  \int_{\bm{x} \in \mathbb{S}_2} 
  \gamma_m(\bm{x})^\ast 
  \cdot 
  \frac{1}{4\pi}
  \hat{\sphharm}_{\nu,\mu}(\bm{x})
  \exp(-\im \wavenumber \bm{x} \dotproduct (\bm{r}_m-\bm{r}_0)) 
  \diff \surf 
  \nonumber\\
  & = 
  \sum_{\nu',\mu'}^{N_m} 
  d_{m,\nu',\mu'}^\ast 
  \cdot 
  \frac{1}{4\pi}
  \int_{\bm{x} \in \mathbb{S}_2}
  \bigl(
  \hat{\sphharm}_{\nu',\mu'}(\bm{x})^\ast \hat{\sphharm}_{\nu,\mu}(\bm{x})
  \nonumber\\ 
  & \quad \hspace{9em}
  \cdot 
  \exp\left(-\im \wavenumber \bm{x} \dotproduct (\bm{r}_m - \bm{r}_0)\right) 
  \bigr)
  \diff \surf
  \nonumber\\ 
  & = 
  \sum_{\nu',\mu'}^{N_m} 
  T_{\nu',\mu'}^{\nu,\mu}(\bm{r}_m - \bm{r}_0) 
  d_{m,\nu',\mu'}^\ast 
  \label{eq:B_element}
\end{align}
using the translation operator $T_{\nu,\mu}^{\nu',\mu'}(\placeholder)$ (see \eqref{eq:def_translation_operator} for definition).
The equivalence between the second and third lines is derived from the following formula~\cite{Ueno:IEEE2021}: 
\begin{align}
  T_{\nu',\mu'}^{\nu,\mu}(\bm{r})
  = 
  \frac{1}{4\pi}
  \int_{\bm{x} \in \mathbb{S}_2} 
  \hat{\sphharm}_{\nu,\mu}(\bm{x}) 
  \hat{\sphharm}_{\nu',\mu'}^\ast(\bm{x}) 
  \exp(-\im \wavenumber \bm{x} \dotproduct \bm{r}) 
  \diff \surf. 
  \label{eq:tl2pw} 
\end{align}
As a special case, when all microphones are omnidirectional, \eqref{eq:B_element} can be simplified to the following form: 
\begin{align}
  \mathcal{F}_m p_{\nu,\mu} = \varphi_{\nu,\mu}(\bm{r}_m - \bm{r}_0), 
\end{align}
which follows from $d_{m,\mu,\nu} = \delta_{\nu, 0}$ and \eqref{eq:def_translation_operator}. 
This simplification can also be easily confirmed from \eqref{eq:derivative_omnidirectional} and \eqref{eq:def_basis_sphwavefun}. 

The modeling using the spherical wave functions requires the parameters $N_0$ and $\bm{r}_0$. 
Although it is challenging to find the optimal values for these parameters from a theoretical standpoint, most of the previous studies suggest setting the global origin $\bm{r}_0$ at the center of the microphone array, particularly when using a spherical array. 
This is because the spherical wave function expansion provides an approximation around $\bm{r}_0$ in a manner similar to the multivariate Taylor expansion\footnote{The coefficients for the spherical wave function expansion up to the $\nu$th order determine those for the Taylor expansion up to the $\nu$th order (\ie, the partial derivatives up to the $\nu$th order). This relationship arises from the fact that all partial derivatives of the spherical wave function $\varphi_{\nu,\mu}(\placeholder)$ of the order less than $\nu$ are zero.}. 
On the other hand, this modeling is invariant to the orientation of the coordinate axes. 
The relationship between the parameter values and estimation performance will be shown in Section~\ref{sec:discrete_simulation}, which indicates that inappropriate settings of the truncation order and global origin can negatively affect the estimation performance. 

\subsubsection{Plane wave functions}
In this case, the sound field is represented as a superposition of the plane wave functions, \ie, 
\begin{align}
  u(\bm{r}) 
  = 
  \sum_{n=1}^N 
  c_n p_n(\bm{r})
\end{align}
with
\begin{align}
  p_n(\bm{r}) = \exp(-\im \wavenumber \bm{x}_n \dotproduct (\bm{r} - \bm{r}_0)).
\end{align}
Here, $\bm{r}_0$ is the global origin and $\bm{x}_1, \ldots, \bm{x}_N \in \mathbb{S}_2$ are the discretized unit directions representing the incoming directions of the plane wave functions. 
Since $p_n$ is represented around $\bm{r}_m$ as 
\begin{align}
  p_n(\bm{r}) = \exp(-\im \wavenumber \bm{x}_n \dotproduct (\bm{r}_m - \bm{r}_0)) \cdot \exp(-\im \wavenumber \bm{x}_n \dotproduct (\bm{r} - \bm{r}_m)),
\end{align}
each element of $\bm{B}$ can then be represented as
\begin{align}
  \mathcal{F}_m p_n 
  = 
  \exp(-\im \wavenumber \bm{x}_n \dotproduct (\bm{r}_m - \bm{r}_0)) \gamma_m(\bm{x}_n). 
\end{align}

The modeling using the plane wave functions requires the parameters $\bm{x}_1, \ldots, \bm{x}_N$. 
Several sampling methods on the sphere are available, such as those described in Section~\ref{sec:boundary_discretization}. 
In contrast to the case of the spherical wave functions, this modeling is invariant to the position of the global origin $\bm{r}_0$ but dependent on the orientation of the coordinate axes. 

\section{Infinite-dimensional Modeling Approach}
\label{sec:discrete_infinite}

The sound field estimation problem defined in Section~\ref{sec:discrete_finite_formulation} can be extended to infinite dimensions with the aid of Hilbert space theories. 
This extension allows for the elimination of parameters related to the basis functions, such as the truncation order and global origin of the spherical wave functions, thereby preventing the potential degradation of estimation performance due to an inappropriate parameter selection.

\subsection{Formulation and solution}
\label{sec:discrete_infinite_formulation}

In this approach, a sound field is modeled directly as an element of the infinite-dimensional function space $\mathscr{H}$ defined as 
\begin{align}
  \mathscr{H} \coloneqq \{u = \mathcal{A} w \in \mathcal{H}(\Omega; \wavenumber) \suchthat w \in L_2(\mathbb{S}_2)\}
\end{align}
with 
\begin{align}
  (\mathcal{A}w)(\bm{r}) 
  = 
  \int_{\bm{x} \in \mathbb{S}_2} 
  w(\bm{x}) 
  \exp(-\im \wavenumber \bm{x} \dotproduct \bm{r}) 
  \diff \surf. 
\end{align}
Here, $L_2(\mathbb{S}_2)$ denotes the set of square-integrable functions from $\mathbb{S}_2$ to $\mathbb{C}$. 
Intuitively, this formulation means that the sound field is modeled as a superposition of an infinite number of plane waves. 
Moreover, we define the inner product $\left<\placeholder,\placeholder\right>_{\mathscr{H}}$ and the norm $\|\placeholder\|_{\mathscr{H}}$ over $\mathscr{H}$ as 
\begin{align}
  \langle u_1, u_2 \rangle_{\mathscr{H}} 
  \coloneqq 
  4\pi
  \int_{\bm{x} \in \mathbb{S}_2} 
  (\mathcal{A}^{-1}u_1)(\bm{x})^\ast (\mathcal{A}^{-1}u_2)(\bm{x}) 
  \diff \surf 
  \label{eq:def_inner_product_H}
\end{align}
and 
\begin{align}
  \|u\|_{\mathscr{H}}
  \coloneqq 
  \sqrt{\langle u, u \rangle_{\mathscr{H}}}, 
  \label{eq:def_norm_H}
\end{align}
respectively. 
The function space $\mathscr{H}$ is used instead of $\mathcal{H}(\Omega; \wavenumber)$ for mathematical convenience, similarly to the commonly used assumption of square integrability in continuous-time signal analysis. 
However, unlike the finite-dimensional modeling, $\mathscr{H}$ is dense in $\mathcal{H}(\Omega; \wavenumber)$ in the sense of uniform convergence on compact sets (see Theorem~\ref{thm:approximation_plane_wave_function}). 
Furthermore, $\mathscr{H}$ and its inner product are invariant under the rotation and translation of the sound fields; that is, if $u \in \mathscr{H}$, then the function $u'(\bm{r}) \coloneqq u(\mathcal{R}(\bm{r} + \bm{r}'))$ with any $\mathcal{R} \in \mathrm{SO}(3)$ and $\bm{r}' \in \mathbb{R}^3$ is also in $\mathscr{H}$ and satisfies $\|u\|_{\mathscr{H}} = \|u'\|_{\mathscr{H}}$. 

The observed signal $\bm{s} \coloneqq [s_1, \ldots, s_M]^\mathsf{T} \in \mathbb{C}^M$ is then represented with the observation noise $\bm{\epsilon} \coloneqq [\epsilon_1, \ldots, \epsilon_M]^\mathsf{T} \in \mathbb{C}^{M}$ as 
\begin{align}
  \bm{s} = \bm{F}u + \bm{\epsilon}, 
\end{align}
where $\bm{F}:\mathscr{H} \to \mathbb{C}^M$ is the vector-valued linear operator defined as 
\begin{align}
  \bm{F}u 
  = 
  \begin{bmatrix}
    \mathcal{F}_1 u
    \\ 
    \vdots 
    \\ 
    \mathcal{F}_M u
  \end{bmatrix}. 
\end{align}
Therefore, the sound field estimation problem is formulated as 
\begin{align}
  \minimize_{u \in \mathscr{H}} 
  \ 
  Q(u) = 
  \| \bm{F} u - \bm{s} \|_{\bm{\Sigma}^{-1}}^2 
  + \lambda \|u\|_{\mathscr{H}}^2 
  \label{eq:inverse_infinite_dimension}
\end{align}
with the regularization parameter $\lambda \in (0, \infty)$. 
Even though this optimization problem is defined in the infinite-dimensional space $\mathscr{H}$, it can be solved in a closed form as follows. 

First, we define $v_m \in \mathscr{H}$ as 
\begin{align}
  v_m(\bm{r}) 
  \coloneqq 
  \frac{1}{4\pi} 
  \int_{\bm{x} \in \mathbb{S}_2} 
  \gamma_m(\bm{x}) 
  \exp(-\im \wavenumber \bm{x} \dotproduct (\bm{r}-\bm{r}_m)) 
  \diff \surf, 
  \label{eq:v_m_def}
\end{align}
so that the observation operator $\mathcal{F}_m$ can be represented as 
\begin{align}
  \mathcal{F}_m u = \langle v_m, u \rangle_{\mathscr{H}}. 
\end{align}
Then, the objective function $Q(\placeholder)$ is rewritten as 
\begin{align}
  Q(u) 
  = 
  \left\|
    \begin{bmatrix}
      \left< v_1, u \right> 
      \\ 
      \vdots 
      \\ 
      \left< v_M, u \right>
    \end{bmatrix}
    - 
    \begin{bmatrix}
      s_1
      \\ 
      \vdots 
      \\ 
      s_M
    \end{bmatrix}
  \right\|_{\bm{\Sigma}^{-1}}^2 
  + 
  \lambda \left<u, u\right>_\mathscr{H}. 
\end{align}
Therefore, we can apply the representer theorem~\cite{Scholkopf:COLT2001, Dinuzzo:NuerIPS2012} to this optimization problem, which guarantees that the optimal solution $\hat{u} \in \mathscr{H}$ has the form of 
\begin{align}
  \hat{u} = 
  \sum_{m=1}^M \hat{\alpha}_m v_m. 
  \label{eq:u_est_infinite}
\end{align}
Here, $\hat{\bm{\alpha}} = [\hat{\alpha}_1 ,\ldots, \hat{\alpha}_M]^\mathsf{T} \in \mathbb{C}^M$ is given as the solution to the following optimization problem: 
\begin{align}
  \minimize_{\bm{\alpha} \in \mathbb{C}^M} 
  \ 
  \tilde{Q}(\bm{\alpha}) = 
  (\bm{K}\bm{\alpha} - \bm{s})^\mathsf{H} \bm{\Sigma}^{-1} (\bm{K}\bm{\alpha} - \bm{s}) 
  + 
  \lambda
  \bm{\alpha}^\mathsf{H} \bm{K} \bm{\alpha}, 
\end{align}
where $\bm{K} \in \mathbb{C}^{M \times M}$ is defined as 
\begin{align}
  \bm{K} 
  = 
  \begin{bmatrix}
    K_{1,1} & \hdots & K_{1,M} 
    \\ 
    \vdots & \ddots & \vdots 
    \\ 
    K_{M,1} & \hdots & K_{M,M} 
  \end{bmatrix}
  \label{eq:K_infinite_def}
\end{align}
with 
\begin{align}
  K_{m_1,m_2} 
  & = 
  \left<v_{m_1}, v_{m_2}\right>_{\mathscr{H}}
  \nonumber\\
  & = 
  \frac{1}{4\pi} 
  \int_{\bm{x} \in \mathbb{S}_2} 
  \gamma_{m_1}(\bm{x})^\ast \gamma_{m_2}(\bm{x})
  \exp(-\im \wavenumber \bm{x} \dotproduct (\bm{r}_{m_1} - \bm{r}_{m_2})) 
  \diff \surf. 
  \label{eq:K_elem_infinite_def}
\end{align}
Note that the optimization problem in \eqref{eq:K_infinite_def} is defined in the finite-dimensional vector space $\mathbb{C}^M$. 
This optimization problem can be solved as 
\begin{align}
  \hat{\bm{\alpha}} = 
  (\bm{K} + \lambda \bm{\Sigma})^{-1} \bm{s}. 
\end{align}
The remaining problems are the calculations of $v_1,\ldots,v_M$ and $\bm{K}$. 

\subsection{Closed-form representation}

The closed-form representation of $v_1,\ldots,v_M$ and $\bm{K}$ can be obtained when the directivities of microphones are modeled using the finite-order spherical harmonic functions as in \eqref{eq:directivity_finite}. 
In this case, by substituting \eqref{eq:directivity_finite} into \eqref{eq:v_m_def} and \eqref{eq:K_infinite_def}, we have 
\begin{align}
  v_m(\bm{r}) 
  & \coloneqq 
  \frac{1}{4\pi} 
  \int_{\bm{x} \in \mathbb{S}_2} 
  \sum_{\nu,\mu}^{N_m} 
  d_{m,\nu,\mu} \hat{\sphharm}_{\nu,\mu}(\bm{x})
  \exp(-\im \wavenumber \bm{x} \dotproduct (\bm{r}-\bm{r}_m)) 
  \diff \surf 
  \nonumber\\
  & = 
  \sum_{\nu,\mu}^{N_m} 
  d_{m,\nu,\mu} \varphi_{\nu,\mu}(\bm{r}-\bm{r}_m)
  \label{eq:v_m_closed_form}
\end{align}
and 
\begin{align}
  K_{m_1,m_2} 
  & = 
  \frac{1}{4\pi} 
  \int_{\bm{x} \in \mathbb{S}_2} 
  \sum_{\nu_1,\mu_1}^{N_{m_1}} 
  \sum_{\nu_2,\mu_2}^{N_{m_2}} 
  d_{m_1,\nu_1,\mu_1}^\ast 
  d_{m_2,\nu_2,\mu_2} 
  \hat{\sphharm}_{\nu_1,\mu_1}(\bm{x})^\ast  
  \hat{\sphharm}_{\nu_2,\mu_2}(\bm{x})
  \nonumber\\
  & \quad \hspace{8em} 
  \cdot
  \exp(-\im \wavenumber \bm{x} \dotproduct (\bm{r}_{m_1} - \bm{r}_{m_2})) 
  \diff \surf
  \nonumber 
  \\ 
  & = 
  \sum_{\nu_1,\mu_1}^{N_{m_1}} 
  \sum_{\nu_2,\mu_2}^{N_{m_2}} 
  d_{m_1,\nu_1,\mu_1}^\ast 
  d_{m_2,\nu_2,\mu_2} 
  T_{\nu_1,\mu_1}^{\nu_2,\mu_2}(\bm{r}_{m_1} - \bm{r}_{m_2}). 
  \label{eq:K_closed_form}
\end{align}
Here, the second lines in \eqref{eq:v_m_closed_form} and \eqref{eq:K_closed_form} are derived using \eqref{eq:sw2pw} and \eqref{eq:tl2pw}, respectively. 

\subsection{Interpretation as kernel ridge regression}
\label{sec:discrete_infinite_kernel}

As a special case, when all microphones are omnidirectional, \ie, $\gamma(\bm{x}) = 1 \ (\bm{x} \in \mathbb{S}_2)$, \eqref{eq:v_m_closed_form} and \eqref{eq:K_closed_form} are simplified respectively to 
\begin{align}
  v_m(\bm{r}) = j_0(k\|\bm{r}-\bm{r}_m\|)
  \label{eq:v_m_kernel}
\end{align}
and 
\begin{align}
  K_{m_1, m_2} = j_0(k\|\bm{r}_1-\bm{r}_2\|). 
  \label{eq:K_kernel}
\end{align}
Notably, these formulations correspond to kernel ridge regression~\cite{Hofmann:IMS2008}, a commonly used function interpolation technique in machine learning. 
In kernel ridge regression, the unknown function $f: \Omega \to \mathbb{C}$ is interpolated from the set of its pointwise evaluation $s_m = f(\bm{r}_m) + \epsilon_m$ at the sampling point $\bm{r}_m \in \Omega$ for each $m \in \{1,\ldots,M\}$ as 
\begin{align}
  \hat{f} = \sum_{m = 1}^M \hat{\alpha}_m \kappa(\bm{r}, \bm{r}_m), 
\end{align}
where $\hat{\bm{\alpha}} \coloneqq [\hat{\alpha}_1, \ldots, \hat{\alpha}_M] \in \mathbb{C}^M$ is given from $\bm{s} \coloneqq [s_1, \ldots, s_M] \in \mathbb{C}^M$ by 
\begin{align}
  \hat{\bm{\alpha}} = 
  (\bm{K} + \lambda \bm{\Sigma})^{-1} \bm{s} 
\end{align}
with 
\begin{align}
  \bm{K}
  = 
  \begin{bmatrix}
    \kappa(\bm{r}_1,\ldots,\bm{r}_1) & \hdots & \kappa(\bm{r}_1,\ldots,\bm{r}_M)
    \\ 
    \vdots & \ddots & \vdots 
    \\ 
    \kappa(\bm{r}_M,\ldots,\bm{r}_1) & \hdots & \kappa(\bm{r}_M,\ldots,\bm{r}_M)
  \end{bmatrix}
\end{align}
Here, $\kappa: \Omega \times \Omega \to \mathbb{C}$ is called a kernel function. 
A typical choice of the kernel function in the general context of machine learning is the Gaussian kernel.
It can be seen that the formulations in \eqref{eq:v_m_kernel} and \eqref{eq:K_kernel} correspond to the kernel ridge regression when the kernel function defined as 
\begin{align}
  \kappa(\bm{r},\bm{r}') \coloneqq j_0(\wavenumber\|\bm{r}-\bm{r}'\|) \quad (\bm{r},\bm{r}' \in \Omega)
  \label{eq:def_kernel}
\end{align}
is used~\cite{Ueno:IWAENC2018}. 
In other words, this means that the function space $\mathscr{H}$ is a reproducing kernel Hilbert space generated by the kernel function defined in \eqref{eq:def_kernel}. 

\subsection{Estimation of expansion coefficients}
\label{eq:discrete_infinite_coefficient}

In many signal processing applications for sound field estimation, including the binaural reproduction described in Section~\ref{sec:applications}, it is often desirable that the estimated sound field is expressed in terms of the expansion coefficients when expanded by the spherical wave functions. 
Although \eqref{eq:u_est_infinite} is not in this form unlike \eqref{eq:boundary_estimated_sound_field} or \eqref{eq:model_finite_dimension_spherical}, its expansion coefficients around an arbitrary expansion center $\bm{r}_0 \in \Omega$, denoted by $\ring{\hat{u}}_{\nu,\mu}(\bm{r}_0)$, \ie,
\begin{align}
  \hat{u}(\bm{r}) = \sum_{\nu,\mu} \ring{\hat{u}}_{\nu,\mu}(\bm{r}_0) \varphi_{\nu,\mu}(\bm{r} - \bm{r}_0)
  \quad \forall \bm{r} \in \Omega, 
\end{align}
can be obtained as follows. 

Since the estimated sound field $\hat{u}$ is expressed as a superposition of $v_1, \ldots, v_M$ as in \eqref{eq:u_est_infinite}, the expansion coefficient $\ring{\hat{u}}_{\nu,\mu}(\bm{r}_0)$, is represented using the expansion coefficients of $v_m$, denoted by $\ring{v}_{m,\nu,\mu}(\bm{r}_0)$, as 
\begin{align}
  \ring{\hat{u}}_{\nu,\mu}(\bm{r}_0) 
  = 
  \sum_{m=1}^{M} 
  \hat{\alpha}_m \ring{v}_{m,\nu,\mu}(\bm{r}_0). 
\end{align}
Here, from \eqref{eq:v_m_closed_form} and \eqref{thm:tranlsation_spherical_wave_function}, $v_m$ is expanded around $\bm{r}_0$ as 
\begin{align}
  v_m(\bm{r})
  & = 
  \sum_{\nu',\mu'}^{N_m} 
  d_{m,\nu',\mu'} \varphi_{\nu',\mu'}(\bm{r}-\bm{r}_m)
  \nonumber\\ 
  & = 
  \sum_{\nu',\mu'}^{N_m} 
  d_{m,\nu',\mu'} 
  \sum_{\nu,\mu}
  T_{\nu,\mu}^{\nu',\mu'}(\bm{r}_0 - \bm{r}_m) 
  \varphi_{\nu, \mu}(\bm{r}-\bm{r}_0),
\end{align}
which means that $\ring{v}_{m,\nu,\mu}(\bm{r}_0)$ is given by 
\begin{align}
  \ring{v}_{m,\nu,\mu}(\bm{r}_0) 
  & 
  = 
  \sum_{\nu',\mu'}^{N_m} 
  T_{\nu,\mu}^{\nu',\mu'}(\bm{r}_0 - \bm{r}_m) 
  d_{m,\nu',\mu'}. 
  \label{eq:v_m_coefficient}
\end{align} 
Therefore, from \eqref{eq:u_est_infinite} and \eqref{eq:v_m_coefficient}, we obtain 
\begin{align}
  \ring{\hat{u}}_{\nu,\mu}(\bm{r}_0) 
  = 
  \sum_{m=1}^M 
  \hat{\alpha}_m 
  \sum_{\nu',\mu'}^{N_m} 
  T_{\nu,\mu}^{\nu',\mu'}(\bm{r}_0 - \bm{r}_m) 
  d_{m,\nu',\mu'}.
\end{align}

\section{Summary and Comparison}
\label{sec:discrete_summary}

The two approaches described in Sections~\ref{sec:discrete_finite} and \ref{sec:discrete_infinite} have a close relationship, as explained below. 
For comparison, the estimated sound fields in these approaches, denoted here by $\hat{u}^{(\text{finite})}$ and $\hat{u}^{(\text{infinite})}$, can be expressed in a unified manner as 
\begin{align}
  \hat{u}^{(\ast)}(\bm{r}) = \sum_{m=1}^M \hat{\alpha}^{(\ast)}_m v^{(\ast)}_m(\bm{r}) \quad (\bm{r} \in \Omega)
\end{align}
with $\hat{\bm{\alpha}}^{(\ast)} \coloneqq [\hat{\alpha}^{(\ast)}_1, \ldots, \hat{\alpha}^{(\ast)}_M] \in \mathbb{C}^M$ defined as 
\begin{align}
  \hat{\bm{\alpha}}^{(\ast)} 
  = 
  (\bm{K}^{(\ast)} + \lambda \bm{\Sigma})^{-1} \bm{s}. 
\end{align}
Here, $(\ast) \in \{(\text{finite}), (\text{infinite})\}$ denotes the estimation methods, \ie, the finite-dimensional and infinite-dimensional modeling approach, and $v^{(\ast)}_m$ and $\bm{K}{(\ast)}$ are given by 
\begin{align}
  v^{(\ast)}_m(\bm{r})
  = 
  \begin{dcases}
    \sum_{n = 1}^N (\mathcal{F}_{m} p_n)^\ast p_n(\bm{r}) & (\ast) = (\text{finite})
    \\ 
    \sum_{\nu,\mu}^{N_m} d_{m,\nu,\mu} \varphi_{\nu,\mu}(\bm{r}-\bm{r}_m) & (\ast) = (\text{infinite}),
  \end{dcases}
\end{align}
and 
\begin{align}
  K^{(\ast)}_{m_1, m_2} 
  = 
  \begin{dcases}
    \sum_{n = 1}^N 
    (\mathcal{F}_{m_1} p_n) 
    (\mathcal{F}_{m_2} p_n)^\ast
    & 
    (\ast) = (\text{finite})
    \\ 
    \begin{aligned}
      & \sum_{\nu_1,\mu_1}^{N_{m_1}} 
      \sum_{\nu_2,\mu_2}^{N_{m_2}} 
      d_{m_1,\nu_1,\mu_1}^\ast 
      d_{m_2,\nu_2,\mu_2} 
      \\
      & \hspace{4em} \cdot T_{\nu_1,\mu_1}^{\nu_2,\mu_2}(\bm{r}_{m_1} - \bm{r}_{m_2})
    \end{aligned}
    & 
    (\ast) = (\text{infinite})
  \end{dcases},
\end{align}
respectively, where $K^{(\ast)}_{m_1, m_2}$ denotes the $(m_1, m_2)$ element of $\bm{K}^{(\ast)}$. 
In particular, when using the spherical wave functions as the basis functions for the finite-dimensional modeling approach, $v^{(\text{finite})}_m$ and $K^{(\text{finite})}_{m_1, m_2}$ are rewritten as 
\begin{align}
  v^{(\text{finite})}_m(\bm{r})
  = 
  \sum_{\nu,\mu}^{N_m}
  d_{m,\nu,\mu}
  \sum_{\nu',\mu'}^{N_0} 
  T_{\nu',\mu'}^{\nu,\mu}(\bm{r}_{m}-\bm{r}_0)^\ast
  \varphi_{\nu',\mu'}(\bm{r}-\bm{r}_0)
\end{align}
and 
\begin{align}
  K^{(\text{finite})}_{m_1,m_2} 
  & = 
  \sum_{\nu_1,\mu_1}^{N_{m_1}} 
  \sum_{\nu_2,\mu_2}^{N_{m_2}} 
  \left(
  \sum_{\nu,\mu}^{N_0}
  T_{\nu_1,\mu_1}^{\nu,\mu}(\bm{r}_{m_1}-\bm{r}_{0})
  T_{\nu_2,\mu_2}^{\nu,\mu}(\bm{r}_{m_2}-\bm{r}_{0})^\ast 
  \right) 
  \nonumber 
  \\ 
  & \quad \hspace{5em} \cdot
  d_{m_1,\nu_1,\mu_1}^\ast 
  d_{m_2,\nu_2,\mu_2}, 
\end{align}
respectively. 
Here, from \eqref{thm:tranlsation_spherical_wave_function} and the formulae~\cite{Martin:Cambridge2006} 
\begin{align}
  T_{\nu_1, \mu_1}^{\nu_2, \mu_2}(\bm{r})^\ast 
  = T_{\nu_2, \mu_1}^{\nu_2, \mu_1}(-\bm{r}), 
\end{align}
\begin{align}
  \sum_{\nu, \mu} T_{\nu_1, \mu_1}^{\nu, \mu}(\bm{r}) T_{\nu, \mu}^{\nu_2, \mu_2}(\bm{r}')
  = T_{\nu_1, \mu_1}^{\nu_2, \mu_2}(\bm{r} + \bm{r}'), 
\end{align}
we can see that 
\begin{align}
  \lim_{N_0 \to \infty} 
  v_m^{(\text{finite})} 
  = 
  v_m^{(\text{infinite})}
\end{align}
and 
\begin{align}
  \lim_{N_0 \to \infty} 
  \bm{K}^{(\text{finite})} 
  = 
  \bm{K}^{(\text{infinite})},
\end{align}
respectively. 
This means that the infinite-dimensional modeling approach can be interpreted as the limit of the finite-dimensional modeling approach when the number of dimensions is increased to infinity. 
Nevertheless, it is noteworthy that the infinite-dimensional modeling approach allows for the computation of the estimated sound field without the need for an infinite sum. 

\section{Simulation Results}
\label{sec:discrete_simulation}

Here, we present the simulation results of sound field estimation using the aforementioned two modeling approaches, denoted here by \textbf{DM-finite} and \textbf{DM-infinite}. 
Throughout the simulations, the speed of sound was set as $c = \SI{340.65}{m/s}$. 
In a three-dimensional free field, $M = 64$ microphones were placed, and four different patterns of their positions were investigated. 
The first two are the spherical configuration with omnidirectional/first-order microphones, where the omnidirectional or first-order (a = 0.5) microphones were placed on an open sphere with a radius of $\SI{1}{m}$ centered at the origin on the basis of the spherical $t$-design with $t = 7$~\cite{Chen:PolyU2004}, in the same manner as in Section~\ref{sec:boundary_simulation}. 
The other two are the irregular configuration with omnidirectional/first-order microphones, where the omnidirectional or first-order (a = 0.5) microphones were placed randomly within a spherical region with a radius of $\SI{1}{m}$ centered at the origin, following the uniform distribution. 
In the case of first-order microphones, each of their orientations was independently determined according to the uniform distribution on $\mathbb{S}_2$. 
The true sound field to be estimated was $u(\bm{r}) = \exp(- \im \wavenumber \bm{x}_\mathrm{inc} \dotproduct \bm{r})$ with the incident direction $\bm{x}_\mathrm{inc} \coloneqq (1, 0, 0)$. 
The observed signals were calculated using \eqref{eq:boundary_observed_signal}, and the observation noises were sampled independently from the circularly symmetric Gaussian distribution with zero mean and a variance of $10^{-3} \times \frac{1}{M} \sum_{m =1}^M |\mathcal{F}_m u|^2$ so that the expected signal-to-noise ratio was $\SI{30}{dB}$. 
In \textbf{DM-finite}, the spherical wave functions were used as the basis functions as in \eqref{eq:model_finite_dimension_spherical} and \eqref{eq:def_basis_sphwavefun}. 
In both methods, the regularization parameter was set as $\lambda = 10^{-3}$. 

As an evaluation criterion, the NMSE, defined as 
\begin{align}
  \mathrm{NMSE}(\hat{u}, u)
  = 
  10 \log_{10}
  \left(
    \frac{\sum_{\mathrm{i} \in I_\mathrm{eval}} |\hat{u}(\bm{r}^{(\mathrm{i})}_\mathrm{eval}) - u(\bm{r}^{(\mathrm{i})}_\mathrm{eval})|^2}{\sum_{\mathrm{i} \in I_\mathrm{eval}} |u(\bm{r}^{(\mathrm{i})}_\mathrm{eval})|^2}
  \right),
\end{align}
was used, where $\hat{u}$ denotes the estimated sound field, and the evaluation points $\{\bm{r}^{(\mathrm{i})}_\mathrm{eval}\}_{\mathrm{i} \in I_\mathrm{eval}}$ were set as all grid points at $\SI{0.1}{m}$ intervals within the spherical region with a radius of $\SI{1}{m}$ centered at the origin. 
Since some parameters and the observation noises were randomly determined, the NMSEs were averaged over 10 trials. 

\begin{figure}
  \centering
  \includegraphics{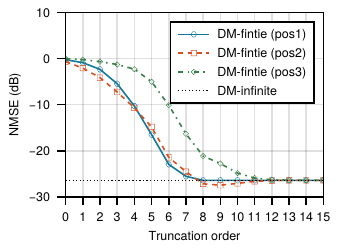}
  \caption{NMSE plotted against truncation order for spherical configuration with first-order microphones at $\SI{300}{Hz}$. The global origin $\bm{r}_0$ was set at $\SI{(0,0,0)}{m}$, $\SI{(1,0,0)}{m}$, and $\SI{(0,1,0)}{m}$ for pos1, pos2, and pos3, respectively. The NMSE for \textbf{DM-infinite} is shown as a horizontal line since it is independent of the truncation order.}
  \label{fig:discrete_measurement_NMSE_param}
\end{figure}

As a preliminary experiment, we first evaluated the relationship between the parameter values and the estimation accuracy for \textbf{DM-finite} as well as comparison with \textbf{DM-infinite} in Fig.~\ref{fig:discrete_measurement_NMSE_param}. 
Here, in the spherical configuration with first-order microphones, the NMSEs were evaluated at $\SI{300}{Hz}$ for each $N_0$ from $0$ to $15$ at three different positions of the global origin, \ie, $\bm{r}_0 \in \{\SI{(0,0,0)}{m}, \SI{(1,0,0)}{m}, \SI{(0,1,0)}{m}\}$. 
As shown in this figure, a higher truncation order resulted in a lower NMSE, but the truncation order at which the NMSE reached its optimal value depended on the position of the global origin. 
In this example, the NMSEs for $\bm{r}_0 = \SI{(0,0,0)}{m}$ and $\bm{r}_0 = \SI{(1,0,0)}{m}$ were similar, but owing to the spherical symmetry, it is expected that the NMSE for $\bm{r}_0 = \SI{(1,0,0)}{m}$ will increase to the same level as that for $\bm{r}_0 = \SI{(0,1,0)}{m}$, depending on the sound field being estimated.
On the other hand, in \textbf{DM-infinite}, which is independent of the truncation order, the NMSEs were comparable to those in \textbf{DM-finite} with a sufficiently large $N_0$. 
From these results, $\bm{r}_0 = \SI{(0,0,0)}{m}$ and $N_0 = 7$ were set in \textbf{DM-finite} for the subsequent evaluations. 

\begin{figure}
  \centering
  \begin{minipage}{0.495\linewidth}
    \centering
    \includegraphics{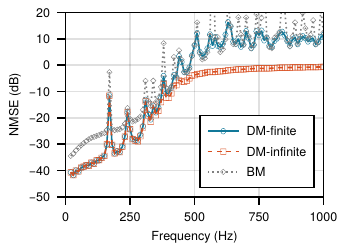}
    \subcaption{Omnidirectional}
  \end{minipage}
  \begin{minipage}{0.495\linewidth}
    \centering
    \includegraphics{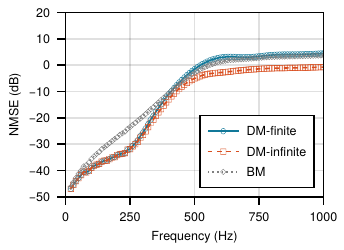}
    \subcaption{First-order ($a = 0.5$)}
  \end{minipage}
  \caption{NMSE plotted against frequency for spherical configuration.}
  \label{fig:discrete_measurement_NMSE_spherical}
\end{figure}

\begin{figure}
  \centering
  \begin{minipage}{\linewidth}
    \centering
    \begin{minipage}{0.4\linewidth}
      \centering
      \includegraphics{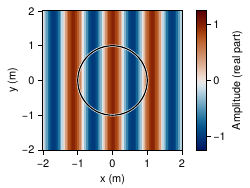}
    \end{minipage}
    \subcaption{True sound field}
    \vspace{1em}
  \end{minipage}
  \begin{minipage}{\linewidth}
    \centering
    \begin{minipage}{0.4\linewidth}
      \centering
      \includegraphics{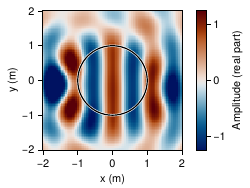}
    \end{minipage}
    \begin{minipage}{0.4\linewidth}
      \centering
      \includegraphics{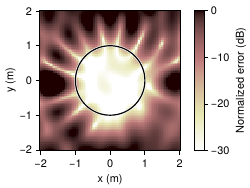}
    \end{minipage}
    \subcaption{Estimated sound field (\textbf{DM-finite})}
    \vspace{1em}
  \end{minipage}
  \begin{minipage}{\linewidth}
    \centering
    \begin{minipage}{0.4\linewidth}
      \centering
      \includegraphics{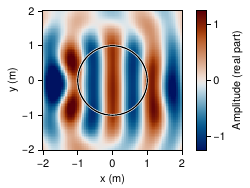}
    \end{minipage}
    \begin{minipage}{0.4\linewidth}
      \centering
      \includegraphics{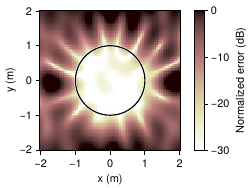}
    \end{minipage}
    \subcaption{Estimated sound field (\textbf{DM-infinite})}
    \vspace{1em}
  \end{minipage}
  \caption{Results of estimation at \SI{300}{Hz} for spherical configuration with first-order microphones: Distributions of sound pressure (in (\textbf{a}) and the left panels of (\textbf{b}) and (\textbf{c})) and normalized error (in the right panels of (\textbf{b}) and (\textbf{c})) plotted in the $xy$-plane ($\bm{r} = \SI{(x, y, 0)}{m}$). The black solid line denotes the boundary of the microphone array.}
  \label{fig:discrete_measurement_300Hz_spherical}
\end{figure}

Next, for the spherical configuration, the estimation accuracies of \textbf{DM-finite}, \textbf{DM-infinite}, and the boundary measurement method (\textbf{BM}) were compared. 
The relationship between frequency and NMSE is plotted in Fig.~\ref{fig:discrete_measurement_NMSE_spherical}. 
Here, \textbf{BM} represents the same values as in Fig.~\ref{fig:boundary_measurement_NMSE}. 
From this figure, it can be seen that \textbf{DM-infinite} achieves NMSEs that are comparable to or lower than those of the other two methods for most frequencies. 
However, the negative effect of forbidden frequencies when using omnidirectional microphones was also observed in \textbf{DM-finite} and \textbf{DM-infinite}, as well as in \textbf{BM}. 
This result supports the fact mentioned in Section~\ref{sec:boundary_forbidden_frequency} that the forbidden frequency problem is inherently due to the observation system, rather than the estimation method itself. 
By comparing \textbf{DM-finite} and \textbf{DM-infinite}, we can see that they show almost the same NMSEs at low frequencies, whereas \textbf{DM-infinite} achieves a lower NMSE at high frequencies. 
From the discussion in Section~\ref{sec:discrete_summary}, it is expected that \textbf{DM-finite} will achieve NMSEs close to those of \textbf{DM-infinite} even at high frequencies if $N_0$ is sufficiently large. 
However, it should be noted that this will result in an increased computational complexity. 
It can be said that \textbf{DM-infinite} achieves a high estimation accuracy without the need for parameter tuning to consider such trade-offs. 
Examples of the true and estimated sound fields and the pointwise normalized error distributions in the case of first-order microphones are also plotted in Fig.~\ref{fig:discrete_measurement_300Hz_spherical}, which indicates that both \textbf{DM-finite} and \textbf{DM-infinite} achieved high estimation accuracies. 

\begin{figure}
  \centering
  \begin{minipage}{0.495\linewidth}
    \centering
    \includegraphics{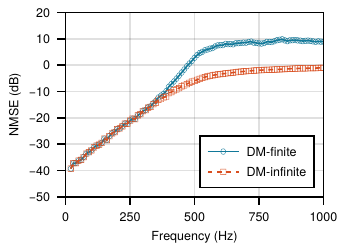}
    \subcaption{Omnidirectional}
  \end{minipage}
  \begin{minipage}{0.495\linewidth}
    \centering
    \includegraphics{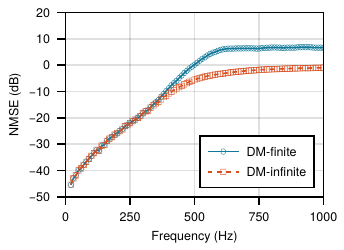}
    \subcaption{First-order ($a = 0.5$)}
  \end{minipage}
  \caption{NMSE plotted against frequency for irregular configuration.}
  \label{fig:discrete_measurement_NMSE_irregular}
\end{figure}

\begin{figure}
  \centering
  \begin{minipage}{\linewidth}
    \centering
    \begin{minipage}{0.4\linewidth}
      \centering
      \includegraphics{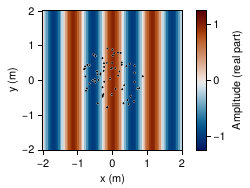}
    \end{minipage}
    \subcaption{True sound field}
    \vspace{1em}
  \end{minipage}
  \begin{minipage}{\linewidth}
    \centering
    \begin{minipage}{0.4\linewidth}
      \centering
      \includegraphics{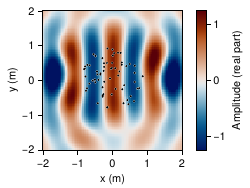}
    \end{minipage}
    \begin{minipage}{0.4\linewidth}
      \centering
      \includegraphics{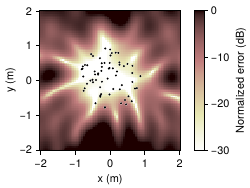}
    \end{minipage}
    \subcaption{Estimated sound field (\textbf{DM-finite})}
    \vspace{1em}
  \end{minipage}
  \begin{minipage}{\linewidth}
    \centering
    \begin{minipage}{0.4\linewidth}
      \centering
      \includegraphics{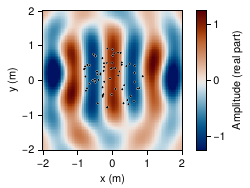}
    \end{minipage}
    \begin{minipage}{0.4\linewidth}
      \centering
      \includegraphics{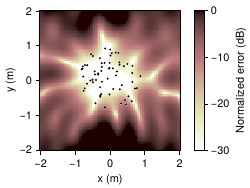}
    \end{minipage}
    \subcaption{Estimated sound field (\textbf{DM-infinite})}
    \vspace{1em}
  \end{minipage}
  \caption{Results of estimation at \SI{300}{Hz} for irregular configuration with first-order microphones: Distributions of sound pressure (in (\textbf{a}) and the left panels of (\textbf{b}) and (\textbf{c})) and normalized error (in the right panels of (\textbf{b}) and (\textbf{c})) plotted in the $xy$-plane ($\bm{r} = \SI{(x, y, 0)}{m}$). The black dots denote the positions of the microphones projected into the $xy$-plane.}
  \label{fig:discrete_measurement_300Hz_irregular}
\end{figure}

Finally, the same evaluations were also conducted for the irregular configuration.
Here, only \textbf{DM-finite} and \textbf{DM-infinite} were evaluated since \textbf{BM} cannot be applied in this configuration. 
The relationship between frequency and NMSE is plotted in Fig.~\ref{fig:discrete_measurement_NMSE_irregular}. 
It can be confirmed from this figure that both \textbf{DM-finite} and \textbf{DM-infinite} were applicable even in the irregular configuration, and the relationship between the frequency and NMSE of these two methods was similar to that observed in Fig.~\ref{fig:discrete_measurement_NMSE_spherical}. 
Examples of the true and estimated sound fields and the pointwise normalized error distributions in the case of first-order microphones are also plotted in Fig.~\ref{fig:discrete_measurement_300Hz_irregular}. 
These also support the applicability of \textbf{DM-finite} and \textbf{DM-infinite} to the irregular configuration. 
As can be seen by comparing Figs.~\ref{fig:discrete_measurement_NMSE_irregular} and \ref{fig:discrete_measurement_300Hz_irregular} to Figs.~\ref{fig:discrete_measurement_NMSE_spherical} and \ref{fig:discrete_measurement_300Hz_spherical}, respectively, the estimation accuracy was lower in the irregular configuration than in the spherical configuration, which indicates that there is room for the optimization of the microphone placement. 
However, it was confirmed that the discrete measurement approach can be flexibly applied to various microphone configurations. 

\section{Experimental Evaluation Using Real-Environment Data}
\label{sec:discrete_experiment}

There have been relatively few reports on evaluation experiments conducted in real-world environments compared to those based on simulations. 
This is because, in addition to the microphone array used for sound field estimation, denser observation data are required for the ``ground truth.'' 
Datasets of spatially dense impulse response measurements in a real-world environment are useful for performance evaluation in realistic scenarios. 
Such datasets are available in, for example, \cite{Koyama:WASPAA2021,Karakonstantis:DTU2021}. 
Here, we show a simple example using the dataset of MeshRIR reported in \cite{Koyama:IEEE2019}. 
In this experiment, the sound field generated by a single loudspeaker was calculated at each point on a grid with $\SI{0.05}{m}$ intervals within a $\SI{1}{m} \times \SI{1}{m}$ square area on the (horizontal) $xy$-plane using the measured room impulse responses. 
The source signal was a low-pass-filtered pulse signal, where the cut-off frequency was $\SI{500}{Hz}$, and the sampling rate was downsampled to $\SI{8000}{Hz}$. 
Among these $21 \times 21 = 441$ points, $18$ points were selected for measurements according to the selection algorithm proposed in \cite{Nishida:EUSIPCO2020}, and the sound field was estimated using the observed signals calculated at those selected points. 
The method described in Section~\ref{sec:discrete_infinite} was used for estimating the sound field in the frequency domain, and the estimation accuracy was evaluated in the time domain using the normalized time-averaged mean square error over the original $441$ grid points. 

\begin{figure}
  \centering
  \begin{minipage}{0.32\linewidth}
    \includegraphics[width = \linewidth]{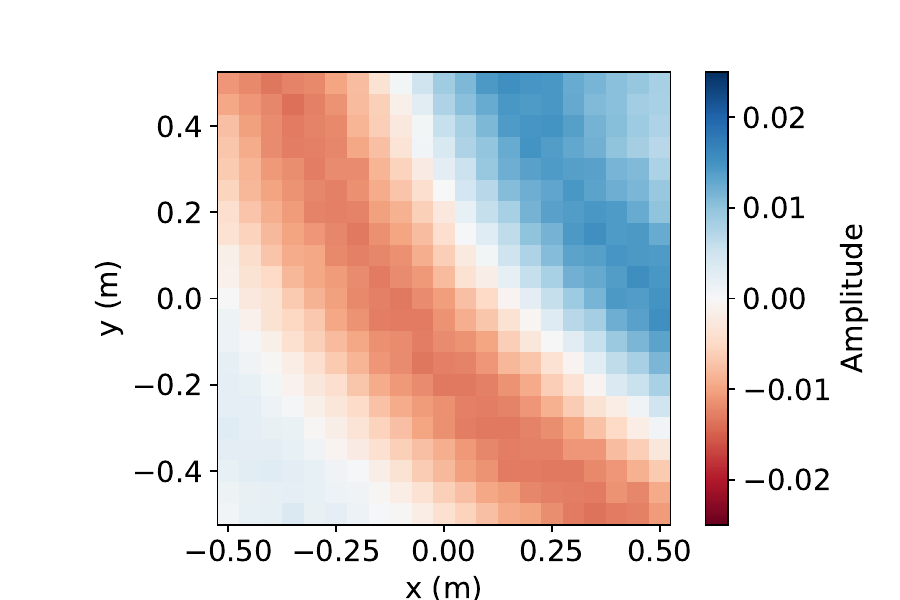}
    \subcaption{True sound field}
  \end{minipage}
  \begin{minipage}{0.32\linewidth}
    \includegraphics[width = \linewidth]{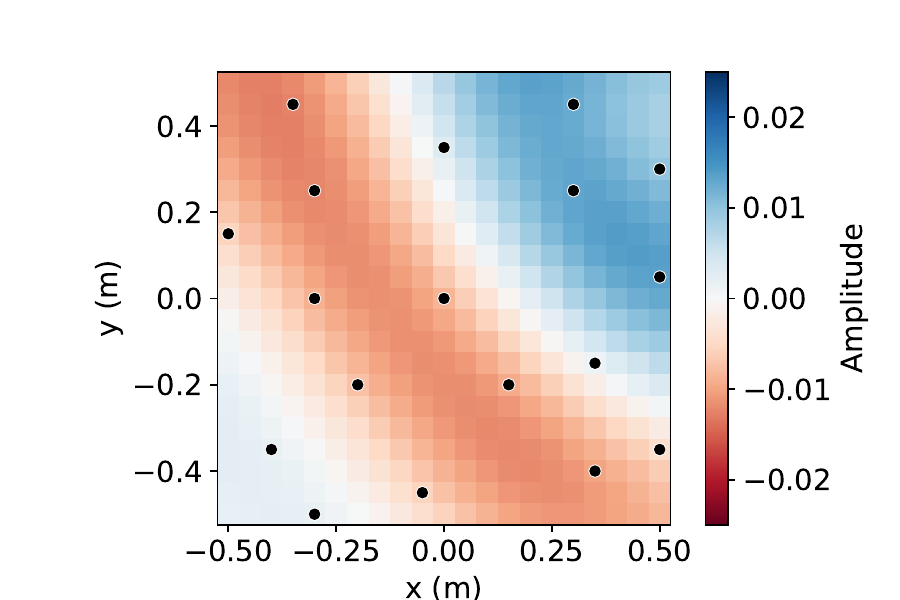}
    \subcaption{Estimated sound field}
  \end{minipage}
  \begin{minipage}{0.32\linewidth}
    \includegraphics[width = \linewidth]{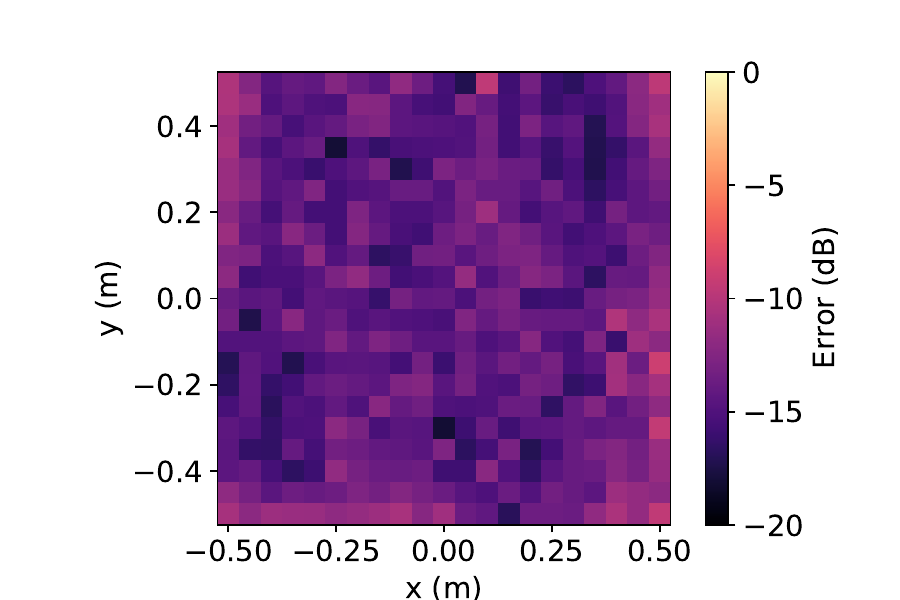}
    \subcaption{Estimation error}
  \end{minipage}
  \caption{Results of estimation using a dataset of room impulse responses: True sound field (\textbf{a}) and estimated sound field (\textbf{b}) at $t = \SI{0.089}{s}$ and pointwise normalized time-averaged square error (\textbf{c}). Black dots in (\textbf{b}) denote the measurement points used for estimating the sound field.}
  \label{fig:estimation_MeshRIR}
\end{figure}

The estimation results are shown in Fig.~\ref{fig:estimation_MeshRIR}. 
It can be observed that the sound field was well estimated with the time-averaged square error of less than $\SI{-10}{dB}$ at many points. 
The normalized time-averaged mean square error, defined as  
\begin{align}
  10 \log_{10}
  \left(
    \frac{\sum_{t \in T_\mathrm{sample}}\sum_{\mathrm{i} \in I_\mathrm{eval}} |\hat{U}(\bm{r}^{(\mathrm{i})}_\mathrm{eval}, t) - U(\bm{r}^{(\mathrm{i})}_\mathrm{eval}, t)|^2}{\sum_{t \in T_\mathrm{sample}}\sum_{\mathrm{i} \in I_\mathrm{eval}, t} |U(\bm{r}^{(\mathrm{i})}_\mathrm{eval}, t)|^2}
  \right),
\end{align}
was $\SI{-13.7}{dB}$, where $U$ and $\hat{U}$ respectively denote the true and estimated sound fields, $T_\mathrm{sample}$ is the set of downsampled times, and $\{\bm{r}^{(\mathrm{i})}_\mathrm{eval}\}_{\mathrm{i} \in I_\mathrm{eval}}$ denotes the original $441$ grid points. 
The details of this experiment are described in \cite{Koyama:WASPAA2021}, and the dataset with a reproducible code is provided at \url{https://sh01k.github.io/MeshRIR/}.

\section{Related Studies}
\label{sec:discrete_related}

Owing to its flexible formulation, many developments of the approaches described in this section have been proposed with modifications in accordance with the target context. 
One such development is the use of the prior information. 
When some prior information or assumption on a target sound field, such as the approximate source direction~\cite{Ueno:IEEE2021} or the sparsity of the source distribution~\cite{Mignot:IEEE2013, Antonello:IEEE2017, Verburg:JASA2018, Murata:IEEE2018}, is available, we can utilize this prior information in the regularization term, \ie, the second term in \eqref{eq:inverse_finite_dimension} or \eqref{eq:inverse_infinite_dimension}. 
In \cite{Ueno:IEEE2021}, the directional weighting function is incorporated into the inner product in \eqref{eq:def_inner_product_H} and the norm in \eqref{eq:def_norm_H} to impose a penalty depending on the incident direction of the sound field. 
The resulting formulation is similar to that derived in Section~\ref{sec:discrete_infinite_formulation}, but the spherical wave functions and translation operator were modified on the basis of the directional weighting function. 
In sparsity-based methods~\cite{Mignot:IEEE2013, Antonello:IEEE2017, Verburg:JASA2018, Murata:IEEE2018, Bertin:Springer2015}, on the other hand, a sound field is typically decomposed into plane wave functions or monopole functions as in \eqref{eq:model_finite_dimension}, and the sparsity criterion of their coefficients, such as the $\ell_p$ norm, is imposed as the regularization term of \eqref{eq:inverse_finite_dimension}. 
Although this type of optimization problem does not allow a closed-form solution in general, various optimization algorithms used in compressed sensing, such as the majorization--minimization algorithm~\cite{Sun:IEEE2017}, the alternating direction method of multipliers~\cite{Boyd:NOW2011}, and other proximal algorithms~\cite{Parikh:NOW2014}, are available to find the solution iteratively. 

Another development is a more accurate modeling of the observation process. 
As described in Section~\ref{sec:boundary}, a microphone array mounted on a rigid surface (i.e., a sound-hard scatterer) is effective in the stable estimation of sound fields. 
However, when multiple such arrays are used, especially at a close distance, inter-array effects, often called \emph{multiple scattering}, occurs as well as single scattering. 
In \cite{Nakanishi:WASPAA2019,Kaneko:JASA2021}, the sound field estimation problem was formulated on the basis of the observation process modeled with multiple scattering taken into account, which achieves a higher estimation performance than a method considering only single scattering. 

Compared to the boundary-measurement approach, direct application of the discrete measurement approach to the estimation of exterior sound fields. 
This is because the design of an appropriate norm (\ie, the regularization term) for exterior sound fields is a non-trivial problem. 
One example of addressing this issue is the regularization proposed by Duraiswami et al.~\cite{Duraiswami:ICASSP2004}, where high-order components are penalized more than low-order ones to prevent excessive unsmoothness in the estimated sound field.

\chapter{Applications}
\label{sec:applications}

Sound field estimation is applicable to various techniques related to spatial sound processing. Among them, we introduce binaural reproduction, sound field synthesis, and spatial active noise control (ANC). 

\section{Binaural Reproduction}

In VR audio systems, it is essential to reproduce spatial sound for listeners. Binaural reproduction techniques reproduce sounds received in both the listener's left and right ears generally through headphones. Binaural signals are typically synthesized using head-related transfer functions (HRTFs), that is, transfer functions from sources to the eardrums in a free field. An HRTF includes the effects of diffraction and reflection caused by the listener's head, torso, and external pinna shape, which contain important features for sound localization in humans. 

When synthesizing a virtual sound space using the given virtual source positions, the binaural signals in a free field can be synthesized by directly applying HRTFs to source signals. On the other hand, synthesizing binaural signals from signals captured by microphones in a real (possibly reverberant) environment is not as simple as that. In this scenario, the synthesized signals should correspond to the signals received in both ears when the listener is present in the recording area. It is also preferable that the binaural signals vary in response to the listener's head movement. Therefore, the sound field estimation methods are necessary for capturing the sound field to reproduce the binaural signals.

For binaural reproduction, a spherical microphone array mounted on an acoustically rigid object is typically used. Since binaural signals are usually reproduced from the expansion coefficients of spherical wave functions at the listener's position, which also makes it possible to adapt the reproduced signals to the listener's head rotation, the spherical microphone array is suitable for estimating them at a specific position. However, to make it possible to adapt the reproduced signals to the listener's head translation, the sound field in a large region must be accurately estimated. Since the size of the reproduced region and the reproducible frequency range are determined by the number of microphones and array size, the use of a large spherical rigid array is impractical in many cases. 

To reproduce binaural signals in a large region, it will be efficient to use distributed microphones or small microphone arrays, such as ambisonic microphones. Moreover, such distributed arrays will have flexibility and scalability in the array placement. The sound field estimation methods introduced in Section~\ref{sec:discrete} are suitable for this purpose. We introduce a method of binaural reproduction from microphone array signals proposed in \cite{Iijima:JASA2021}.

\subsection{Problem formulation}

Our objective is to reproduce the binaural signals received by a listener at a remote location by using the measurements obtained by multiple microphones in a recording area. By using the listener's HRTFs (or their proxies), we can synthesize the binaural signals such that they are identical to those received when the listener is present in the recording area. The reproduced binaural signals should vary depending on the listener's head position and direction. To reproduce the spatial sound including reverberations, it is necessary to capture the sound field inside a target listening region by using multiple microphones. Then, the binaural signals are synthesized using HRTFs. Therefore, this binaural reproduction problem is twofold: sound field estimation by multiple microphones and binaural rendering using the estimated sound field and given HRTFs. 

\subsection{Estimation of expansion coefficients of spherical wave functions using multiple microphones}

The method for estimating the expansion coefficients of spherical wave functions from multiple distributed microphones is described in Section~\ref{eq:discrete_infinite_coefficient}. The expansion coefficients of the estimated sound field $\hat{u} \in \mathcal{H}(\Omega, k)$ around $\bm{r}\in\Omega$, denoted by $\ring{\hat{u}}_{\nu,\mu}(\bm{r})$, are obtained from the observation $\bm{s}$ in the frequency domain as
\begin{align}
  \ring{\hat{u}}_{\nu,\mu}(\bm{r}_0) 
  = 
  \sum_{m=1}^M 
  \hat{\alpha}_m 
  \ring{v}_{\nu,\mu}(\bm{r})
\end{align}
with $\hat{\bm{\alpha}} = [\hat{\alpha}_1, \ldots, \hat{\alpha}_M]^\mathsf{T} \in \mathbb{C}^M$ defined as 
\begin{align}
  \hat{\bm{\alpha}} 
  = 
  (\bm{K} + \lambda \bm{I})^{-1} \bm{s}, 
\end{align}
where $\ring{v}_{\nu,\mu}(\bm{r})$ and $\bm{K}$ are defined in \eqref{eq:v_m_coefficient} and \eqref{eq:K_closed_form}, respectively. 

\subsection{Binaural rendering based on spherical wave decomposition}

The estimated expansion coefficients, which are also referred to as ambisonic coefficients, can be used to reproduce the spatial sound using loudspeakers or headphones. We here consider reproducing the binaural signals with the estimated $\ring{\hat{u}}_{\nu,\mu}(\bm{r})$ and given HRTFs $\{h_{\mathrm{L,R}}(\bm{r}_{\mathrm{s},j})\}_{j=1}^J$, when the listener is present in the recording area. We also define the spherical harmonic expansion of $\{h_{\mathrm{L,R}}(\bm{r}_{\mathrm{s},j})\}_{j=1}^J$, $\{H_{\mathrm{L,R},\nu,\mu}\}_{\nu,\mu}$, as
\begin{align}
  h_{\mathrm{L,R}}(\bm{r}_{\mathrm{s},j}) = \sum_{\nu,\mu} H_{\mathrm{L,R},\nu,\mu} \hat{\sphharm}_{\nu,\mu}(\theta_{\mathrm{s},j},\phi_{\mathrm{s},j}).
\end{align}

We assume that the HRTFs are transfer functions from point sources on a spherical surface $\partial D$ with a radius of $R_\mathrm{s} \in (0, \infty)$ centered at the origin to the listener's ears. Therefore, the sound field $u(\bm{r})$ is represented as the weighted integral of spherical waves from a point source on $\partial D$ as
\begin{align}
  u(\bm{r}) = \int_{\bm{r}_{\mathrm{s}} \in \partial D} w_{\mathrm{sph}}(\bm{r}_{\mathrm{s}}) G(\bm{r}; \bm{r}_{\mathrm{s}}) \diff \surf,
  \label{eq-ch6-1:sw_exp}
\end{align}
where $G(\bm{r}; \bm{r}_{\mathrm{s}})$ is the transfer function of the point source, which is equivalent to the free-field Green's function defined as
\begin{align}
  G(\bm{r}; \bm{r}_{\mathrm{s}}) &= 
  \frac{\exp \left( \im \wavenumber \|\bm{r}-\bm{r}_{\mathrm{s}}\| \right)}{4\pi \|\bm{r}-\bm{r}_{\mathrm{s}}\|} \notag\\
  &= \sum_{\nu,\mu} \frac{\im \wavenumber}{4\pi} j_{\nu}(\wavenumber \|\bm{r}\|) h_{\nu}(\wavenumber R_{\mathrm{s}}) \hat{\sphharm}_{\nu,\mu}(\bm{r}/\|\bm{r}\|) \hat{\sphharm}_{\nu,\mu}(\bm{r}_{\mathrm{s}}/\|\bm{r}_{\mathrm{s}}\|)^{\ast}.
  \label{eq-ch6-1:green_mono}
\end{align}
The spherical wave weight $w_{\mathrm{sph}}$ can be related to the expansion coefficients $\ring{\hat{u}}_{\nu,\mu}(\bm{r})$ as
\begin{align}
  w_{\mathrm{sph}}(\bm{r}_{\mathrm{s}}) = 
  \sum_{\nu,\mu} \frac{1}{\im^{\nu+1} \wavenumber R_{\mathrm{s}}^2 h_{\nu}(\wavenumber R_{\mathrm{s}})} \ring{\hat{u}}_{\nu,\mu}(\bm{r}) \hat{\sphharm}_{\nu,\mu}(\bm{r}_{\mathrm{s}}/\|\bm{r}_{\mathrm{s}}\|). 
  \label{eq-ch6-1:sw_weight}
\end{align}
This relation can be confirmed by substituting \eqref{eq-ch6-1:green_mono} and \eqref{eq-ch6-1:sw_weight} into \eqref{eq-ch6-1:sw_exp} and using the orthogonality of the spherical harmonic function. 

Finally, the binaural signals at $\bm{r}$, $y_{\mathrm{L,R}}(\bm{r})$, are obtained using $\ring{\hat{u}}_{\nu,\mu}(\bm{r})$ and $H_{\mathrm{L,R},\nu,\mu}$ as 
\begin{align}
  y_{\mathrm{L,R}}(\bm{r}) &= \int_{\bm{r}_{\mathrm{s}} \in \partial D} w_{\mathrm{sph}}(\bm{r}_{\mathrm{s}}) h_{\mathrm{L,R}}(\bm{r}_{\mathrm{s}}) \diff \surf \notag\\
  &= \sum_{\nu,\mu} \frac{4\pi}{\im^{\nu + 1} \wavenumber h_{\nu}(\wavenumber R_{\mathrm{s}})} \ring{\hat{u}}_{\nu,\mu}(\bm{r}) H_{\mathrm{L,R},\nu,\mu}. 
\end{align}
The infinite sum is truncated up to a finite order in practice. 

The adaptation to the listener's head translation and rotation is achieved by transforming the estimated expansion coefficients by the translation operator and Wigner-$D$ matrix given in Section~\ref{eq:preliminary_translation_rotation}. 

\subsection{Experiments}

One of the benefits of the method based on infinite-dimensional analysis is its applicability to a wide range of microphone array geometries. Therefore, we consider using multiple small microphone arrays. Such a composite array system will have flexibility in the array placement and scalability for adding/removing small microphone arrays, depending on the setting of the region of interest. We evaluate the method described in this section by numerical simulation. 

\begin{figure}
  \centering
  \includegraphics[width=0.5\linewidth]{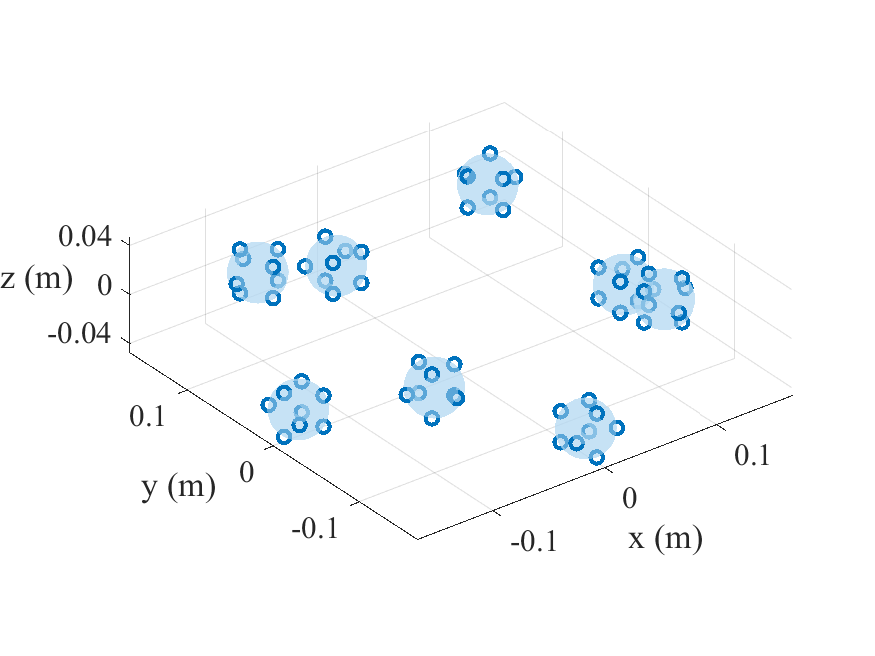} 
  \caption{Composite microphone array}
  \label{fig-ch6-1:micarray}
\end{figure}

For each small microphone array, we use one array of eight unidirectional microphones, whose arrangement is identical to that of commercially available second-order ambisonic microphones (the shape of a tetragonal trapezohedron). The composite microphone array consists of eight small microphone arrays as shown in Fig.~\ref{fig-ch6-1:micarray}, so the total number of microphones is 64. To cover the listening area, which should be larger than the average size of the listener's head, four small arrays were equiangularly placed on the circle with a radius of $\SI{0.145}{m}$ at two heights $z=\SI{\pm0.025}{m}$.

We evaluate the binaural reproduction accuracy using the composite microphone array when a single point source is positioned at $\bm{r}_{\mathrm{ps}}=(r_{\mathrm{ps}},\pi/2,\phi_{\mathrm{ps}})$ in the free field. HRTFs from the sources at $\bm{r}_{\mathrm{ps}}$ and $\{\bm{r}_{\mathrm{s},j}\}_{j=1}^J$ are obtained by numerical computation using Mesh2HRTF~\cite{Ziegelwanger:ICSV2015,Ziegelwanger:JASA2015}. The evaluation measure is the NMSE defined as
\begin{align}
  &\mathrm{NMSE}(r_{\mathrm{ps}},\phi_{\mathrm{ps}},\omega_f) = \notag\\
  &\quad \quad 10 \log_{10} \frac{|y_{\mathrm{L}}^{\mathrm{est}}(\omega_f;r_{\mathrm{ps}},\phi_{\mathrm{ps}}) - y_{\mathrm{L}}^{\mathrm{true}}(\omega_f;r_{\mathrm{ps}},\phi_{\mathrm{ps}})|^2}{|y_{\mathrm{L}}^{\mathrm{true}}(\omega_f;r_{\mathrm{ps}},\phi_{\mathrm{ps}})|^2},
  \label{eq-ch6-1:nmse}
\end{align}
where $y_{\mathrm{L}}^{\mathrm{est}}$ and $y_{\mathrm{L}}^{\mathrm{true}}$ are the estimated and true binaural signals, respectively, and $\omega_f$ is the $f$th angular frequency. 

\begin{figure}
  \centering
  \begin{minipage}{0.4\linewidth}
    \includegraphics[width=1.0\linewidth]{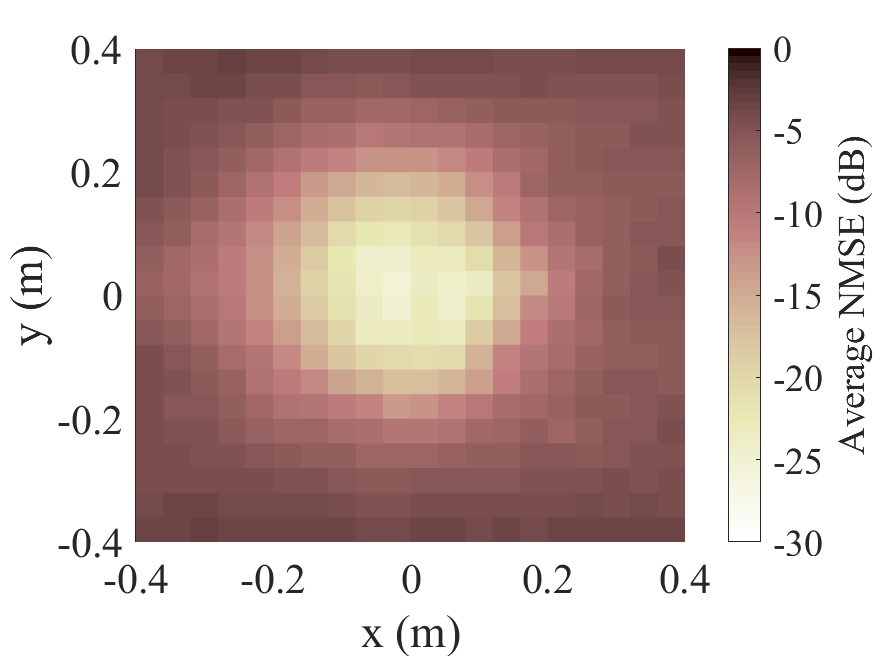} 
    \subcaption{$xy$-plane}
  \end{minipage}
  \begin{minipage}{0.4\linewidth}
    \includegraphics[width=1.0\linewidth]{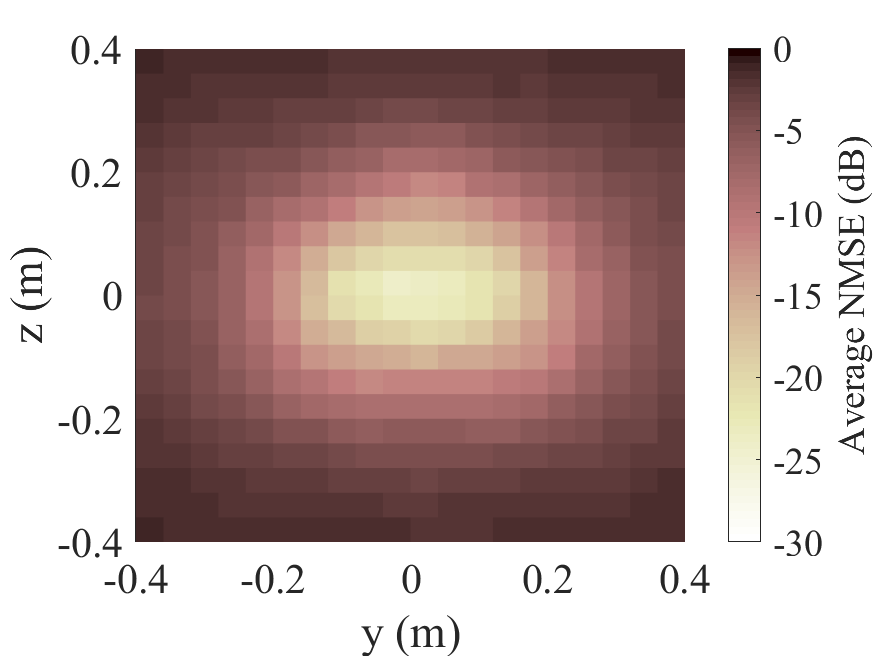}
    \subcaption{$yz$-plane} 
  \end{minipage}
  \caption{Distribution of average NMSE. The figure is taken from \cite{Iijima:JASA2021}.}
  \label{fig-ch6-1:NMSE_dist}
\end{figure}

The point source is set at $\SI{(1.5, 0.0, 0.0)}{m}$ in the recording area. We generated binaural signals at the origin using the observed signals of the microphone arrays, changing their center positions. The center positions of the microphone array were investigated for $21 \times 21$ grid positions inside the square region of $\SI{0.8}{m} \times \SI{0.8}{m}$, and 60 directions were sampled at $\theta=\pi/2$ from the direction of HRTFs. The NMSEs defined in \eqref{eq-ch6-1:nmse} are averaged for directions and frequencies of up to $\SI{1.6}{kHz}$ and plotted for each position in Fig.~\ref{fig-ch6-1:NMSE_dist}. The NMSEs are particularly low in the region where the microphones are placed. Further experimental evaluations including listening tests can be found in \cite{Iijima:JASA2021}.

\section{Sound Field Synthesis}

Sound field synthesis is aimed at synthesizing a desired spatial sound using multiple loudspeakers. Techniques of sound field synthesis can be applied to VR audio systems that do not require the wearing of headphones and personal audio systems generating multiple different sound zones. 

There are two major categories of sound field synthesis methods. One category includes analytical methods based on the boundary integral representations derived from the Helmholtz equation, e.g., \textit{wave field synthesis} and \textit{higher-order ambisonics}~\cite{Berkhout:JASA1993,Spors:AESconv2008,Ahrens:Acustica2008,Wu:IEEE2009,Koyama:IEEE_J_ASLP2013}. The other category includes numerical methods based on the minimization of a certain cost function defined for the synthesized and desired sound fields inside a target region, e.g., \textit{pressure matching} and \textit{mode matching}~\cite{Kirkeby:JASA1993,Daniel:AESConvention2003,Poletti:JAES2005}. Many analytical methods require the array geometry of loudspeakers to have a simple shape, such as a sphere, plane, circle, or line, and driving signals are obtained from a discrete approximation of an integral equation. The numerical methods allow the use of a loudspeaker array of arbitrary geometry, and the driving signals are generally derived as a closed-form least-squares solution. Since the region in which the loudspeakers can be placed is limited in practical situations, a flexible loudspeaker array geometry in numerical methods will be preferable. 

Among the numerical methods, pressure matching is widely used because of its simplicity of implementation, which is based on synthesizing the desired pressures at a discrete set of control points placed over the target region. However, the region between the control points is not taken into consideration because of discrete approximation. Therefore, its reproduction accuracy can deteriorate when the distribution of the control points is not sufficiently dense. On the other hand, a smaller number of control points is better in practice because the transfer functions between the loudspeakers and the control points are generally measured in advance. We introduce \textit{weighted pressure matching} proposed in \cite{Koyama:JAES2023}, which is a combination of pressure matching and sound field estimation techniques to evaluate the synthesis error in the continuous target region. 

\subsection{Problem formulation}

\begin{figure}
  \centering
  \includegraphics[width=0.5\linewidth]{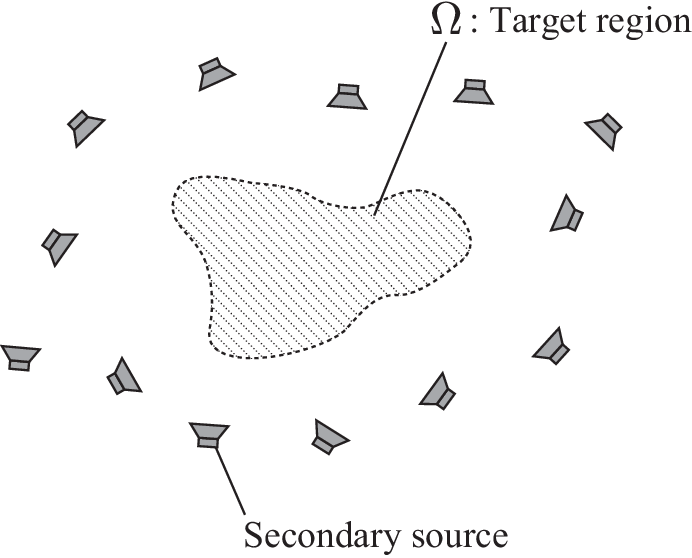} 
  \caption{Sound field synthesis}
  \label{fig-ch6-2:sfc}
\end{figure}

Suppose that $L$ secondary sources (i.e., loudspeakers) are placed around a target region $\Omega \subset \mathbb{R}^3$, as shown in Fig~\ref{fig-ch6-2:sfc}. The sound field $u_{\mathrm{syn}}(\bm{r},\omega)$ at the position $\bm{r}\in\mathbb{R}^3$ and the angular frequency $\omega \in \mathbb{R}$ synthesized using the secondary sources is represented as 
\begin{align}
  u_{\mathrm{syn}}(\bm{r}) = \sum_{l=1}^L d_l g_l(\bm{r}),
\end{align}
where $d_l$ is the driving signal of the $l$th secondary source ($l\in\{1,\ldots,L\}$) and $g_l(\bm{r})$ is the transfer function from the $l$th secondary source to the position $\bm{r}$. The transfer functions $\{g_l(\bm{r})\}_{l=1}^L$ are assumed to be known by measuring or modeling them in advance. 

The goal of sound field synthesis is to obtain the $\{d_l\}_{l=1}^L$ of the $L$ secondary sources so that $u_{\mathrm{syn}}(\bm{r})$ coincides with the desired sound field, denoted by $u_{\mathrm{des}}(\bm{r})$, inside $\Omega$. We define the cost function to determine the driving signal $\{d_l\}_{l=1}^L$ as
\begin{align}
  J = \int_{\bm{r} \in \Omega} \left| \sum_{l=1}^L d_l g_l(\bm{r}) - u_{\mathrm{des}}(\bm{r}) \right|^2 \mathrm{d} \vol.
  \label{eq-ch6-2:cost}
\end{align}
The optimal driving signal can be obtained by solving the minimization problem of $J
$; however, it is necessary to approximate $J$ to make the minimization problem including the regional integration solvable. 

\subsection{Pressure matching}

Pressure matching is one of the simplest strategies for approximately solving the minimization problem of $J$. First, the target region $\Omega$ is discretized into $N$ control points whose positions are denoted by $\bm{r}_{\mathrm{c},n}$ ($n\in\{1,\ldots,N\}$). The cost function $J$ is approximated as the error between the synthesized and desired pressures at the control points. The optimization problem of pressure matching is written as
\begin{align}
  \minimize_{\bm{d}\in\mathbb{C}^L} \ \| \bm{Gd} - \bm{u}^{\mathrm{des}} \|_2^2 + \eta \|\bm{d}\|_2^2,
\label{eq-ch6-2:cost_pm}
\end{align} 
where $\bm{d}=[d_1, \ldots, d_L]^{\mathsf{T}}\in\mathbb{C}^L$ is the vector of the driving signals, $\bm{u}^{\mathrm{des}}=[u_{\mathrm{des}}(\bm{r}_{\mathrm{c},1}), \ldots, u_{\mathrm{des}}(\bm{r}_{\mathrm{c},N})]^{\mathsf{T}}\in\mathbb{C}^N$ is the vector of the desired sound pressures, and $\bm{G}\in\mathbb{C}^{N \times L}$ is the matrix consisting of the transfer functions $\{g_l(\bm{r}_{\mathrm{c},n})\}_{n,l}$ between $L$ secondary sources and $N$ control points. The second term is the regularization term that prevents an excessively large value of $\bm{d}$, and $\eta\in (0, \infty)$ is a constant parameter. The closed-form solution of \eqref{eq-ch6-2:cost_pm} is obtained as
\begin{align}
  \hat{\bm{d}} = (\bm{G}^{\mathsf{H}}\bm{G} + \eta \bm{I})^{-1} \bm{G}^{\mathsf{H}} \bm{u}^{\mathrm{des}}. 
\end{align} 
  
\subsection{Weighted pressure matching}

The cost function of pressure matching is formulated such that the synthesized pressure corresponds to the desired pressure only at the control points owing to the discrete approximation. Therefore, the region between the control points is not taken into consideration. When the distribution of the control points is sufficiently dense, the sound field in the target region can be represented with a sufficient degree of accuracy by the pressure values only at the control points. However, because the transfer functions at the control points $\{g_l(\bm{r}_{\mathrm{c},n})\}_{l,n}$ are normally measured by microphones in practice, a small number of control points is preferable. To overcome this drawback of pressure matching, in weighted pressure matching, the cost function $J$ is approximated by interpolating the sound field from the pressures at the control points. On the basis of the kernel interpolation in Section~\ref{sec:discrete_infinite_kernel}, $g_l(\bm{r})$ and $u_{\mathrm{des}}(\bm{r})$ are interpolated from those at the control points as
\begin{align}
  \hat{g}_l(\bm{r}) &= \bm{\kappa}(\bm{r})^\mathsf{T} (\bm{K} + \lambda \bm{I})^{-1} \bm{g}_l \coloneqq \bm{z} (\bm{r})^{\mathsf{T}} \bm{g}_l, \\
  \hat{u}_{\mathrm{des}}(\bm{r}) &= \bm{\kappa} (\bm{r})^{\mathsf{T}} (\bm{K} + \lambda \bm{I})^{-1} \bm{u}^{\mathrm{des}} \coloneqq \bm{z} (\bm{r})^{\mathsf{T}} \bm{u}^{\mathrm{des}},
\end{align}
where $\bm{g}_l\in\mathbb{C}^L$ is the $l$th column vector of $\bm{G}$, $\bm{z}(\bm{r})$ is defined as $\bm{z}(\bm{r})^{\mathsf{T}}=\bm{\kappa}(\bm{r})^{\mathsf{T}}(\bm{K}+\lambda\bm{I})^{-1}$, and $\bm{\kappa}(\bm{r})\in\mathbb{C}^N$ and $\bm{K}\in\mathbb{C}^{N \times N}$ are respectively the vector and matrix consisting of the kernel function defined with the positions $\{\bm{r}_{\mathrm{c},n}\}_{n=1}^N$. Then, the cost function $J$ can be approximated using $\hat{\bm{g}}(\bm{r}) =[\bm{z}(\bm{r})^{\mathsf{T}}\bm{g}_1, \ldots, \bm{z}(\bm{r})^{\mathsf{T}}\bm{g}_L]^{\mathsf{T}}$ as 
\begin{align}
  J &\approx \int_{\bm{r} \in \Omega} \left| \sum_{l=1}^L d_l \hat{g}_l(\bm{r}) - \hat{u}_{\mathrm{des}}(\bm{r}) \right|^2 \diff \vol \notag\\
  &= \int_{\bm{r} \in \Omega} \left| \bm{z}(\bm{r})^{\mathsf{T}} \left( \bm{G}\bm{d} -  \bm{u}^{\mathrm{des}} \right) \right|^2 \diff \vol \notag\\
  &= \left( \bm{Gd} - \bm{u}^{\mathrm{des}} \right)^{\mathsf{H}} \bm{W} \left( \bm{Gd} - \bm{u}^{\mathrm{des}} \right),
\end{align}
where 
\begin{align}
  \bm{W} = \int_{\bm{r} \in \Omega} \bm{z}(\bm{r})^{\ast} \bm{z}(\bm{r})^{\mathsf{T}} \diff \vol.
\end{align}
Therefore, the optimal driving signal $\bm{d}$ is obtained by solving
\begin{align}
\minimize_{\bm{d}\in\mathbb{C}^L} \ \left( \bm{Gd} - \bm{u}^{\mathrm{des}} \right)^{\mathsf{H}} \bm{W} \left( \bm{Gd} - \bm{u}^{\mathrm{des}} \right) + \eta \|\bm{d}\|_2^2.
\end{align}
Again, the regularization term is added. Thus, the optimal driving signal is derived as
\begin{align}
  \hat{\bm{d}} = \left( \bm{G}^{\mathsf{H}}\bm{W}\bm{G} + \eta \bm{I}  \right)^{-1} \bm{G}^{\mathsf{H}} \bm{W} \bm{u}^{\mathrm{des}}.
\end{align}
This driving signal can be regarded as the solution of the minimization problem of the weighted mean square error between the synthesized and desired pressures at the control points. Weighted pressure matching increases the reproduction accuracy of pressure matching only by introducing the weighting matrices. Note that the matrices $\bm{W}$ can be computed only with the positions of the control points and the target region $\Omega$ by defining the kernel function using \eqref{eq:def_kernel}.
  
\subsection{Experiments}
  
We evaluate weighted pressure matching compared with pressure matching. The experiments are performed by using the impulse responses measured in a practical environment included in the MeshRIR dataset~\cite{Koyama:WASPAA2021}. 32 loudspeakers are regularly placed along the borders of two squares with dimensions of $\SI{2.0}{m} \times \SI{2.0}{m}$ at heights of $\SI{\pm 0.2}{m}$. The measurement region is a square with dimensions of $\SI{1.0}{m} \times \SI{1.0}{m}$ at $z=\SI{0.0}{m}$. The measurement region is discretized at intervals of $\SI{0.05}{m}$, and $21 \times 21$ evaluation points were obtained. 36 control points are regularly chosen from the evaluation points.

\begin{figure}
  \centering
  \begin{minipage}{0.4\linewidth}
    \includegraphics[height=0.75\linewidth]{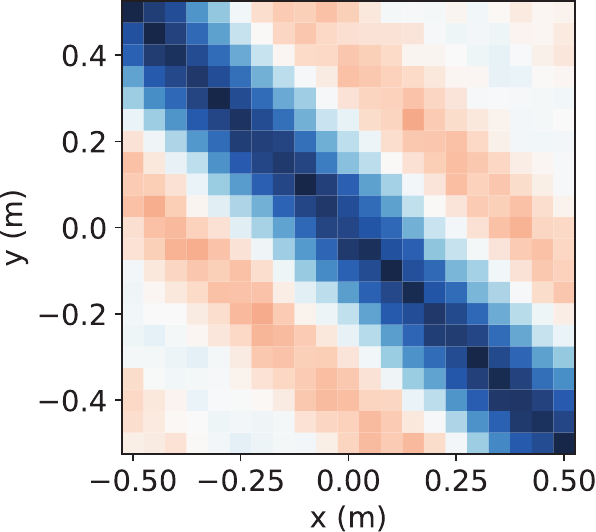}
    \subcaption{Pressure matching}
  \end{minipage}
  \begin{minipage}{0.4\linewidth}
    \includegraphics[height=0.75\linewidth]{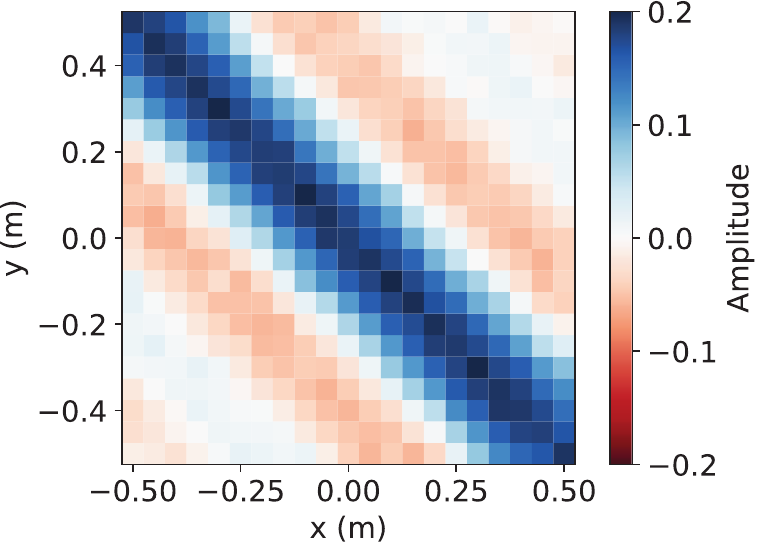}
    \subcaption{Weighted pressure matching}
  \end{minipage}
  \caption{Reproduced pressure distribution.}
  \label{fig-ch6-2:syn}
  \vspace{1.5em}
  \begin{minipage}{0.4\linewidth}
    \includegraphics[height=0.75\linewidth]{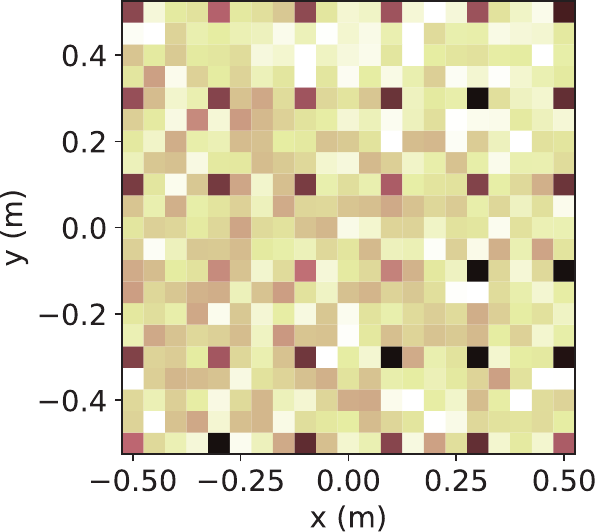}
    \subcaption{Pressure matching}
  \end{minipage}
  \begin{minipage}{0.4\linewidth}
    \includegraphics[height=0.75\linewidth]{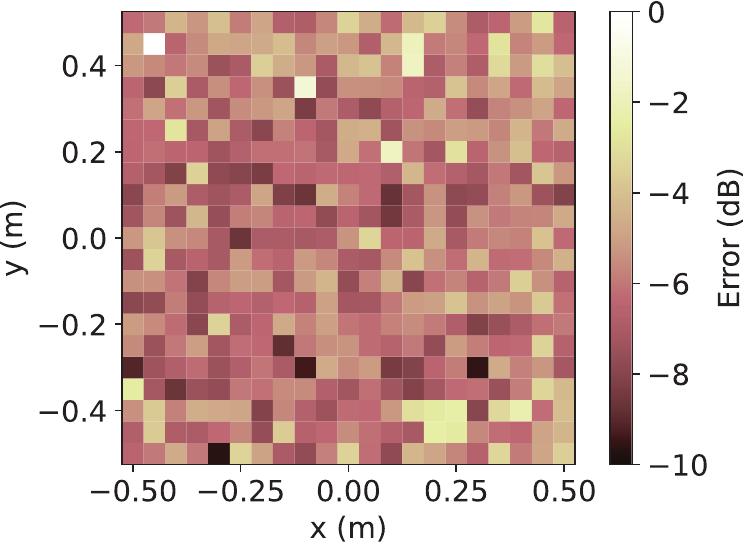}
    \subcaption{Weighted pressure matching}
  \end{minipage}
  \caption{Time-averaged square error distribution.}
  \label{fig-ch6-2:error}
\end{figure}

Figure~\ref{fig-ch6-2:syn} shows the pressure distribution reproduced by the two methods when the desired sound field is set to a plane wave propagating to the direction of $(\pi/2, -\pi/4)$. The source signal is a pulse signal whose frequency band is low-pass-filtered up to $\SI{900}{Hz}$. Figure~\ref{fig-ch6-2:error} is the time-averaged square error distribution. In pressure matching, a small error is observed at the positions of the control points, but the region between them contains large errors. The error of weighted pressure matching is small over the target region. Additional evaluation results can be found in \cite{Koyama:JAES2023}. 

\section{Spatial Active Noise Control}

ANC is aimed at suppressing unwanted noise by generating antinoise with secondary sources, i.e., loudspeakers. Achieving effective ANC over a three-dimensional space generally requires multiple microphones and loudspeakers to capture and synthesize a three-dimensional sound field. Adaptive filters are typically used to minimize the cost function, which in conventional multipoint pressure control is defined as the power of the residual noise signals captured by error microphones~\cite{Kuo:ProcIEEE1999,Kajikawa:APSIPA_J_2012,Elliott:IEEE1987}. Thus, the multipoint-pressure-control-based methods primarily reduce noise at error microphone positions. Therefore, the noise field between the error microphone positions is not taken into consideration.  

On the other hand, spatial ANC aims to control noise over a continuous three-dimensional target region~\cite{Zhang:IEEE2018,Maeno:IEEE2020}. We introduce a spatial ANC approach based on the kernel interpolation of a sound field presented in Section~\ref{sec:discrete_infinite_kernel}. As opposed to the conventional multipoint pressure control, the cost function is defined as the acoustic potential energy inside the target region. To minimize this cost function with an adaptive filter, it is necessary to estimate the noise field from discrete microphone measurements. We apply the kernel interpolation described in Section~\ref{sec:discrete}. We also derive a time-domain adaptive filtering algorithm, which is an extension of the well-known filtered-x least mean squares (FxLMS) algorithm~\cite{Kuo:ProcIEEE1999} for minimizing the regional noise power. 

\subsection{Problem formulation}

\begin{figure}
  \centering
  \includegraphics[width=0.8\linewidth]{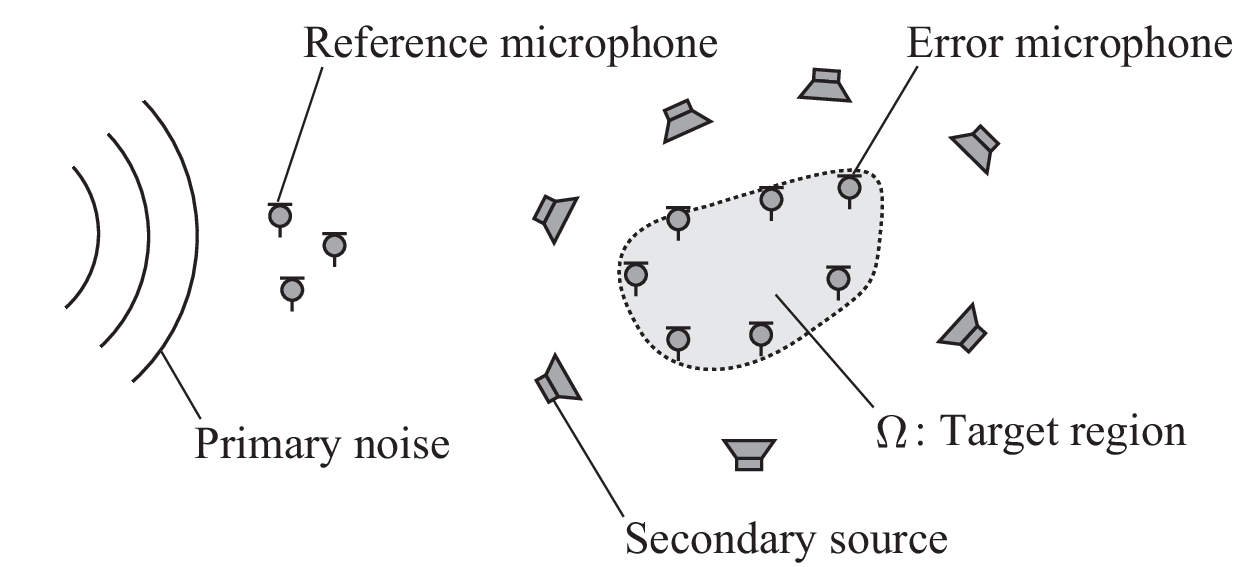} 
  \caption{Spatial active noise control}
  \label{fig-ch6-3:spatial_anc}
\end{figure}

Suppose that the target control region $\Omega \subset \mathbb{R}^3$ is set and undesired noise arrives from outside of $\Omega$. The objective of spatial ANC is to reduce incoming noise over the region $\Omega$ by controlling the output of the loudspeakers (secondary sources) placed around $\Omega$. The incoming noise is called the primary noise. As shown in Fig.~\ref{fig-ch6-3:spatial_anc}, error microphones are placed inside $\Omega$ to measure the error of the adaptation process. Although they can be arbitrarily placed inside $\Omega$, it is usually desirable that the error microphones are placed near the boundary of $\Omega$ to save space for ANC users. The secondary sources placed around $\Omega$ are used to generate an antinoise field of the primary noise from the signals captured by the reference microphones. 

The signals of the reference microphones and secondary sources in the discrete time domain are denoted by $\bm{x}(n)\in\mathbb{R}^R$ and $\bm{y}(n)\in\mathbb{R}^L$ with the time index $n$, where the numbers of reference microphones and secondary sources are $R$ and $L$, respectively. The error microphone signals and primary noise at the error microphone positions are denoted by $\bm{e}(n)$ and $\bm{d}(n)$ ($\in\mathbb{R}^M$), respectively. The number of error microphones is $M$. The total pressure field at the position $\bm{r}\in\Omega$ and time $n$ is denoted by $u(\bm{r},n)$. The pressure field at the position of the $m$th microphone $\bm{r}_m$ is equal to the $m$th element of $\bm{e}(n)$, meaning that $e_m(n)=u(\bm{r}_m,n)$. The signals in the frequency domain are distinguished by the angular frequency $\omega$ as an argument, for example, $\bm{x}(\omega)$ ($\in\mathbb{C}^R$).

The impulse response between the secondary sources and the error microphones (secondary paths) is represented by a FIR filter $\bm{G}(i)\in\mathbb{R}^{M \times L}$ of length $J$, where each $i$ indicates a matrix of filter coefficients. The error microphone signal $\bm{e}(n)$ can be expressed as a superposition of the primary noise and the secondary source signals as 
\begin{align}
    \bm{e}(n) = \bm{d}(n) + \sum_{i=0}^{J-1} \bm{G}(i) \bm{y}(n-i).
    \label{eq-ch6-3:error-time}
\end{align}
We assume that the true secondary paths are known. The control filter is a FIR filter $\bm{W}(i)\in\mathbb{R}^{L \times R}$ of length $I$. The driving signals of the secondary sources, $\bm{y}(n)$, are obtained by filtering the reference signal $\bm{x}(n)$ with the control FIR filter $\bm{W}(i)$ as
\begin{align}
    \bm{y}(n) = \sum_{i=0}^{I-1} \bm{W}(i) \bm{x}(n-i).
    \label{eq-ch6-3:drv-time}
\end{align}
It is assumed that the effect of the secondary source signals is negligible at the reference microphone positions. 

The above-described ANC setup is a typical multichannel feedforward ANC system. However, the adaptive filtering algorithm is derived with the aim of reducing the primary noise at the error microphone positions. Thus, the cost function for updating the control filter $\bm{W}(i)$ is typically formulated as the power of the error microphone signals as
\begin{align}
\mathcal{J} = \mathbb{E}\left[ \| \bm{e}(n) \|_2^2 \right],
\label{eq-ch6-3:cost_mpc}
\end{align}
where $\mathbb{E}[\cdot]$ represents the expectation with respect to the time. However, the noise power in the region between the error microphones is not taken into consideration in this cost function. 

The objective of spatial ANC is to reduce the power of $u(\bm{r},n)$ over the region $\Omega$. Therefore, the cost function is formulated as the acoustic potential energy inside $\Omega$, 
\begin{align}
\mathcal{L} = \mathbb{E}\left[ \int_{\bm{r} \in \Omega} u(\bm{r},n)^2 \diff \vol 
\right].
\label{eq-ch6-3:cost}  
\end{align}
Since it is not possible to directly obtain the continuous distribution $u(\bm{r},n)$, it is necessary to derive an adaptive filtering algorithm for minimizing $\mathcal{L}$ by predicting $u(\bm{r},n)$ from the error microphone signals $\bm{e}(n)$. 

\subsection{Kernel-interpolation-based spatial ANC}

\subsubsection{Formulation in frequency domain}

The relationships given in \eqref{eq-ch6-3:error-time} and \eqref{eq-ch6-3:drv-time} are respectively represented in the frequency domain as 
\begin{align}
\bm{e}(\omega) &= \bm{d}(\omega) + \bm{G}(\omega) \bm{y}(\omega), \\
\bm{y}(\omega) &= \bm{W}(\omega) \bm{x}(\omega).
\end{align}
The pressure distribution $u(\bm{r},\omega)$ can be estimated from the error signals $\bm{e}(\omega)$ as 
\begin{align}
u(\bm{r},\omega) &= \bm{z}(\bm{r},\omega)^{\mathsf{T}} \bm{e}(\omega), \\ 
\bm{z}(\bm{r},\omega) &= \left[ (\bm{K}(\omega) + \lambda \bm{I})^{-1} \right]^{\mathsf{T}} \bm{\kappa}(\bm{r}, \omega).
\end{align}
Here, $\bm{z}(\bm{r},\omega)$ is the interpolation filter in the frequency domain, which is obtained from the formulation of kernel ridge regression in Section~\ref{sec:discrete_infinite_kernel}. 

The cost function $\mathcal{L}$ defined in \eqref{eq-ch6-3:cost} is rewritten in the frequency domain as
\begin{align}
\bar{\mathcal{L}} &= \mathbb{E}\left[ \int_{\bm{r} \in \Omega} u(\bm{r},n)^2 \diff \vol \right] \notag\\
&= \mathbb{E}\left[ \bm{e}(\omega)^{\mathsf{H}} \bm{A}(\omega) \bm{e}(\omega) \right],
\end{align}
where
\begin{align}
\bm{A}(\omega) &= \int_{\bm{r} \in \Omega} \bm{z}(\bm{r},\omega)^{\ast} \bm{z}(\bm{r},\omega)^{\mathsf{T}} \diff \vol \notag\\
&= \bm{P}(\omega)^{\mathsf{H}} \left[ \int_{\bm{r} \in \Omega} \bm{\kappa}(\bm{r},\omega)^{\ast} \bm{\kappa}(\bm{r},\omega)^{\mathsf{T}} \diff \vol \right] \bm{P}(\omega).
\label{eq-ch6-3:weight-freq}
\end{align}
Here, $\bm{P}(\omega)=(\bm{K}(\omega) + \bm{I})^{-1}$ is defined. Thus, the cost function $\bar{\mathcal{L}}(\omega)$ is the power of the error signals $\bm{e}(\omega)$ subject to the weighting matrix $\bm{A}(\omega)$.

Therefore, the least-mean-square algorithm for updating $\bm{W}(\omega)$ is derived in the frequency domain as
\begin{align}
\bm{W}_{n+1}(\omega) &= \bm{W}_n(\omega) - \mu \frac{\partial}{\partial \bm{W}(\omega)^{\ast}} \bm{e}(\omega)^{\mathsf{H}} \bm{A}(\omega) \bm{e}(\omega) \notag\\
&= \bm{W}_n(\omega) - \mu \bm{G}(\omega)^{\mathsf{H}} \bm{A}(\omega) \bm{e}(\omega) \bm{x}(\omega)^{\mathsf{H}},
\end{align}
where the subscript $n$ represents the time index. 

\subsubsection{Adaptive filtering algorithm in time domain}

Next, we formulate the time-domain algorithm. The time-domain interpolation filter $\bm{z}(\bm{r},i)$ is obtained by the inverse discrete-time Fourier transform of $\bm{z}(\bm{r},\omega)$ as
\begin{align}
\bm{z}(\bm{r},i) = \mathrm{DFT}^{-1} \left[ \bm{z}(\bm{r},\omega) \right].
\end{align}
Then, the pressure distribution in the time domain is estimated as
\begin{align}
u(\bm{r},n) = \sum_{i=-\infty}^{\infty} \bm{z}(\bm{r},i)^{\mathsf{T}} \bm{e}(n-i).
\end{align}
Thus, the cost function \eqref{eq-ch6-3:cost} is reformulated as
\begin{align}
\mathcal{L} = \mathbb{E} \left[ \sum_{i,j=-\infty}^{\infty} \bm{e}(n-i)^{\mathsf{T}} \bm{\Gamma}(i,j) \bm{e}(n-j) \right],
\end{align}
where $\bm{\Gamma}(i,j)$ is the coefficient matrix of the interpolation filter defined as
\begin{align}
    \bm{\Gamma}(i,j) = \int_{\bm{r} \in \Omega} \bm{z}(\bm{r},i) \bm{z}(\bm{r},j)^{\mathsf{T}} \diff \vol.
\end{align}
The gradient of $\mathcal{L}$ with respect to the filter coefficients $\bm{W}(i)$, denoted as $\bm{\Delta}(i)$, is obtained as
\begin{align}
\bm{\Delta}(i) &= \frac{\partial \mathcal{L}}{\partial \bm{W}(i)} \notag\\
&= 2 \sum_{k=-\infty}^{\infty} \sum_{j=0}^{J-1} \bm{G}(j)^{\mathsf{T}} \bm{A}(k)^{\mathsf{T}} \mathbb{E} \left[ \bm{e}(n) \bm{x}(n-i-j-k)^{\mathsf{T}} \right],
\end{align}
where $\bm{A}(k)$ is the inverse discrete-time Fourier transform of $\bm{A}(\omega)$ obtained using $\bm{\Gamma}(i,j)$. Therefore, the FxLMS algorithm for $\bm{W}(i)$ is obtained as 
\begin{align}
\bm{W}_{n+1}(i) = \bm{W}_n(i) - \mu \sum_{j=0}^{J+2K-1} \bm{H}(j)^{\mathsf{T}} \bm{e}(n-K) \bm{x}(n-i-j)^{\mathsf{T}},
\end{align}
where 
\begin{align}
\bm{H}(i) = \sum_{j=0}^{2K} \bm{A}(j) \bm{G}(i-j).
\end{align}
Note that $\bm{A}(k)$ is truncated to the length $2K+1$.

\subsection{Experiments}
  
Experiments are conducted to evaluate the proposed method in comparison with the multipoint pressure control (MPC)-based ANC, which is based on the cost function \eqref{eq-ch6-3:cost_mpc}. In a 2D free field, the target region $\Omega$ is set to a square with dimensions of $1.0~\mathrm{m} \times 1.0~\mathrm{m}$ with its center at the origin. The numbers of error microphones and secondary sources are $M=24$ and $L=12$, respectively. The error microphones are regularly placed along the boundary of $\Omega$, but half of them are shifted $0.03~\mathrm{m}$ outward to alleviate the effect of the forbidden frequency problem. The secondary sources are regularly arranged along the square with dimensions of $2.0~\mathrm{m} \times 2.0~\mathrm{m}$. The reference signal is directly obtained from the primary noise source. 

\begin{figure}
    \centering
    \begin{minipage}{0.4\linewidth}
        \includegraphics[width=\linewidth]{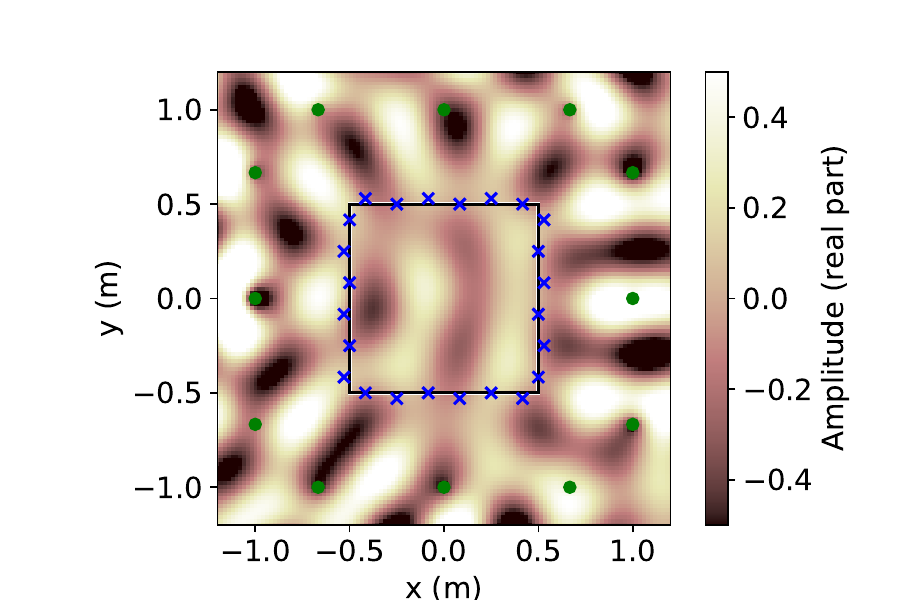}
        \subcaption{MPC}
    \end{minipage}
    \begin{minipage}{0.4\linewidth}
        \includegraphics[width=\linewidth]{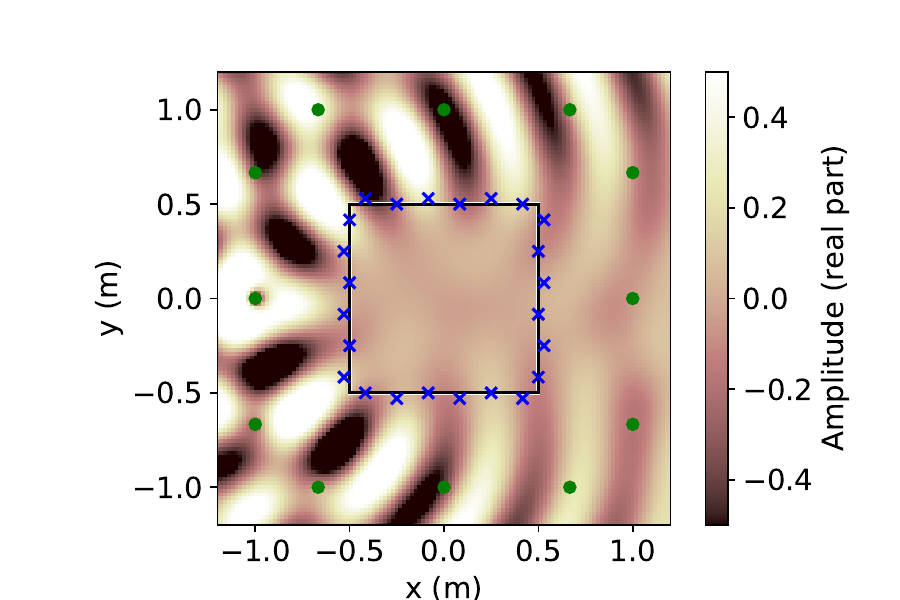}
        \subcaption{KI}
    \end{minipage}
    \caption{Pressure distribution at $700~\mathrm{Hz}$.}
    \label{fig-ch6-3:amp}
    \vspace{1.5em}
    \centering
    \begin{minipage}{0.4\linewidth}
        \includegraphics[width=\linewidth]{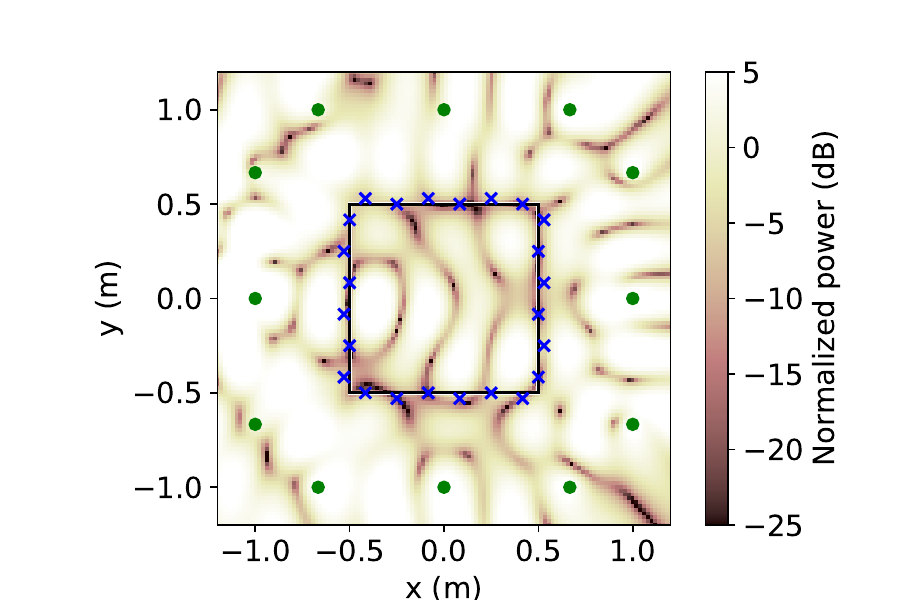}
        \subcaption{MPC}
    \end{minipage}
    \begin{minipage}{0.4\linewidth}
        \includegraphics[width=\linewidth]{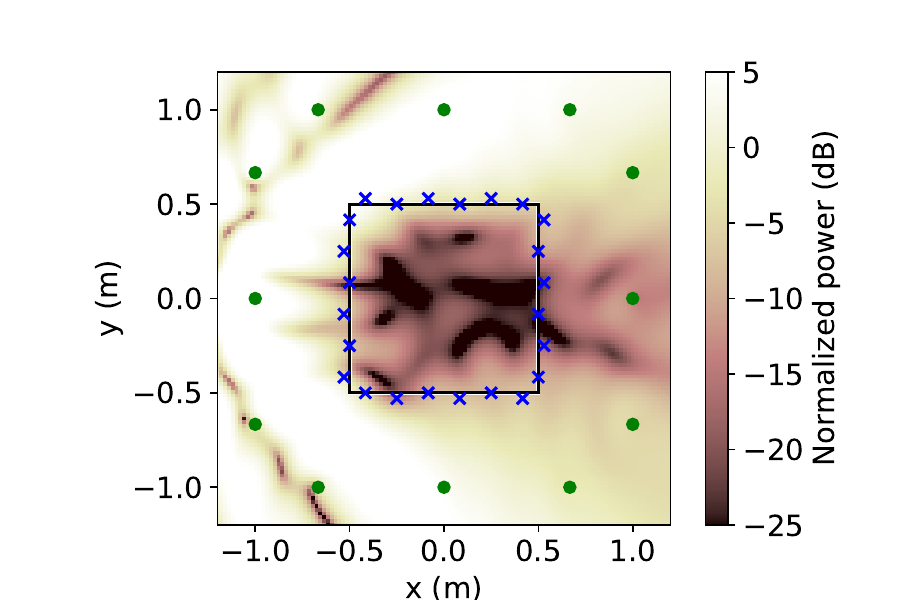}
        \subcaption{KI}
    \end{minipage}
    \caption{Normalized power distribution at $700~\mathrm{Hz}$.}
    \label{fig-ch6-3:pw}
\end{figure}

Figure~\ref{fig-ch6-3:amp} shows the pressure distribution at $700~\mathrm{Hz}$. The power distribution normalized by the average power of the primary noise field $u_{\mathrm{p}}$ inside $\Omega$, i.e., $|u(\bm{r},\omega)|^2/\int_{\bm{r} \in \Omega} |u_{\mathrm{p}}(\bm{r},\omega)|^2 \diff \vol$, is shown in Fig.~\ref{fig-ch6-3:pw}. The acoustic power inside $\Omega$ is reduced by the kernel-interpolation-based method, but remains the same or is amplified for MPC. Further experimental results including time-domain adaptive filtering can be found in \cite{Koyama:IEEE2021}.

\chapter{Conclusions}
\label{sec:conclusions}

In this paper, we provided a tutorial overview of sound field estimation methods. 
First, we defined the scope of the sound field estimation problem using mathematical expressions to clarify the goal of this paper. 
After presenting the mathematical preliminaries, we reviewed two major approaches of sound field estimation: the boundary measurement approach and the discrete measurement approach. 
Our focus was on conveying the fundamental ideas in an easy-to-understand manner while also briefly introducing advanced topics with reference to the literature. 
Finally, we described several applications of sound field estimation. 
We believe that the systematic theory presented in this paper will be valuable for devising or understanding advanced methods of sound field estimation in the future. 

\chapter*{Acknowledgments}

This work was supported by JSPS KAKENHI Grant Number JP23K24864.

\printbibliography

\end{document}